\documentclass[twocolumn]{aastex63}

\def\aj{{AJ}}
\def\apj{{ApJ}}

\def\apjl{{ApJL}}
\def\araa{{ARA\&A}}
\def\aap{{A\&A}}
\def\pasp{{PASP}}

\def\mnras{{MNRAS}}
\def\nat{{Nature}} 

\def\icarus{{Icarus}}

\def\psj{{Plan. Sci. Journal}}

\usepackage{threeparttable}

\begin{document}

\title{Mantle Degassing Lifetimes through Galactic Time and the Maximum Age Stagnant-lid Rocky Exoplanets can Support Temperate Climates}

\author[0000-0001-8991-3110]{Cayman T. Unterborn}
\affiliation{Southwest Research Institute, San Antonio, TX 78238, USA}
\email{cayman.unterborn@swri.org}

\author[0000-0002-6943-3192]{Bradford J. Foley}
\affiliation{Department of Geosciences, Pennsylvania State University,
  University Park, PA 16802, USA}
  
\author[0000-0002-1571-0836]{Steven J. Desch}
\affiliation{School of Earth and Space Exploration, Arizona State University, Tempe, AZ 85004}

\author[0000-0003-1705-5991]{Patrick A. Young}
\affiliation{School of Earth and Space Exploration, Arizona State University, Tempe, AZ 85004}

\author[0000-0002-0984-4117]{Gregory Vance}
\affiliation{School of Earth and Space Exploration, Arizona State University, Tempe, AZ 85004}
 
\author{Lee Chieffle}
\affiliation{School of Earth and Space Exploration, Arizona State University, Tempe, AZ 85004}

\author[0000-0002-7084-0529]{Stephen R. Kane}
\affiliation{Department of Earth and Planetary Sciences, University of
  California, Riverside, CA 92521, USA}

\begin{abstract}
    
    The ideal exoplanets to search for life are those within a star's habitable zone. However, even within the habitable zone planets can still develop uninhabitable climate states. Sustaining a temperate climate over geologic ($\sim$Gyr) timescales requires a planet contain sufficient internal energy to power a planetary-scale carbon cycle. A major component of a rocky planet’s energy budget is the heat produced by the decay of radioactive elements, especially $^{40}$K, $^{232}$Th, $^{235}$U and $^{238}$U. As the planet ages and these elements decay, this radiogenic energy source dwindles. Here we estimate the probability distribution of the amount of these heat producing elements (HPEs) that enter into rocky exoplanets through Galactic history, by combining the system-to-system variation seen in stellar abundance data with the results from Galactic chemical evolution models. Using these distributions, we perform Monte-Carlo thermal evolution models that maximize the mantle cooling rate. This allows us to create a pessimistic estimate of lifetime a rocky, stagnant-lid exoplanet can support a global carbon cycle and temperate climate as a function of its mass and when it in Galactic history. We apply this framework to a sample of 17 likely rocky exoplanets with measured ages, 7 of which we predict are likely to be actively degassing today despite our pessimistic assumptions. For the remaining planets, including those orbiting TRAPPIST-1, we cannot confidently assume they currently contain sufficient internal heat to support mantle degassing at a rate sufficient to sustain a global carbon cycle or temperate climate without additional tidal heating or undergoing plate tectonics.
\end{abstract}

\section{Introduction}
The climate state of an exoplanet is a primary determining factor of whether it is likely to be habitable to life as we know it. While life may manifest within subglacial oceans such as those on icy moons, surface life is more easily detected remotely. Thus, a temperate surface is often considered a first-order requirement for detectable life to develop on a planet. The likelihood of a temperate climate is typically assessed based on the stellar radiation a planet receives, particularly whether it lies within its host-star's so-called ``habitable zone'' \citep[e.g., ][]{Kasting1993, kopparapu2013}.  Lying within a star's habitable zone, however, does not guarantee that a planet will have a temperate surface suitable for liquid water or life. Indeed, to truly be considered habitable, a planet must be neither too hot to evaporate the entirety of its water nor sterilize the planet's surface \citep[e.g., ][]{Abbot2012,Foley2016_review}, or too cold to undergo global glaciation \citep{Kadoya2014,Menou2015,Haqq2016}; both extremes can develop even within the nominal habitable zone. 

Whether a planet is capable of consistently remaining in this temperate state over geologic ($\sim$Gyr) timescales relies, in part, on its ability to regulate the abundance of greenhouse gases in its atmosphere. Of the major greenhouse gases, CO$_2$ is known to be regulated by the carbonate-silicate cycle, at least on Earth. The balance between the rate of delivery of CO$_2$ to the atmosphere via melt-induced degassing of carbon in the mantle or crust, weathering and return of C in the form of carbonates to the mantle determines atmospheric CO$_2$ concentrations; negative feedbacks involved in surface weathering and degassing help to stabilize climate \citep[e.g., ][]{walker1981,Sleep2001b,Kasting2003,Foley2016_review}. However, this stabilizing feedback can fail if the input rate of CO$_2$ to the atmosphere from degassing drops too low, as weathering can draw down atmospheric CO$_2$ levels low enough for the planet to potentially fall into a snowball climate state. A planet without active degassing will lack an active carbonate-silicate cycle, and the stabilizing climate feedbacks it provides. Specifically, a lack of degassing is likely to lead to frozen snowball climates on most habitable zone planets \citep{Foley2018_stag,Foley2019_stag}, except for planets with large C budgets, where hot-house climates are likely to form instead \citep{Foley2018_stag,Foley2019_stag}. While temperate climates may be possible on water-worlds without active mantle degassing \citep{Kite18}, it is not known whether this extends to more Earth-like planets, with Earth-like levels of surface volatiles. 

How long planets can maintain mantle degassing is a critical determinate in whether a planet is potentially habitable \textit{today}, when we observe it. The age of an exoplanet is roughly the age of its host-star. If the host-star is older than the degassing lifetime of the planet in question, it is possible the planet is unable to sustain the feedbacks necessary for climate regulation. Mantle degassing is ultimately caused by the surface-to-interior interplay between mantle volcanism and surface tectonic processes, both of which are powered by the internal heat budget of the planet \citep{Foley2018_stag,Foley2019_stag}. As a planet ages, it cools as its heat budget decreases over time. Planetary heat budgets are composed from a variety of sources: secular cooling after planet formation, the gravitational potential energy released during core and mantle differentiation, the crystallization of any inner core, tidal heating induced by the host-star or other planets in the system, and the radioactive decay of the long-lived radionuclides $^{235}$U, $^{238}$U, $^{232}$Th and $^{40}$K. In all, these elements account for $\sim100$ TW, or 30--50\%, of the Earth's current surface heat flow. Due to the radioactive nature of these elements, however, this current heat flow represents only $\sim$20\% of the Earth's radiogenic heat budget when it formed $\sim$4.5 Gyr ago.

\begin{figure*}[t]  
    \centering
    \includegraphics[width=\linewidth]{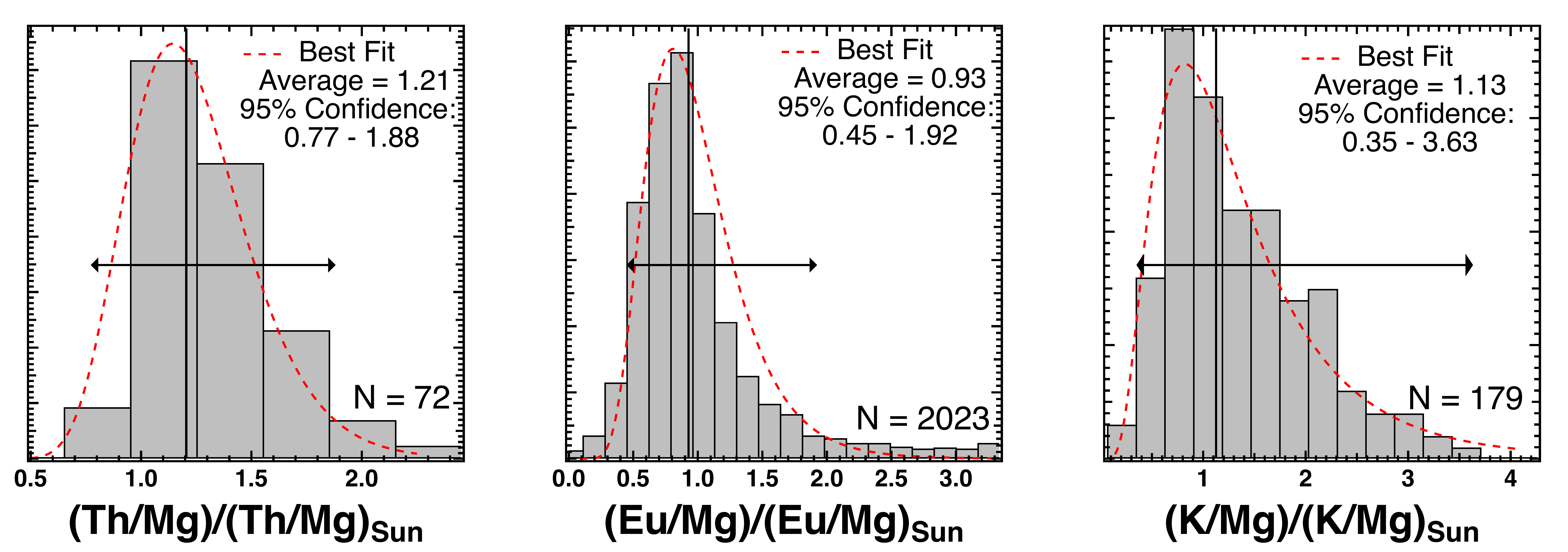}
    \caption{Histograms of measured abundances of Th (left), Eu (as a proxy for U, center) and K (right) for our samples of stars described in Appendix \ref{sec:obs_vals}. Total numbers of stars in each sample are noted. Best fits assuming a log-normal distribution are shown in red with the average value and 95\% confidence interval shown for each element in each panel.}
    \label{fig:abund_histograms}
\end{figure*}

Planetary radiogenic heat budgets are dependent only on the total abundances of the heat producing elements (HPEs) U, Th and K within their interior. Like all elements present in a planet, radionuclide concentrations are set by the abundance of these elements in the protoplanetary nebula, their subsequent fractionation relative to the primary rocky-planet elements (Mg, Si, Fe) during planet formation and the final distribution of these elements between exoplanet mantle, core and crust. \citet{Unte15} proposed that the concentration of Th in an exoplanet's host-star can provide a rough estimate for the Th concentration in an orbiting rocky planet's mantle due to its long half-life and refractory nature. \citet{Unte15} and \citet{Bote18} observed over a factor-of-two variation in Th relative to the Sun in a sample of Solar twins. Furthermore, the Hypatia catalog \citep{Hink14} shows significant variation in the abundances of Th, bulk K, and bulk U, the latter inferred from its nucleosynthetic proxy Eu, suggesting that there may be significant system-to-system variation in the abundances of the HPEs (Figure \ref{fig:abund_histograms}), meaning some exoplanets may form with a higher or lower concentration of HPEs than the Earth and Sun. Previous work has highlighted the importance of HPEs for the longevity of potentially habitable climates on stagnant-lid planets and planets with plate tectonics \citep{Foley2018_stag,Foley2019_stag,Nimmo2020,Ooster2021}. These previous studies, however, treated HPE abundance as a free parameter, rather than constrained by observations of the HPEs themselves \citep{Foley2018_stag,Foley2019_stag,Nimmo2020,Ooster2021}. In this work, we quantify the degree of variation in each of the HPEs through Galactic history and explore the effects of this range of radiogenic heat budgets on the lifetime of mantle degassing. 

\section{Estimating Mantle Degassing Lifetime}

To estimate the lifetime of mantle degassing, we utilize an updated version of the thermal evolution models of \citet{Foley2018_stag} and \citet{Foley2019_stag} (Appendix \S\ref{sec:thermal_mod}--\S\ref{sec:thermal_mod2}). We define the lifetime of mantle degassing as planet's age when its degassing rate first falls below 10\% of the Earth's present day degassing rate, scaled linearly by planet surface area \citep[$\approx 6 \times 10^{12}$ mol yr$^{-1}$, ][Appendix \ref{sec:Methods}]{Marty1998}, scaled linearly by planet surface area. Below this 10\% threshold, CO$_2$-poor snowball climates are expected to form, as at these low degassing rates a steady-state between weathering and degassing results in CO$_2$ levels too low to prevent global glaciation \citep{Kadoya2014,Haqq2016,Foley2019_stag}. To estimate the range of planetary degassing lifetimes we adopt a Monte-Carlo approach, randomly sampling within the best-fit distributions from Figure \ref{fig:abund_histograms} to determine a planet's initial budget of Th, U (as Eu) and K, accounting for the volatility in each element during planet formation and applying corrections for the production and decay of the HPEs through time from Galactic Chemical Evolution (GCE) models \citep[][Appendix \S\ref{sec:radio_abunds}--\ref{sec:obs_vals}]{Frank14}. Additionally, we randomly sample within uniform distributions of mantle reference viscosities and initial temperatures, both key geophysical parameters that affect a planet's thermal history (Table \ref{tab:params}, Appendix \S\ref{sec:final_structure}). For a given planet mass between 1--6 M$_\oplus$ and planet formation times, $t$, between the birth of the Milky Way ($t=0$ Gyr) and today ($t=12.5$ Gyr), we perform $\sim10^{5}$ thermal evolution models using these randomly determined values as inputs. Individual thermal evolution models are then run until the degassing rate falls below 10\% of the Earth's current value. 

We focus our modeling on stagnant-lid exoplanets. Stagnant-lid tectonics may be the most likely dynamic state for rocky exoplanets as the Earth is the only rocky planet we know of that exhibits plate tectonics \citep[e.g., ][]{Breuer2015} and the special conditions thought needed for plate-like mantle convection to develop \citep[e.g., ][]{Berco2015_treatise}. While plate tectonics has many advantages over stagnant-lid tectonics when it comes to mantle degassing \citep[e.g., ][]{Foley2016_review, Nimmo2020}, whether a planet is in the plate-tectonic or stagnant-lid regime is highly uncertain and difficult to predict from first principles \citep[e.g., ][]{Valencia2007b,ONeill2007,vanheck2011,Foley2012,noack2014}. Critically, however, by assuming a stagnant-lid state of tectonics, our thermal evolution models makes pessimistic assumptions regarding mantle cooling rate. Specifically, we assume all melt produced contributes to mantle cooling by release of latent heat and cooling of the hot melt at the planet's surface. This implicitly assumes that all melt produced is erupted, when in reality up to 90\% of melt may intrude and solidify at depth \citep{Crisp1984}. We also ignore heating from the cooling of the core, and thus assume planets with mantles that are entirely internally heated. The effect is to produce the fastest reasonable mantle cooling rate. That is, our models will produce the shortest reasonable degassing lifetime for the rocky planets modeled. The pessimistic nature of our models means that our most robust predictions are for planets that could still be degassing today; relaxing assumptions in our models would only act to increase degassing lifetime. This means those planets we estimate could be degassing with stagnant-lids would be even more likely to be degassing today if they instead experience plate tectonics. 

\begin{figure*}
    \centering
    \includegraphics[width=\linewidth]{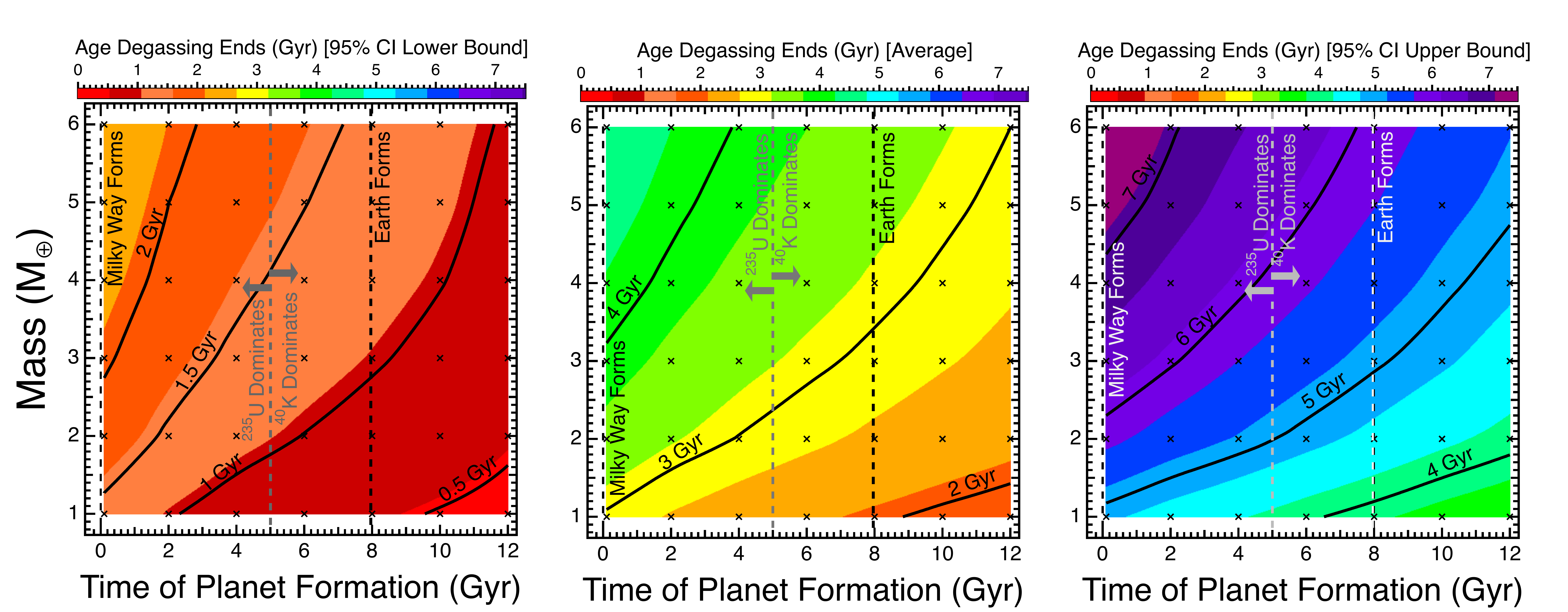}
    \caption{Interpolated color maps of average (center) and lower (left) and upper (right) 95\% confidence interval bounds on mantle degassing lifetime for stagnant-lid exoplanets as a function of planet mass and age of the system. The data interpolated are represented by crosses with averages and confidence intervals being determined by 50,000 random samplings of our parameter space (see Appendix \ref{sec:Methods}). We define the cessation of degassing as occurring when the degassing rate drops below 10\% of the Earth's current value, scaled linearly by planet surface area (see Appendix \ref{sec:Methods}). }
    \label{fig:lifetimes}
\end{figure*}

For a stagnant-lid rocky exoplanet, we find that the lifetime of mantle degassing increases with planet mass but has decreased as the Galaxy has aged (Figure \ref{fig:lifetimes}). We find this lifetime is primarily a function of a planet's initial radiogenic heat budget, $Q_0$, and the reference mantle viscosity, with little dependence on the initial mantle temperature or the planet's central Fe-core mass fraction (Figures \ref{fig:all_params} and \ref{fig:core}), similar to \citet{Foley2018_stag}. With higher $Q_0$, the planet has more heating power, and can thus stay warm enough to melt and degas for longer. As the Galaxy aged, the individual HPEs were produced and decayed at different rates, meaning the concentration a planet would inherit upon its formation and $Q_0$ are both a function of when it formed in Galactic history \citep[Appendix \S\ref{sec:GCE}; Figure \ref{fig:rat_frank}; ][]{Frank14}. We also find a higher reference viscosity slows mantle cooling, thereby allowing high temperatures, and therefore a greater degree of volcanism, thus allowing mantle degassing to be sustained for longer (Figure \ref{fig:all_params}). A larger reference viscosity also makes the stagnant lid thicker, which reduces the rate of mantle melting, and hence outgassing. However, this latter effect is less important than the effect of greater retention of interior heat. 

From these degassing lifetimes, we estimate the distribution of maximum current ages, Age$_{\rm{max}}$, for which a stagnant-lid exoplanet will be actively degassing \textit{today} ($t = 12.5$ Gyr, Figure \ref{fig:TimeMass}). Because of the pessimistic assumptions adopted in our model, planets younger than Age$_{\rm{max}}$ very likely contain sufficient radiogenic heat to be degassing today, regardless of their tectonic state. Planets older than than Age$_{\rm{max}}$, however, would require additional compositional or geophysical complexity to be included in our model, some of which we explore below. Adopting the average degassing lifetime for a given mass and formation time, we estimate an average maximum age, Age$_{\rm{max}}^{\rm{Avg}}$, for planets between 1 and 6 M$_{\oplus}$ to be:
\begin{equation}
    \rm{Age^{\rm{Avg}}_{max}} = -7.1 + 8.9\left(\frac{M_p}{\rm{M_{\oplus}}}\right)^{0.09} \rm{Gyr}.
\end{equation}
This equation incorporates the effects of Galactic chemical evolution on HPE abundance and the potential system-to-system variation in HPE concentration based on stellar abundance measurements (Figures Figure \ref{fig:rat_frank} and \ref{fig:abund_histograms}), reference viscosity and initial mantle temperature. From this equation, we estimate the average maximum age, Age$_{\rm{max}}^{\rm{Avg}}$, to be $\sim$1.8 Gyr for 1 M$_\oplus$ stagnant-lid exoplanets, increasing to $\sim$3.3 Gyr for 6 M$_{\oplus}$ planets (Figure \ref{fig:TimeMass}, center). 

Using the upper-limit of the 95\% confidence interval of our predicted degassing lifetimes for a given mass and formation time (Figure \ref{fig:TimeMass}, right), we estimate the upper-limit of the 95\% confidence interval in the maximum age, Age$^{\rm{UL95\%CI}}_{\rm{max}}$, for planets between 1 and 6 M$_{\oplus}$ to be:
\begin{equation}
    \rm{Age^{\rm{UL95\%CI}}_{max}} = -1.7 + 5.4\left(\frac{M_p}{\rm{M_{\oplus}}}\right)^{0.21} \rm{Gyr}. 
\end{equation}
This yields Age$_{\rm{max}}^{\rm{UL95\%CI}}$ values of $\sim$3.7 and 6.2 Gyr for 1 and 6 M$_{\oplus}$ planets, respectively (Figure \ref{fig:TimeMass}, right). These longer degassing lifetimes are primarily only relevant for planets with reference viscosities 10-100 times that of the Earth (Figure \ref{fig:h0visc}). Conversely, at the lower-limit of the 95\% confidence interval of our predicted degassing lifetimes, 1 M$_{\oplus}$ planets would only be degassing today if younger than $\sim$500 Myr, with this maximum age increasing to 1 Gyr for 6 M$_{\oplus}$ planets (Figure \ref{fig:TimeMass}, left). 

Of all exoplanets discovered to date, 694 have reported measurements of mass, radius, and host-star age, as well as the respective uncertainties in each,  according to the NASA exoplanet archive (\citet[][]{Akeson13},\dataset[10.26133/NEA1]{https://doi.org/10.26133/NEA1}). Of these, we find 17 planets (Table \ref{tab:planets}) between 1 and 6 M$_{\oplus}$ with average bulk densities $>5$ ${\rm{g}}\,{\rm cm}^{-3}$, a density high enough to maximize the likelihood these planets are rocky without significant H$_2$/He atmospheres \citep{Schulze21}, and equilibrium surface temperatures below the zero-pressure melting curve of dry peridotite \citep[$\sim$1300 K, ][]{Katz2003}, thus allowing a solid surface. These planets represent a range of ages between 1.4 and 11 Gyr. Assuming a planet's mass and age form a bivariate normal distribution, we estimate the probability that these likely rocky exoplanets are younger than Age$_{\rm{max}}$ for their mass, and thus the likelihood they are actively degassing today. Because of the dependence of degassing lifetime on both the planet's HPE budget and mantle reference viscosity (Figures \ref{fig:all_params}, \ref{fig:h0visc},\ref{fig:hpe_plot2}--\ref{fig:sixME_hpe_plot2}), and the lack of empirical constraints on mantle viscosity, we  calculate Age$_{\rm{max}}$ using the average degassing lifetime from Figure  \ref{fig:lifetimes} (center; Age$^{\rm{Avg}}_{\rm{max}}$) and the upper-bound of the 95\% confidence interval (Figure  \ref{fig:lifetimes}, right; Age$_{\rm{max}}^{\rm{UL95\%CI}}$), which represents our least pessimistic estimate of Age$_{\rm{max}}$.

\begin{figure*}[t!]
    \centering
    \includegraphics[width=0.9\linewidth]{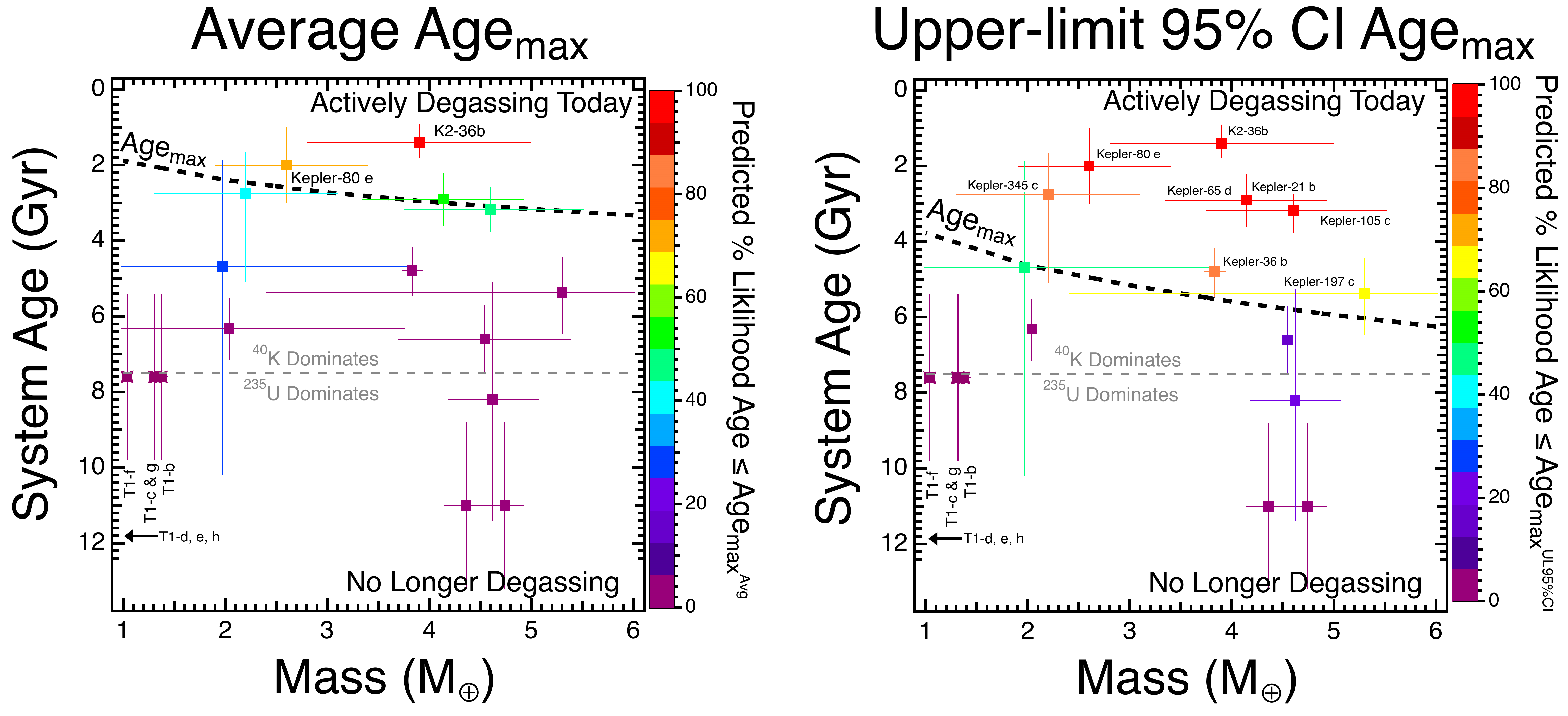}
    \caption{Probability that host-star/planet age is $\leq$Age$^{\rm{Avg}}_{\rm{max}}$ (left) and $\leq$Age$^{\rm{UL95\%CI}}_{\rm{max}}$ (right) for the 17 planets within our sample (Table \ref{tab:planets}) as a function of their mass and system age. Those planets with probabilities greater than 1$\sigma$ (67\%) confidence are labeled and TRAPPIST-1 is included for reference. We define active degassing as having degassing rates greater than 10\% of the Earth's current value,scaled linearly by planet surface area (see Appendix \ref{sec:Methods}).}
    \label{fig:plan_probs}
\end{figure*}

We estimate only one planet, K2-36b, is younger than Age$^{\rm{Avg}}_{\rm{max}}$ for its mass at greater than 2$\sigma$ (95\%) confidence (Figure \ref{fig:plan_probs}; Table \ref{tab:planets}). Additionally, at the $1\sigma$ ($>67\%$) confidence level, Kepler 80-e is younger than Age$^{\rm{Avg}}_{\rm{max}}$. These planets would still require a sufficient HPE abundance to support degassing lifetimes longer than their current age, but no requirement for high mantle reference viscosity $\gtrsim10^{23}$ Pa s (Figures \ref{fig:h0visc}, \ref{fig:hpe_plot2} and \ref{fig:sixME_hpe_plot2}). Adopting the more optimistic Age$_{\rm{max}}^{\rm{UL95\%CI}}$, three additional planets are younger than this maximum age for their mass at $\geq2\sigma$ confidence (Kepler-80 e, Kepler-65 d and Kepler-105 c) and with $>1\sigma$ confidence (Kepler-245 c and Kepler-36 b). We note though that those stagnant-lid exoplanets younger than Age$_{\rm{max}}^{\rm{UL95\%CI}}$ would require both a sufficient HPE budget and mantle reference viscosities 10-100 times greater than the Earth's based on our pessimistic thermal evolution models (Figures \ref{fig:h0visc}, \ref{fig:hpe_plot2} and \ref{fig:sixME_hpe_plot2}). 

The remaining 10 planets in this sample all have probabilities below $\sim$50\% of being younger than Age$_{\rm{max}}^{\rm{UL95\%CI}}$, including the TRAPPIST-1 system. 
We therefore cannot confidently assume that these planets are actively degassing at a sufficient rate to sustain a temperate climate \textit{today}. Our model, however, is intentionally pessimistic in its determination of Age$_{\rm{max}}$. There are a number of factors, aside from the radiogenic HPE budget and planet size, that can change a planet's degassing lifetime not included in our model. By examining the effects of these additional parameters on our model results, we can gauge the degree Age$_{\rm{max}}$ can change as we relax the pessimistic assumptions of our model.

\section{Factors that Extend Stagnant-lid Degassing Lifetimes}
Of the HPEs, $^{40}$K is the dominant element controlling the lifetime of mantle degassing (Figures \ref{fig:hpe_plot2} and \ref{fig:sixME_hpe_plot2}), and thus sets Age$_{\rm{max}}$ for those planets likely degassing today. Unlike Th and U, a planet's concentrations of K and $^{40}$K are not directly inferrable from abundance determinations of the host star (Appendix \S\ref{sec:planet_K}). $^{40}$K is a moderately volatile element, meaning its concentration relative to the rock-building elements (e.g., Mg) in a planet will be reduced compared to the host-star due to volatilization effects during planet formation. Furthermore, the host-star reveals only information on the bulk K abundance, and not the $^{40}$K/K ratio; whether the Earth's initial $^{40}$K/K ratio of 0.1419\% is universal for all rocky exoplanets is unknown. We estimate that simply doubling $^{40}$K/K would increase the distribution of Age$_{\rm{max}}$ by $\sim$1 Gyr (Figure \ref{fig:Kchange}). An effective method for altering a planet's $^{40}$K/K is supernova injection of $^{40}$K-rich material into the  protoplanetary disk prior to planet formation. Little work, however, has been done to estimate the range of possible exoplanetary $^{40}$K/K via this process. We therefore modeled the production and distribution of $^{40}$K and K in a 15 M$_{\odot}$ supernovae progenitor (Appendix \S\ref{sec:planet_K}, Figures \ref{fig:warmjets}--\ref{fig:40K}). We find that while the production of extremely $^{40}$K-rich material is possible, the total mass injected into the disk is unlikely to increase a protoplanet's $^{40}$K/K by any appreciable amount, except potentially for those orbiting in low-mass, M-dwarf disks. 

Mantle reference viscosity is the other key factor controlling degassing lifetime and Age$_{\rm{max}}$. On average, we find degassing lifetime increases by $\sim0.66$ Gyr per factor of 10 increase in reference viscosity for a 1 M$_{\oplus}$ planet, and by $\sim0.92$ Gyr for a 6 M$_{\oplus}$ planet. These would then correspond to increases in Age$^{\rm{Avg}}_{\mathrm{max}}$ of $\sim0.71$ and $\sim1.03$ Gyr per factor of 10 increase in mantle reference viscosity, respectively. Because viscosity increases with pressure \citep[e.g., ][]{karato1993,hirth2003}, super-Earth lower mantles can have very large viscosities \citep[e.g., ][]{Stamenkovic2011,Tackley2013,noack2014,Schaefer2015} due to their larger core-mantle boundary pressure \citep{Unter19}. As a result, mantle reference viscosity may, on average, increase with increasing planet size, leading to a corresponding increase in Age$^{\rm{Avg}}_{\mathrm{max}}$, such that Age$^{\rm{Avg}}_{\mathrm{max}}$  may increase more sharply with planet size than our model results indicate. However, viscosity extrapolations to such extreme temperature-pressure conditions are highly uncertain \citep{Karato2010}. Moreover, if the increase in viscosity with depth is large enough, the lower mantles of super-Earths may cease convecting entirely. 

\citet{Dorn2018} found that planets larger than 3-4 M$_\oplus$ may not experience outgassing at all, due to melt forming at a high enough pressure that it would be to dense to rise to the surface. In contrast, our models find that melt forms at low enough pressures to be buoyant regardless of planet size, unless the reference viscosity is increased to $\sim 10^{24}-10^{25}$ Pa s, a factor of $\sim 10-100$ above the upper bound in our models. Testing this very high viscosity case, we find volcanism and degassing is precluded on planets larger than $>2-3$ M$_{\oplus}$, consistent with \citet{Dorn2018}. A further increase in reference viscosity would prevent volcanism from ever occurring on planets of all sizes. Determining the exact cause of the difference between our results and \citet{Dorn2018} would require a detailed study with numerical convection models, and is therefore beyond this paper's scope. However, it is likely related to the inclusion of pressure-dependent viscosity in \citet{Dorn2018}; this causes higher viscosity in the lower mantle as planet size increases, potentially leading to overall less vigorous convection and thicker lithospheres that suppress melting. Mantle viscosity is therefore important in determining whether degassing is prevented on large planets due to dense melt. An additional implication of our finding on the role of mantle viscosity is that it would be difficult to extend degassing lifetimes, or Age$_{\rm{max}}$, much beyond our 95\% confidence interval upper limit by increasing mantle viscosity, especially for larger planets; these high viscosities would instead prevent degassing entirely. 

Reference viscosity is also influenced by planet composition, including both the relative abundances of the major rock-forming elements (Mg, Si, Fe) and the oxidation state of these elements. Fayalite (Fe$_2$SiO$_4$) has a lower viscosity by a factor of $\sim1000$ than forsterite (Mg$_2$SiO$_4$ and the dominant component of Earth's mantle), meaning a more FeO-rich mantle will have a lower viscosity than an Mg-rich one \citep{Zhao2009}, likely lowering Age$_{\rm{max}}$. Mantle FeO content, however,  also decreases the melting temperature of the mantle, making melting and degassing easier. More in-depth models are needed to analyze these counter-acting effects. Viscosity will also depend on silica (SiO$_2$) content, with a lower Mg/Si leading to a higher viscosity \citep{Ballmer2017, Spaargaren2020}. However, the effect is limited, as even a very silica-rich planet with Mg/Si = 0.5 only has a factor of $\sim10$ higher viscosity than the Earth \citep{Spaargaren2020}. Silica-rich planets may therefore have a slightly higher distribution of Age$_{\rm{max}}$ than those with an Earth-like planet composition considered in our model. Studies of stellar abundances, however, predict mantles with such a low mantle Mg/Si due to silica enrichment are very rare \citep{Unter19, Spaargaren2020}. 

Finally, mantle volatile content, in particular water, is known from rock deformation experiments to have a significant effect on viscosity \citep[e.g., ][]{hirth1996}, with the viscosity of mantle rock decreasing with increasing water content. At the same time, water content also affects the mantle solidus, with melting temperatures decreasing as mantle water content increases \citep[e.g., ][]{kushiro1968}. These two effects counteract each other. The same is true for other compositional factors, such as mantle Fe content \citep{Dorn2018}. Ultimately, to fully capture these competing effects, new models incorporating the combined compositional effects on viscosity and solidus would be needed. However, we provide a first-order estimate of how Age$_{\rm{max}}$ changes with mantle water content, based on model suites where mantle solidus and reference viscosity are systematically varied using the solidus parameterization of \citet{Katz2003} and the diffusion creep viscosity flow laws of \citet{hirth2003} (Appendix \ref{sec:water}). 

For water contents $\gtrsim0.05-0.1$ wt\%, the effect of water lowering mantle viscosities dominates, causing  Age$^{\rm{Avg}}_{\rm{max}}$ to decrease compared to a dry-planet baseline as the lower viscosity leads to more rapid cooling (Figure \ref{fig:water_effect}). Above these water contents, the decrease in solidus temperature begins to dominate, and  Age$^{\rm{Avg}}_{\rm{max}}$ begins to increase in comparison to a dry planet baseline as mantle melting can occur at lower temperatures. For a 1 M$_\oplus$ planet, we find  Age$^{\rm{Avg}}_{\rm{max}}$ can be lowered by at most $\sim1.4$ Gyr for moderate water contents ($\sim0.05-0.1$ wt\%), and raised by up to $\sim1.6$ Gyr for higher water contents ($\sim0.3$ wt\%), with the exact change depending on the chosen water-viscosity dependency exponent, $r$ (Figure \ref{fig:water_effect}). These effects are more pronounced for a 6 M$_{\oplus}$ planet. These estimates are likely upper limits, as we do not include overburden pressure of surface water, which will lower the rate of, or entirely prevent, volcanism \citep{Kite09,Cowan2014,Kriss21}, as well as the evolution of in- and outgassing rates of water over time \citep[e.g., ][]{mcgovern1989,Crow11,Spaargaren2020}. 

The addition of tidal heating \citep[e.g., ][]{Barnes2009,driscoll2015b} or magnetic induction heating \citep[e.g., ][]{Kislyakova2017} would also prolong mantle degassing and increase Age$_{\rm{max}}$ beyond our model estimates. Both of these sources of heat depend on a planet's orbit, and will persist as long as the orbital configuration allows, unlike radiogenic heat sources which decay over time. Planets heated by tides or magnetic induction can therefore sustain degassing for very long times, potentially even indefinitely. Exoplanetary orbit parameters can be constrained observationally, so the likelihood of significant tidal or magnetic induction heating can be estimated in most cases. 

As explained above, our model makes pessimistic assumptions that err on the side of hastening the end of volcanism, including neglecting mantle plumes and assuming all melt erupts at the surface. We found that varying melt intrusion did not significantly change our results; 90\% melt intrusion increases our estimated degassing lifetime by only $\sim 100$ Myrs for an Earth mass planet. Mantle plumes, however, can prolong volcanism, even after the upper mantle has cooled below the point where it can melt through passive upwelling, depending on the temperature difference between plume and surrounding mantle. While the effect of plumes can only be fully captured with higher dimensional dynamic models, we can roughly approximate how much they might extend degassing lifetime based on our models where mantle solidus temperature was varied (Figure \ref{fig:solidus}). There we found a 100$^{\circ}$C decrease in solidus temperature increases degassing lifetime by $\approx 0.4$ Gyrs for a 1 M$_{\oplus}$ planet. Therefore if upwelling plumes are 100-300$^{\circ}$C hotter than surrounding mantle, as estimated for Earth \citep{White1995,Yang1998,Thompson2000}, degassing could last up to $\approx 0.4-1.2$ Gyr longer. If plume volcanism is as active as Earth, then it likely could maintain sufficient CO$_2$ outgassing for maintaining a temperate climate, as plume volcanism accounts for $\approx 1/3$ of the CO$_2$ degassing on Earth \citep{Marty1998}. Plume activity may be more muted on stagnant lid planets, however, as inefficient mantle cooling leads to a smaller temperature anomaly for plumes \citep{ORourke2015}, so the estimates given here for how plumes prolong volcanism should be considered upper bounds. 

Our model predictions can also be applied to Venus, assuming it is a stagnant-lid planet. Venus is a 0.82 M$_\oplus$ planet, meaning our model predicts average and upper-limit Age$_{\rm{max}}$  values of 1.6 and 3.5 Gyr, respectively. That is, it should not exhibit significant enough volcanism and outgassing today to support a temperate climate ($>$ 10\% Earth's present rate).  There is some evidence, however, for volcanism and outgassing in the last $<$ 1 Myr \citep{Smrekar2010,Filiberto2020,Byrne2022}, likely plume related \citep{Gulcher2020}. If corroborated, this would not necessarily invalidate the model predictions, as rates of Venusian volcanism are probably too low to support outgassing rates above our threshold rate of degassing to support a temperate climate. Estimates of the rate of volcanism on Venus, if present, are uncertain, but most fall in the range of $\sim$0.1--1 km$^{3}$ yr$^{-1}$ with a highest estimate of $\sim$10 km$^{3}$ yr$^{-1}$ \citep[][ and references therin]{Fegl89, Byrne2022}. These volcanism rates are therefore $\sim$30--300 times lower (with a minimum of 2 times lower for the highest rate for Venus) than Earth's estimated 26-34 km$^{3}$ yr$^{-1}$ \citep{Crisp1984}, and would lead to comparably lower CO$_2$ outgassing rates. Only the upper end estimate of $\sim10$ km$^{3}$ yr$^{-1}$ would be large enough to drive CO$_2$ outgassing at a rate $>$ 10\% the modern Earth's. Moreover, it is possible that Venus is experiencing volcanism due to lying in a tectonic regime intermediate to stagnant lid and plate tectonic end members, and therefore having a thinner lithosphere than expected for a purely stagnant-lid planet. There is evidence for limited plate-tectonics-like subduction \citep{Sandwell1992,Davaille2017} and relative movement of crustal blocks on the surface \citep{Byrne2021}, as well a relatively thin lithosphere and high heat flux, compared to pure stagnant-lid models \citep{Borrelli2021}. Venus, therefore, demonstrates that planets may operate in a regime intermediate to the end-member plate tectonics and stagnant-lid regimes, leading to longer-lived volcanism than our pessimistic stagnant-lid models predict.

Plate tectonics can drastically increase the lifetime of degassing on a rocky exoplanet to ages beyond Age$_{\rm{max}}$ due to lithospheric thinning allowing for mantle material to melt at lower temperatures \citep[e.g.][]{Kite09}. For planets significantly older than $Age_{\rm{max}}$ where tidal or induction heating are unlikely, our stagnant-lid framework would not be able to explain any atmospheric observations showing atmospheric chemistries indicative of active mantle degassing or the presence of a temperate climate. These older planets, then, may provide an ideal sample to search for rocky exoplanets undergoing plate tectonics similar to Earth, or indicate planets with complex tectonic histories that have potentially transitioned between different tectonic modes over time. 

Our definition of Age$_{\rm{max}}$ represents a pessimistic upper-limit of the ``temporal habitable zone'' for rocky, stagnant-lid exoplanets not undergoing tidal heating. For planets older than this ``zone'', we estimate they will have exhausted their internal radiogenic heat budget to the point where interior melting is limited and mantle degassing rates are no longer sufficient to support a temperate climate \textit{today}, when we observe it. Other compositional and dynamical factors may increase Age$_{\rm{max}}$ for these stagnant-lid planets, as described above, however, they are often fraught with other complications that may impact other aspects of planetary habitability.

\section{Conclusion}
An individual rocky exoplanet provides us with a sparseness of direct data with which to  understand its evolution. Host-star age and radionuclide abundance, while indirectly telling us about the planet, are critical, and currently underutilized, observables that will allow us to better understand both an exoplanet's history and its current likelihood of being temperate \textit{today}, regardless of tectonic state. The framework we present here that combines direct and indirect observational data with dynamical models not only provide us with a pessimistic baseline for understanding which parameter(s) most control a stagnant-lid exoplanet's ability to support a temperate climate, but also where more lab-based and computational work is needed to quantify the reasonable range of these parameters (e.g., mantle reference viscosity). As we move to more in-depth characterization of individual targets in the James Webb Space Telescope era, these direct and indirect astronomic observables coupled with laboratory data and models from the geoscience community will allow us to better estimate whether a rocky exoplanet planet in both the canonical and temporal habitable zones, or has exhausted its internal heat and is simply too old to be ``Earth-like.''

\section*{Acknowledgements}
CTU acknowledges the support of Arizona State University through the SESE Exploration fellowship. The results reported herein benefited from collaborations and/ or information exchange within NASA's Nexus for Exoplanet System Science (NExSS) research coordination network sponsored by NASA's Science Mission Directorate, and gratefully acknowledge support from grant NNX15AD53G awarded to SD. We thank the anonymous reviewer for their helpful comments which improved the manuscript's clarity considerably. 

 \newcommand{\noop}[1]{}

\appendix
\restartappendixnumbering
\section{Methods}
\label{sec:Methods}
We adopt the stagnant-lid thermal evolution model of \citet{Foley2018_stag,Foley2019_stag} updated to account for a planet's mass, core mass fraction (CMF), metamorphic degassing rate, individual HPE contents, fractionation of HPEs into the crust and solidus changes due to mantle depletion (Appendices \S\ref{sec:thermal_mod} and \S\ref{sec:thermal_mod2}). In these models, initial radioactive heat production budgets are determined in a Monte-Carlo fashion sampling within the observed variability of HPEs in FGK stars, corrected for fractionation and volatility effects (Appendix \S\ref{sec:radio_abunds}). We define the cessation of degassing as the moment when a planet's degassing rate first falls below 10\% of the Earth's present day degassing rate \citep[$\approx 6 \times 10^{12}$ mol yr$^{-1}$, ][]{Marty1998}, scaled linearly by planet surface area. \cite{Foley2019_stag} confirmed that degassing rates must be $> 10$ \% the modern day Earth's for a temperate climate on stagnant-lid planets, even if the planet is mostly ocean covered and thus dominated by seafloor weathering. Seafloor weathering has an overall slower rate than continental weathering on the modern Earth, and thus with only seafloor weathering active atmospheric CO$_2$ levels (and surface temperatures) would be higher, for a given degassing rate \citep[e.g., ][]{Krissansen-Totton2017,Hayworth2020,Glaser20}. Planets with higher land fractions would thus require higher degassing rates than our chosen threshold to remain temperate. Moreover, our degassing rate threshold was determined for a planet receiving a stellar radiative flux equal to what Earth receives today. Lower incoming radiative fluxes would also require higher rates of degassing than our assumed threshold in order to sustain temperate climates.  

We scale our degassing rate threshold with planet surface area, because the total weathering rate increases linearly with the area of weatherable rock. Thus a planet with a larger surface area will need a proportionally higher degassing rate to sustain a temperate climate. Ultimately, planets with degassing rates below our threshold for temperate climates may not be the best targets for atmospheric characterization or detectable surface life, as they are likely to lie in snowball climate states. Finally, another climate extreme is possible if the rate of CO$_2$ degassing overwhelms the surface weathering rate; this will produce a hot-house, Venus-like climate. 

\subsection{Updates to Thermal Evolution Model}
\label{sec:thermal_mod}
\citet{Foley2018_stag} and \citet{Foley2019_stag} give a thorough description of our model, and all of the key governing equations are listed in Appendix \S\ref{sec:thermal_mod2}. Here we will only highlight the differences between the model of \citet{Foley2018_stag} and the model presented in this paper. For an Earth-like CMF = 0.33, we use the scaling laws from \citet{Vale06} \& \citet{Valencia2007b} to determine average mantle density, planet radius, mantle thickness, and surface gravity as a function of planet mass:
\begin{equation}
    \rho = \rho_{\oplus}\left(\frac{{M_p}}{\mathrm{M}_{{\oplus}}} \right)^{0.2}
\end{equation}
\begin{equation}
    R_p = R_{\oplus} \left(\frac{{M_p}}{\mathrm{M}_{{\oplus}}} \right)^{0.27}
\end{equation}
\begin{equation}
    d = d_{\oplus}\left(\frac{{M_p}}{\mathrm{M}_{{\oplus}}} \right)^{0.28}
\end{equation}
\begin{equation}
\label{eq:grav}
    g = \frac{G{M_p}}{R_p^2}
\end{equation}
where $\mathrm{M_p}$ is the planet mass and $G$ is the gravitational constant. Reference Earth values are $\rho_{\oplus} = 4450$ kg m$^{-3}$, $R_{\oplus}=6378$ km, and $d_{\oplus}=2890$ km.  

To vary the core mass fraction, we use the scaling laws from \citet{Noack2016} (See also: \citet{Foley2020_div}), which assume that all iron resides in the core. The resulting equations for $R_p$, $R_c$, and $\rho$, as a function of CMF and planet mass, are: 
\begin{equation}
    R_p  = (7121 - 2021\times \mathrm{CMF})\left(\frac{{M_p}}{\mathrm{M_\oplus}} \right)^{0.265} \rm{km}
\end{equation}
\begin{equation}
    R_c = (19200\times \mathrm{CMF}-31760\times \mathrm{CMF}^2 +18100\times \mathrm{CMF}^3)\left(\frac{{M_p}}{\mathrm{M_\oplus}} \right)^{0.252} \rm{km}
\end{equation}
\begin{equation}
    \rho = \frac{(1-\mathrm{CMF}) \mathrm{M_p}}{\frac{4}{3}\pi(R_p^3-R_c^3)}.
\end{equation}
Gravity is still calculated from \ref{eq:grav}. 

We assume all other material properties are independent of planet size and CMF. This is a simplification because thermal conductivity, expansivity, and viscosity are all functions of pressure, and the larger the planet or larger the core, the higher the pressures reached in the mantle. However, robust parameterizations for how to incorporate these pressure effects are currently lacking. There is still significant uncertainty about how key material properties change at the extreme temperature and pressure conditions of super-Earth lower mantles, which are not yet experimentally accessible  \citep[e.g., ][]{Duffy2015_treatise}. Moreover, how significant variation of key material properties with pressure, and hence depth in a super-Earth mantle, modifies the dynamics of the convecting mantle has not been extensively studied; scaling laws for convective heat flux and velocity that take these pressure effects into account have not yet been developed.  

Lacking robust parameterizations for viscosity, thermal expansivity, and thermal conductivity pressure effects, we instead randomly vary the mantle reference viscosity, $\mu_{\mathrm{ref}}$, in our models over a four-orders-of-magnitude range. We focus on viscosity because, of the key mantle material properties, it shows the strongest dependence on pressure and mantle composition, as it can vary by orders of magnitude \citep[e.g., ][]{karato1993,hirth2003}. We use a standard Arrhenius temperature-dependent viscosity law in our models: 
\begin{equation}
\label{mu_i}
    \mu_i = \mu_n \exp{\left(\frac{E_v}{RT_p}\right)},
\end{equation}
where $\mu_i$ is mantle interior viscosity, $T_p$ is mantle potential temperature, $E_v = 300$ kJ mol$^{-1}$ \citep[e.g., ][]{karato1993} is activation energy, and $R$ is the universal gas constant. The reference viscosity is defined at reference temperature $T_{\mathrm{ref}} = 1623$ K, or approximately Earth's present day mantle potential temperature. The constant $\mu_n$ is then adjusted in each model run to match the chosen reference viscosity. Our results therefore explicitly demonstrate how degassing lifetime depends on mantle reference viscosity, which itself may vary with planet size or composition. 

The equations for calculating the rate of metamorphic degassing due to crustal burial, from \citet{Foley2018_stag}, are reformulated in terms of pressure, rather than depth (see Appendix \S\ref{sec:outgassing_mod}). As such they can then be applied to planets with variable size and CMF, and hence different surface gravities. Finally, we have improved the melting model to include depletion of the mantle and a subsequent increase in the melting temperature. We follow the method of \citet{Tosi2017}, and assume the solidus can increase by up to 150 K upon full depletion of the mantle, the difference in the solidi of harzburgite and peridotite. The degree of mantle depletion is calculated based on the volume of crust present at each timestep (see Appendix \S\ref{sec:melting}). The model also tracks each of the four major HPEs separately, rather than treating them together with an average decay constant as in \citet{Foley2018_stag}. Here we are interested in observationally constrained variations in each of the four major HPEs, so we naturally must treat each HPE separately in the model (see Appendix \S\ref{sec:melting}). The remainder of the model equations are general and can be applied to planets with different masses and core mass fractions. 

\subsection{Thermal Evolution Model}
\label{sec:thermal_mod2}

Here we give a complete description of the coupled thermal evolution and volatile cycling model. Assuming pure internal heating, and that all melt produced contributes to cooling of the mantle, mantle thermal evolution is given by:
\begin{equation}
\label{thermal_evol}
V_{man} \rho c_p \frac{dT_p}{dt} = Q_{man} - A_{man} F_{man} - f_m \rho_m \left (c_p \Delta T_m + L_m \right) ,    
\end{equation}
where $V_{man}$ is the volume of the actively convecting mantle, $\rho$ the average density of the mantle, $c_p$ the heat capacity, $T_P$ the potential temperature of the mantle, $t$ time, $Q_{man}$ the total radiogenic heat production rate of the mantle, $A_{man}$ the surface area of the top of the convecting mantle (base of the stagnant-lid), $F_{man}$ the heat flux from the mantle, $f_m$ the volumetric melt production rate, $\rho_m$ the density of melt, $\Delta T_m$ the temperature difference between erupted melt and the surface temperature, and $L_m$ the latent heat of the mantle \citep[e.g., ][]{Stevenson1983,Hauck2002,Reese2007,Fraeman2010,Morschhauser2011,Driscoll2014,Foley2018_stag,Foley2019_stag}. The volume of the convecting mantle is $V_{man} = (4/3) \pi ((R_p-\delta)^3 - R_c^3)$, where $R_p$ is the planet radius, $R_c$ is the core radius, and $\delta$ the thickness of the stagnant-lid. The surface area of the top of the convecting mantle is then $A_{man} = 4 \pi (R_p-\delta)^2$. Finally, the temperature difference between erupted melt and the surface temperature is $\Delta T_m = T_p - P_i \gamma - T_s$, where $\gamma$ is the average adiabatic temperature gradient of mantle melt, estimated as $\gamma \approx 2 \times 10^{-8}$ K Pa$^{-1}$ in \citet{Foley2018_stag}, and $P_i$ is the pressure where melting begins (derived below in \S\ref{sec:melting}).   

The thickness of the stagnant-lid, $\delta$, is then \citep[e.g., ][]{Schubert1979,Spohn1991}:
\begin{equation}
\label{delta1}
\rho c_p (T_p - T_l) \frac{d \delta}{dt} = -F_{man} - k \frac{\partial T}{\partial z} \Bigr |_{z=R_p-\delta},    
\end{equation}
where $T_l$ is the temperature at the base of the stagnant-lid, $k$ is the thermal conductivity (assumed to be the same throughout the crust and mantle for simplicity), and $z$ is the height above the planet's center. The mantle heat flux, $F_{man}$, and lid base temperature, $T_l$, are calculated from the following scaling laws for stagnant-lid convection 
\citep{Reese1998,Reese1999,Solomatov2000b,korenaga2009}: 
\begin{equation}
\label{heat_flux}
F_{man} = \frac{c_1 k(T_p - T_s)}{d} \Theta^{-4/3} Ra_i^{1/3}
\end{equation}
and 
\begin{equation}
\label{T_l}
T_l = T_p - \frac{a_{rh} R T_p^2}{E_v} ,
\end{equation}
where $c_1$ and $a_{rh}$ are constants (assumed to be $c_1 = 0.5$ and $a_{rh} = 2.5$), and $T_s$ is the surface temperature, here fixed to 273 K, as surface temperature fluctuations of order 100 K, that could result from changes in atmospheric CO$_2$, do not significantly impact the evolution of the underlying mantle in these models \citep{Foley2019_stag}. The mantle thickness is $d=R_p - R_c$, and $\Theta$ is the Frank-Kamenetskii parameter, $\Theta = E_v (T_p - T_s)/(RT_p^2)$, where $E_v$ is the activation energy for mantle viscosity and $R$ is the universal gas constant. The internal Rayleigh number, $Ra_i$, is defined as $Ra_i =  \rho g \alpha (T_p - T_s) d^3/(\kappa \mu_i)$, where $g$ is gravity, $\alpha$ the thermal expansivity of the mantle, $\kappa$ the thermal diffusivity of the mantle, and $\mu_i$ the viscosity at the mantle potential temperature of $T_p$ (see Equation \ref{mu_i}). 
    
\subsubsection{Melting and Crustal Evolution}
\label{sec:melting}
Large-scale mantle melting, and subsequent volcanism, takes place when passively upwelling mantle is hot enough to cross the solidus beneath the lid. As in \citep[e.g., ][] {Fraeman2010,Foley2018_stag}, the pressure at which melting begins, $P_i$, is calculated from the intersection of the mantle adiabat, with adiabatic gradient $\gamma_{mantle}$, and the dry peridotite solidus from \citet{Takahashi1983}:
\begin{equation}
\label{p_melt}
P_i = \frac{T_p - T_{sol0}}{120 \times 10^{-9} - \gamma_{mantle}} , 
\end{equation}
where $T_{sol0}$ is the melting solidus temperature at 0 pressure, and the adiabatic gradient in the mantle is $\gamma_{mantle} \approx 2 \times 10^{-8}$ K Pa$^{-1}$. The solidus temperature at 0 pressure depends on the degree of mantle depletion, which we estimate based on the volume of crust present, as explained below. We also do not allow the pressure where melting begins to exceed 10 GPa, the approximate pressure where silicate melts become denser than solids \citep[e.g.][]{Dorn2018}. Melting stops at the base of the lid, which occurs at pressure
\begin{equation}
\label{pf}
P_f = \rho_l g \delta ,
\end{equation} 
where $\rho_l$ is the average density of the crust and lithosphere (assumed to be $\rho_l = 3300$ kg m$^{-3}$). The melt fraction, $\phi$, is 
\begin{equation}
\label{phi}
\phi = \frac{P_i - P_f}{2} \left(\frac{d\phi}{dP} \right)_S,
\end{equation}
where $(d\phi/dP)_S \approx 1.5 \times 10^{-10}$ Pa$^{-1}$. The melt production rate, $f_m$, is calculated as (see \citet{Foley2018_stag} for a derivation)  
\begin{equation}
\label{fm}
f_m = 17.8 \pi R_p v (d_{melt} - \delta) \phi,
\end{equation}
where $v$ is the characteristic convecting mantle velocity and $d_{melt}=P_i/(\rho_lg)$ is the depth where melting begins. The convecting mantle velocity is \citep{Reese1998,Reese1999,Solomatov2000b,korenaga2009}:
\begin{equation}
\label{vel1}
v = c_2 \frac{\kappa}{d} \left(\frac{Ra_i}{\Theta} \right)^{2/3} ,
\end{equation}
where $c_2$ is a constant.  

Melting produces a crust whose thickness, $\delta_c$, and volume, $V_{crust}$, evolve over time. To calculate the evolution of the $V_{crust}$ we assume that all melt produced contributes to growth of the crust, and that all crust buried to depths below the lithospheric thickness, $\delta$, founders into the mantle. The resulting equation is
\begin{equation}
\label{crust2}
\frac{d V_{crust}}{dt} = f_m - \left(f_m-4\pi(R_p-\delta)^2\textrm{min}\left(0,\frac{d\delta}{dt}\right) \right)(\tanh{((\delta_c-\delta)20)}+1),
\end{equation} 
The second term on the right-hand side of \ref{crust2} describes the rate of crust loss due to foundering of the crust; the hyperbolic tangent function formulation allows this crust loss rate to go to 0 when $\delta_c<\delta$. The term $4\pi(R_p-\delta)^2\textrm{min}(0,d\delta/dt)$ captures the loss of crust when the lid thickness is decreasing and $\delta_c = \delta$, and is 0 otherwise (that is, if either the lid thickness is growing or the crust ends before the base of the stagnant-lid). 
The crustal thickness is calculated from the volume of crust as 
\begin{equation}
 \delta_c = R_p - \left(R_p^3-\frac{3V_{crust}}{4\pi} \right)^{1/3}   .
\end{equation}

To incorporate how depletion of the mantle influences the solidus, and thus later melt production, we increase $T_{sol0}$ linearly with crust thickness following \citep{Tosi2017}: 
\begin{equation}
    T_{sol0} = 1423 + \Delta T_{sol} \times \max{\left(1,\frac{\delta_c}{\delta_{ref}}\right)} .
\end{equation}
Here 1423 K is the dry peridotite solidus temperature at 0 pressure from \citet{Takahashi1983}. $\Delta T_{sol} = 150$ K is the increase in the solidus upon full depletion (which is set here to the difference in the zero pressure solidus temperatures for peridotite and harzburgite), and $\delta_{ref} = 0.2 V_{man}^0/A_{surf}$ is the reference crust thickness produced upon full depletion of the mantle. Here $V_{man}^0 = (4/3)\pi(R_p^3-R_c^3)$ and $A_{surf} = 4 \pi R_p^2$ is the surface area of the planet. The solidus can thus increase by up to 150 K due to mantle depletion. We explore other compositional effects that affect the solidus (e.g., water) in the main text.
 
HPEs are preferentially partitioned into the crust during mantle melting, due to their incompatible nature. We track this partitioning for all 4 long-lived HPEs assuming accumulated fractional melting. The evolution of crustal heat production rate, for a given HPE, is thus: 
\begin{equation}
\label{Q_crust_U238}
\frac{d Q_{c,i}}{dt} = \frac{x_{m,i} f_m}{\phi} [1 - (1-\phi)^{1/D_{i}} ] -  x_{c,i} \left(f_m-4\pi(R_p-\delta)^2\textrm{min}\left(0,\frac{d\delta}{dt}\right) \right)(\tanh{((\delta_c-\delta)20)}+1) - \frac{Q_{c,i}}{\tau_{i}} .
\end{equation} 
Here, $Q_{c,i}$ is the heat production rate in the crust resulting from one of the four HPEs tracked in the model. The total crustal heat production rate, $Q_{crust} = Q_{c,U238} + Q_{c,U235} + Q_{c,Th} + Q_{c,K}$. Each heat producing element (HPE) has a specified decay constant, $\tau_i$, distribution coefficient, $D_i$, and crustal and mantle heat production rate per unit volume, $x_{c,i}$ and $x_{m,i}$, respectively (e.g., $\tau_{U238}$ is the decay constant for $^{238}$U, $D_{U238}$ the distribution coefficient for $^{238}$U, and $x_{c,U238}$ the heat production per unit volume in the crust due to $^{238}$U). The well known half-lives taken from \citet{turc1982}, are used to calculate the decay constants, $\tau_i$. The chosen distribution coefficients of $D_{U238}=D_{U235} = 0.0012$ and $D_{Th} = 0.0029$ are from \citet{Beattie1993}, and $D_{K} = 0.0011$ is from \citet{Hart1974} assuming $60$ \% olivine and 40 \% pyroxene in the mantle. The evolution of mantle heat production from a given radioactive isotope is:   
\begin{equation}
\label{Q_man_u238}
\frac{d Q_{m,i}}{dt} = x_{c,i} \left(f_m-4\pi(R_p-\delta)^2\textrm{min}\left(0,\frac{d\delta}{dt}\right) \right)(\tanh{((\delta_c-\delta)20)}+1) -  \frac{x_{m,i} f_m}{\phi} [1 - (1-\phi)^{1/D_{i}} ] - \frac{Q_{m,i}}{\tau_{i}}.
\end{equation} 
As before, the total heat production rate of the mantle, $Q_{man} = Q_{m,U238} + Q_{m,U235} + Q_{m,Th}+ Q_{m,K}$. The total heat production rates per unit volume in the crust and mantle are also sums of the heat production rates of the four HPEs: $x_c = x_{c,U238} + x_{c,U235} + x_{c,Th} + x_{c,K}$ and $x_m = x_{m,U238} + x_{m,U235} + x_{m,Th} + x_{m,K}$. 

\subsubsection{Crustal Geotherm}
Equation \ref{delta1} requires as an input 
the conductive heat flux at the base of the lid. The temperature profile for steady-state one-dimensional heat conduction with constant heat production rates in the crust and mantle, neglecting advection, is used to determine this heat flux \citep[for details see: ][]{Foley2018_stag}: 
\begin{equation}
\label{flux_lid}
{ -k \frac{\partial T}{\partial z} \Bigr |_{z=R_p-\delta} = \frac{k(T_l-T_c)}{\delta-\delta_c} - \frac{x_m(\delta-\delta_c)}{2} }
\end{equation}      
when $\delta > \delta_c$, and 
\begin{equation}
\label{flux_crust}
-k \frac{\partial T}{\partial z} \Bigr |_{z=R_p-\delta} = -\frac{x_c \delta_c}{2} + \frac{k(T_l-T_s)}{\delta_c} 
\end{equation} 
when $\delta = \delta_c$. The temperature at the base of the crust, $T_c$, is 
\begin{equation}
\label{T_c}
{T_c = \frac{T_s(\delta-\delta_c)+T_l\delta_c}{\delta} + \frac{x_c \delta_c^2(\delta-\delta_c) + x_m \delta_c(\delta-\delta_c)^2}{2 k \delta}.}
\end{equation} 
The mantle radiogenic heating rate, per unit volume, $x_m = Q_{man}/(V_{man}+V_{lid})$, where $V_{lid} = (4/3)\pi((R_p-\delta_c)^3-(R_p-\delta)^3)$ is the volume of the sub-crustal stagnant-lid. The crustal radiogenic heat rate per unit volume is $x_c = Q_{crust}/V_{crust}$, where $Q_{crust}$ is the total radiogenic heating rate in the crust, and $V_{crust}$ is the volume of the crust. 

\subsubsection{CO$_2$ Outgassing Rates}
\label{sec:outgassing_mod}
CO$_2$ is outgassed to the atmosphere due to both mantle melting, and subsequent volcanism, and metamorphic breakdown of carbonated minerals as the crust is buried \citep{Foley2018_stag,Foley2019_stag}. \citet{Foley2018_stag} showed that the temperature, as a function of depth, where metamorphic decarbonation occurs can be approximated as a simple linear relationship,
\begin{equation}
\label{T_dcarb}
T_{decarb} = A \rho_l g (R_p-z) + B,
\end{equation} 
where $A=9.66 \times 10^{-8}$ K Pa$^{-1}$, $B=835.5$ K, $T_{decarb}$ is temperature in Kelvin, and $\rho_l = 3300$ kg m$^{-3}$ is the lithosphere density. The depth where decarbonation occurs, $\delta_{carb}$, is \citep[for details see:][]{Foley2018_stag} 
\begin{equation}
\label{d_carb}
\delta_{carb} = \frac{\delta_c}{2} + \frac{k(T_c-T_{s})}{\delta_c x_c} - \frac{A\rho_l g k}{x_c} - \frac{k}{x_c} \sqrt{\left(\frac{x_c \delta_c}{2k} + \frac{T_c-T_{s}}{\delta_c} - A\rho_l g \right)^2 + \frac{2 x_c}{k}(T_{s} - B)} .
\end{equation} 
With the decarbonation depth determined by \ref{d_carb}, the metamorphic outgassing flux is given by 
\begin{equation}
\label{f_meta}
F_{meta} = \frac{R_{crust}f_m}{2 V_{carb}}(\tanh{((\delta_c-\delta_{carb})20)}+1) 
\end{equation}
where $R_{crust}$ is the size of the crustal CO$_2$ reservoir (in moles), and the hyperbolic tangent function allows $F_{meta}$ to go to zero when $\delta_{c} < \delta_{carb}$, and to $R_{crust} f_m/(V_{carb})$ when $\delta_c > \delta_{carb}$. The outgassing rate due to mantle melting is
\begin{equation}
\label{Fd}
F_d = \frac{f_m R_{man} [1 - (1-\phi)^{1/D_{CO_2}} ]}{\phi (V_{man}+V_{lid})}, 
\end{equation}
where $R_{man}$ is the size of the mantle CO$_2$ reservoir, and $D_{CO_2} = 10^{-4}$ is the distribution coefficient for CO$_2$ \citep{Hauri2006}. The evolution of the mantle and crustal reservoirs, when crustal decarbonation is active (e.g., $\delta_{carb} < \delta_c$), follow 
\begin{equation}
    \frac{dR_{man}}{dt} = -F_d
\end{equation}
\begin{equation}
    \frac{dR_{crust}}{dt} = F_d,
\end{equation}
while when crustal decarbonation is inactive ($\delta_c < \delta_{carb}$)
\begin{equation}
\label{rcrust2}
\frac{dR_{crust}}{dt} =F_d - \frac{R_{crust}}{V_{crust}}\left(f_m-4\pi(R_p-\delta)^2\textrm{min}\left(0,\frac{d\delta}{dt}\right) \right)(\tanh{((\delta_c-\delta)20)}+1)
\end{equation} 
\begin{equation}
\label{rman2}
\frac{dR_{man}}{dt} = -F_d+ \frac{R_{crust}}{V_{crust}}\left(f_m-4\pi(R_p-\delta)^2\textrm{min}\left(0,\frac{d\delta}{dt}\right) \right)(\tanh{((\delta_c-\delta)20)}+1).
\end{equation}

The total CO$_2$ budget of the mantle and crust, $C_{tot}$, is conserved, such that $C_{tot} = R_{man} + R_{crust}$. As in \citet{Foley2018_stag}, the planet starts with all of the CO$_2$ residing in the mantle, and CO$_2$ is outgassed over time to the surface. The entire allotment of each HPE also initially resides in the mantle, as before planetary evolution begins we assume that there is no crust present (crust formation does not take place until mantle convection, and subsequent volcanism, begins). The abundance of each HPE is linked to the stellar observed abundances, as explained in Appendix \S\ref{sec:radio_abunds}. 

Finally, an important limit for the carbon cycle and habitability is the global weathering supply limit. This limit is the upper bound on weathering rate, and is set by the rate at which CO$_2$ drawdown would occur if all available fresh surface rock is completely carbonated as soon as it is brought to the surface. On a stagnant-lid planet, this limit is assumed to be set by volcanism. Using the average composition of basalt on Earth, the total amount of CO$_2$ that can be drawn down by crustal carbonation is $\chi = 5.8$ mol kg$^{-1}$ of basalt \citep{Foley2019_stag}. We assume the crust will have a basaltic composition, as it is a result of primary mantle melting of a peridotitic mantle. Exoplanets, however, could have different mantle bulk compositions, that would lead to different crustal compositions upon melting. However, the weathering demand, $\chi$, only increases by about a factor of two for the extreme ultramafic composition end-member, peridotite, and  only decreases by about a factor of two if the crust is felsic, like Earth's continental crust. Average compositions from \citet{taylor1985}, for continental crust, and \citet{Warren2016}, for peridotite, were used in making these estimates. The total variation in weathering demand is thus approximately a factor of 4, from ultramafic to felsic end-members. This would not significantly change the planetary conditions (e.g., mantle CO$_2$ budget, size, internal heat production rate, etc.) that control when planets enter a supply-limited weathering regime and thus develop hothouse climates \citep{Foley2018_stag, Foley2019_stag}.  

The weathering supply limit, $F_{sl}$, assuming all erupted basalt is available for weathering, is 
\begin{equation}
F_{sl} = \epsilon f_m \chi \rho_l
\end{equation}
with units of mol yr$^{-1}$; $\epsilon$ is the fraction of mantle melt produced that erupts at the surface ($\epsilon = 0.1$ is assumed). In the models, supply-limited weathering is assumed to lead to an inhospitable, hothouse climate when $F_d + F_{meta} > F_{sl} + 10^{14}$ mol yr$^{-1}$ at any point during planetary evolution, as in \citet{Foley2018_stag}. The total outgassing rate exceeding the weathering supply limit by $10^{14}$ mol yr$^{-1}$ means that hot climates, with $T_s \ge 400$ K, would form in $\sim 100$ Myr, well within the typical degassing lifetimes of modeled planets. 

\subsection{Input Radionuclide Abundances}
\label{sec:radio_abunds}
We define the initial total heat production rate of the mantle, $Q_{0}$, as a function of the specific power ($P_X$) produced by each HPE ($X$), their concentration within the planet ($C_X$) and the mass of the mantle ($m_{\rm{man}}$). We calculate the initial HPE abundance as $Q_{0} = P_X \times C_X \times m_{\rm{man}}$. In order to quantify the range of HPE concentrations in rocky exoplanets ($C_{X}^{\rm{planet}}$), we randomly sample within the observationally-constrained stellar abundance distributions for each HPE (Figure 1 of main text) and apply the scaling relationship to convert from stellar abundance to mantle concentration as a function of planet formation time, $t$, after the birth of the Milky Way:
\begin{equation}
\label{eq:conc}
\frac{C_{X}^{\rm{planet}} (t)}{C_{X}^{\rm{Earth-like}} (t)} = \left(\frac{f_X^{\rm{planet}}}{ f_X^{\rm{Earth}}}\right)\times\frac{C_X^{\rm{star}}}{C_X^{\rm{Sun}}}
\end{equation}
where $C_{X}^{\rm{star}}$, $C_{X}^{\rm{Sun}}$ are the concentrations of the element in the randomly selected stellar abundance and the Sun, respectively, and $f_X$ is a correction for volatility effects during planet formation. $C_{X}^{\rm{Earth-like}} (t)$ represents the predicted initial concentration of an HPE if it was a ``coscmochemically Earth-like'' planet \citep{Frank14} forming at some $t$ (Figure \ref{fig:rat_frank}). We re-scale the results of \citet{Frank14} such that the Earth-like abundance of each HPE at the time the Earth formed ($t = 8$ Gyr) are that of \citet{palme2003}. This lowers our choice of $C_{X}^{\rm{Earth-like}}$ by $\sim25$\% for each HPE compared to those presented in \citet{Frank14}, who adopted the initial Earth HPE abundances of \citet{turc1982}. 

We define the concentration of an HPE in both stars and the Sun as its molar ratio with Mg (e.g., X/Mg). We normalize relative to Mg as it is more likely to remain in the mantle, as opposed to Si which may partition into the core \citep{Hiro13}. Equation \ref{eq:conc} then becomes:
\begin{equation}
    C_{X}^{\rm{planet}} (t) =  \left(\frac{f_X^{\rm{planet}}}{ f_X^{\rm{Earth-like}}}\right)*\frac{\left(X/\rm{Mg}\right)_{\rm{Star}}}{\left( X/\rm{Mg}\right)_{\rm Sun}}* C_{X}^{\rm{Earth-like}}(t).
\end{equation}

We define $f_X$ as the fraction an element is enriched or depleted relative to Mg during planet formation relative to the host-star:
\begin{equation}
    f_X = \frac{\left(X/\rm{Mg}\right)_{\rm planet}}{\left(X/\rm{Mg}\right)_{\rm star}}.
\label{eq:vol}
\end{equation}
For those elements that fractionate relative to Mg during planet formation, $f_X$ will be greater than one in the case of enrichment and less than one in the case of depletion. Of the HPEs, U and Th are both refractory and not expected to fractionate between star and planet relative to Mg. That is to say $f_{\rm{U}}$ and $f_{\rm{Th}}$ are $\sim1$ for both a rocky exoplanet and its host star, as well as the Earth and the Sun. For the current Bulk Silicate Earth (mantle + crust), the present-day molar ratios of Th/Mg and U/Mg are $3.4\times10^{-8}$ and 9$\times10^{-9}$, respectively \citep{McD03}. Comparatively, the Sun's current composition \citep{Lodd09}, which is usually defined to be equal to the abundances in CI chondrites, has molar values of Th/Mg = $3.5\times10^{-8}$ and U/Mg = $9.8\times10^{-9}$. Among all chondrites, our best proxies for planet-forming materials, their whole-rock Th/Mg and U/Mg abundance ratios are within $<$10\% of the CI value, except for CV chondrites, which are 1.4 times the solar value \citep{Wass88}. The model of \citet{Desc18}, which computes how refractory elements redistribute themselves in protoplanetary disks, predicts small deviations ($<20$\%) of the molar ratios of elements at least as refractory than Mg in planetary materials (50\% condensation temperature $T_{c}^{\rm{Mg}}$ = 1336 K; \citet{Lodd03}). Th and U have 50\% condensation temperatures of 1659 and 1610 K, respectively \citep{Lodd03}. Based on these models, along with chondrite abundances and the Earth's abundances, we expect that Th/Mg and U/Mg ratios in a rocky exoplanet should match within tens of percent of the ratios in the star. We therefore set both the Earth-like and planet values of $f_{\rm{Th}}$ and $f_{\rm{U}}$ to 1.

In contrast to U and Th, K is substantially depleted in the Earth relative to Sun and CI chondrites, with $f^{\rm{Earth}}_{\rm{K}}$ = 0.19 \citep{Lodd09,McD03}. This is part of the well-known planetary volatility trend observed in the compositions of the Earth and other planets: elements less refractory than Mg ($T_{c} < T_{c}^{\rm{Mg}}$) are depleted in the Earth, relative to Mg and CI chondrites, by amounts that increase with decreasing condensation temperature \citep{McD03}. The 50\% condensation temperature of K is $T_{\rm{c}}^{\rm{K}} = 1006$ K \citep{Lodd03}, and K is thus ``moderately volatile'' in comparison with Mg, Th and U. The relative volatility of K is reflected in the range of K/Mg among chondrites, which is wider than the spread in Th/Mg or U/Mg. The Solar value of K/Mg is closest to the CI value of $3.6\times10^{-3}$; however it can vary from as low as $1.3\times10^{-3}$ in CV chondrites, to as high as $4.7\times10^{-3}$ in EH chondrites and $3.2\times10^{-3}$ in EL chondrites, i.e., from 0.4 times CI in CV, to 1.3 times CI and 0.9 times CI in EH and EL chondrites. This demonstrates that fractionation of K occurs among planetary materials, although the Earth's depletion by a factor of 5 remains unexplained \citep[see, however,][]{Desch20}. Without knowledge of the mechanism that depletes K relative to Mg during planet formation, we cannot constrain $f_{\rm{K}}^{\rm{planet}}$. Changes in $f_{\rm{K}}^{\rm{planet}}$ relative to $f_{\rm{K}}^{\rm{Earth}}$, will directly change $C_{\rm{K}}^{\rm{planet}}$ by an equal amount. Initially, we set $f_{\rm{K}}^{\rm{planet}} = f_{\rm{K}}^{\rm{Earth}} = 0.19$ and discuss the consequences of variable $f_{\rm{K}}^{\rm{planet}}$ in the main text. $^{40}$K is not expected to fractionate relative to K in any planet formation scenario; therefore it will be depleted by the same amount as bulk K between a star and planet.

For our models then, $C_X^{\rm{planet}}$ is simply a function of the ratio of $(X/\rm Mg)$ between the star and the Sun:
\begin{equation}
\label{eq:CX_fin}
    C_{X}^{\rm{planet}} (t) =  \frac{\left(X/Mg\right)_{Star}}{\left(\rm X/Mg\right)_{\rm Sun}}*\rm C_{X}^{Earth-like} (t)
\end{equation}

This method is similar to that used by \citet{Frank14} in their determination of a ``cosmochemically Earth-like'' planet, however our model is able to account for the system-to-system variation in HPE concentrations due to inefficient mixing within the Galaxy. For all HPE concentrations relative to the Sun we adopt the Solar composition model of ref. \citet{Lodd09}.

\subsection{Observational Range of HPE Concentrations}
\label{sec:obs_vals}
We compile measured stellar Th abundances from \citet{Unterborn2015} and \citet{Bote18} for a sample of 72 Solar twins and analogs (Main Text Figure 1, Left). Solar twins and analogs are stars of similar metallicity (that is, iron abundance), mass and surface temperature to that of the Sun. Because of these similarities to the Sun, systematic uncertainties in abundance measurements due to the assumed stellar atmosphere model are minimized, particularly of trace elements like Th. The reported Th abundances do not include abundance information for Mg, instead providing only Si. To correct for this, we assume a constant Si/Mg molar ratio equal to that of the Sun (Si/Mg = 0.95; \citet{Lodd09}) for each of these stars. The distribution of stellar Si/Mg abundances are between $0.9\pm0.2$ by mole \citep{Hink18}, thus the uncertainty introduced in our conversion from Th/Si to Th/Mg does not drastically increase the range of planetary Th concentrations that we explore across our Monte-Carlo thermal models. Assuming these abundances follow a log-normal distribution, we calculate an average current-day value ($t = 12.5$ Gyr after the birth of the Milky Way) of $\left(\rm Th/Mg\right)_{\rm star} = 1.21$ times our chosen Solar value (Table \ref{tab:vals}) with the 95\% confidence interval between 0.77 and 1.88 times Solar (Main Text Figure 1, Left). 

Unlike thorium, uranium has yet to be measured in young Sun-like stars. Additionally, the isotopic ratio of $^{235}\rm U/^{238}\rm U$ has not been measured in any system outside of the Solar System. In the absence of direct observational constraints on U, we adopt Eu as its proxy. The ratios of r-process elements (i.e., Eu, Th, and U) in stars are remarkably well correlated for extremely old, metal-poor stars with r-process enhancements \citep[][ and references therein]{Beer05, Roed09,Barb11,hansen17}. The nucleosynthetic origins of third-peak r-process elements are observationally and theoretically correlated \citet{Gori01,Freb07}. For the Sun, log(U/Eu) $\approx$ -1 (by mole); that is, U is depleted by a factor of 10 relative to Eu. Given that ultra- and hyper-metal poor stars in particular may reflect element production and enrichment from single or a few events, we would expect a similarly narrow variation in both U and Eu, with little change in their abundances relative to each other due to their co-production. The concentration of Europium, defined as Eu/Mg, then is a viable proxy for predicting the system-to-system variation in U abundances. Europium too is as refractory as U, with a 50\% condensation temperature of 1356 K \citep{Lodd03}. Therefore the concentration of Eu as determined from Eu/Mg is, like U, not expected to vary between host-star and exoplanet by more than tens of percent, i.e., less than the total observed range. For comparison, the Bulk Silicate Earth has Eu/Mg by mole of $10.5\times10^{-8}$ \citep{McD03}, while the Sun's value is nearly the same at $9.6\times10^{-8}$ \citep{Lodd09}. We, therefore, adopt Eu/Mg as a proxy for the distribution of U stellar abundances and set $f_{\rm{Eu}}$ to 1 for both the planet and Earth-like values. Both $^{235}$U and $^{238}$U isotope abundances are sampled independently from the Eu distribution.

Though still difficult to measure, Eu is observed in stars with reasonable frequency. We adopt a data set of 2040 FGK stars with measured Eu and Mg abundances in the Hypatia Catalog (Figure 1 of main text, center; \citet{Hink14}). We assume this data set to be an upper-limit of the range of Eu abundances relative to the Sun, as these measurements are inherently less precise than abundance determinations from Solar twins and analogs. Assuming these abundances follow a log-normal distribution and that U/Eu is constant throughout the Galaxy (U/Eu = 1/10), we calculate an average current-day  (Eu/Mg)$_{\rm {star}}$ = (U/Mg)$_{\rm {star}}$ = 0.93 times the Solar ratio (Table \ref{tab:vals}), with 95\% of of our sample falling between 0.45 and 1.92 times Solar (Main Text Figure 1, center). 

Bulk K/Mg ratios show a larger range of variation than Th/Mg and Eu/Mg ratios (Main Text Figure 1, right). There are 179 FGK stars with both K and Mg reported in the Hypatia catalog \citep{Hink14}. Assuming these abundance ratios follow a log-normal distribution, we calculate an average current-day  (K/Mg)$_{\rm {star}}$ of 1.13 times the Solar ratio (Table \ref{tab:vals}) with 95\% of all data falling between 0.35 and 3.63 times Solar (Main Text Figure 1, right). Only the single isotope $^{40}$K is radioactive, but no data exist for $^{40}$K/K ratios outside of the Solar System. For this model setup, a variation in a planet's $^{40}$K/K will effectively act as an increase or decrease in bulk K, similar to our discussion of $f_K$ above. We discuss the consequences of variable volatility of K on our model results in the main text and Appendix B.

By taking our data from the Hypatia catalog, we are implicitly combining abundances from different sources with different measurement uncertainties. Because we adopt the median abundance value if multiple sources are available for the same star, we likely overestimate the range of any abundance ratio in Figure \ref{fig:abund_histograms}. 

Our Monte-Carlo models described below randomly sample within each distribution of Figure \ref{fig:abund_histograms} independently. There is observational evidence from Solar Twins that stellar Th/Eu is roughly constant through Galactic time \cite{Bote18}, suggesting that these abundances are correlated. It is not observationally known, however, whether this extends to U/Th in metal-rich stars. Both U and Th are produced via the r-process, and thus their abundances in stars do correlate somewhat. $^{40}$K, however, is produced via the s-process and its correlation with the r-process elements is not known. Bulk K is produced via explosive oxygen burning \citep{Shimansky03}, and has been observationally shown to correlate with the $\alpha$-elements \citep[e.g., Mg;][]{Zhang06}, and is therefore unlikely to correlate with Eu, Th or U through Galactic time. The models of \citet{Frank14} do capture each of these behaviors, thus our treatment of $C_{X}^{\rm{planet}} (t)$ does somewhat capture the correlation between Eu, U and Th (Figure \ref{fig:rat_frank}). Our treatment of the system-to-system variations implied from Figure \ref{fig:abund_histograms}, causes our model to explore some areas of U and Th parameter space that is unlikely. Given that we find these elements to have a minor effect on the longevity of mantle degassing, however, these inclusions of Eu, Th and U correlations is not likely to substantially change our determinations of Age$_{\rm{max}}$.

\subsection{Monte-Carlo Model}
\label{sec:final_structure}

\begin{figure}[h!]
    \centering
    \includegraphics[width=\linewidth]{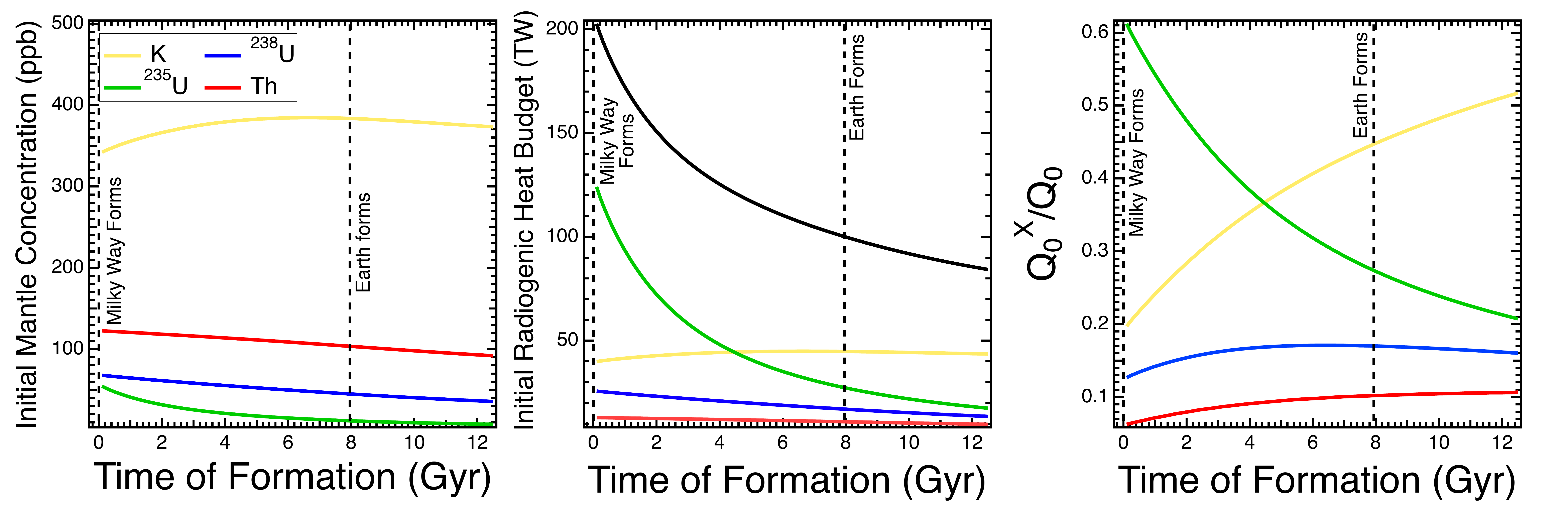}
    \caption{\textbf{Left: }Initial mantle abundances (left), initial radiogenic heat budget ($Q_0$; center) and fraction of total heat budget (right) for a cosmochemically Earth-like planet the as function of time of planet formation from \citet{Frank14} and adopting the current Earth HPE abundances of \citet{palme2003} for $^{40}$K (yellow), $^{238}$U (blue), $^{235}$U (green) and Th (red). The time of Earth formation ($t$ = 8 Gyr) is shown for reference.}
    \label{fig:rat_frank}
\end{figure}

For this work we adopt a Monte-Carlo method for determining the degassing lifetime of planets with a fixed size, CMF, and formation age (in terms of time after Galaxy formation). In each Monte-Carlo suite, $\sim 10^4-10^5$ models are run. In each run  $C_{X}^{planet}$ is determined using equation \ref{eq:CX_fin} by independently sampling from the log-normal distributions of each HPE as shown in Figure 1 of the main text and multiplying by the $C_{X}^{Earth-like}$ corresponding to its formation time after the birth of the Milky Way ($t = 0$, Figure \ref{fig:rat_frank}). This method allows us to simulate both the galactic chemical evolution of the HPEs over Galactic history as well as system-to-system variation of these HPEs due to their stochastic distribution due to inefficient mixing of these elements throughout the Galaxy. 

Unlike the HPE abundances, which can be constrained by stellar observations, other factors that influence a planet's thermal evolution can not be constrained observationally. We account for variation in two of the most important of these factors, the initial mantle temperature and mantle reference viscosity, in our models by randomly sampling from uniform distributions across reasonable uncertainty ranges. For initial mantle temperature, we use a range of 1700--2000 K, and for mantle reference viscosity we use a range of $10^{19}-10^{23}$ Pa s. The uniform distribution for reference viscosity is sampled in log space, or, in other words, we sample from a uniform distribution of $\log_{10}({\mu_{\mathrm{ref}}}) = 19-23$. The range of mantle viscosities considered covers two orders of magnitude above and below typical estimates for the average viscosity of Earth's mantle, based on post-glacial rebound studies \citep[e.g., ][]{Mitrovica2004}. Initial mantle temperature is not well constrained for Earth or any planet, but our assumed range covers the spread typically used in studies of solar system planets \citep[e.g., ][]{Hauck2002,Fraeman2010,Morschhauser2011,Breuer2015,Foley2020_div}. With the reference viscosity, initial mantle temperature, and radiogenic heat production rate set, the initial stagnant lid thickness is calculated assuming that the conductive heat flux at the base of the lid matches the advective heat flux supplied by the convecting mantle. The last remaining parameter to set is the total carbon budget of the mantle and surface reservoirs, C$_{\rm{tot}}$. 

For our suites of Monte-Carlo models, the concentration of CO$_2$ in the mantle at the model start time is set to $\sim5\times 10^{-3}$ wt\%, regardless of planet mass or other assumed properties. We examined the effects of higher or lower mantle C contents and found that the lifetime of mantle degassing was insensitive to the total mantle C concentration above $\sim10^{-5}$ wt\%, in agreement with \citep{Foley17}. Below this threshold, the mantle is more likely to exhaust its entire C budget causing degassing to end even if there is a sufficient HPE budget to support volcanism for longer. Above $\sim10^{-5}$ wt\%, there is sufficient mantle CO$_2$ to support degassing until the planet's dwindling heat budget no longer supports mantle melting and volcanism. We also assume no CO$_2$ in the atmosphere at model start time, and, since the models start with no crust present, no CO$_2$ in the crust either. Whether a planet's C would initially lie entirely in the mantle, as we assume, in the atmosphere, or a combination of the two, is not known. However, \citet{Foley2019_stag} found the initial distribution of C between surface and interior does not significantly affect subsequent climate evolution, at least when liquid water is present and silicate weathering can occur. For an Earth-sized planet, $\approx 5 \times 10^{-3}$ wt\% CO$_2$ scales to a total C budget of $5 \times 10^{21}$ mol; this value is about a factor of 2 lower than the estimate for the Earth given by \citet{sleep2001}. 

Since models are presented in terms of a CO$_2$ concentration in wt\%, larger planets will have larger total C budgets than smaller planets, for the same CO$_2$ concentration. In calculating rates of CO$_2$ outgassing, our models assume the concentration of CO$_2$ in the source region to mantle melts is set by the (time-evolving) bulk mantle CO$_2$ concentration. We use melt-solid partition coefficients to then calculate the CO$_2$ concentration in the resulting melt, and hence rate of CO$_2$ outgassing. In doing so we implicitly assume the mantle oxidation state of the planets we model is the same as the present day Earth. A more reduced mantle would favor production of more reduced gases at the expense of CO$_2$, and lower the CO$_2$ outgassing rate \citep[e.g., ][]{Tosi17}. More reduced conditions also reduce the solubility of C species in mantle melt \citep[e.g., ][]{Grew20}. A more reduced mantle would thus be equivalent to lower mantle CO$_2$ concentrations in our models.  

All parameters varied in our model are shown in Table \ref{tab:params}.

\clearpage

\begin{table}[htbp]
    \caption{Range and sampled distribution type for Monte-Carlo models}

    \centering
    \begin{tabular}{l|c|c}
        Parameter & Sampled Range & Distribution Type \\
        \hline
         K/Mg/(K/Mg)$_\odot$&  Avg:1.38; 95\% CI: 0.37--3.67& log-normal distribution \\
         Th/Mg /(Th/Mg)$_\odot$ & Avg:1.24; 95\% CI: 0.77--1.88 & log-normal distribution \\
         U/Mg/(U/Mg)$_\odot$&  Avg:0.99; 95\% CI: 0.45--1.92 & log-normal distribution \\
         Mantle Reference Viscosity & $10^{19}$--$10^{23}$ Pa s & flat distribution \\
         Initial Mantle Temperature & 1700--2000 K & flat distribution \
    \end{tabular}
    \label{tab:params}
\end{table}

\begin{table}[htbp]
  \centering
  
  \caption{Sample of likely solid rocky planets}
    \resizebox{\linewidth}{!}{
    \begin{tabular}{lcccccc}
      & Mass & Density & Age & T$_{eq}$ &\% Probability & \% Probability \\
        Planet &  (M$_{\oplus}$)& (g cm$^{-3}$) & (Gyr) & (K) & Age $\leq$ Age$^{\rm{Avg}}_{\rm{max}}$ & Age $\leq$ Age$^{\rm{UL} \rm{95\%} \rm{CI}}_{\rm{max}}$ \\
    \hline
    K2-36 b & 3.9$\pm$1.1  & 7.3$\pm$2.1& 1.4$\pm$0.45  & 1224 & 100   & 100 \\
    Kepler-80 e & 2.6$\pm$0.75 &6.5$\pm$1.9& 2$\pm$1 & 629 & 70  & 99 \\
    Kepler-65 d & 4.14$\pm$0.8 &6.5$\pm$1.2 & 2.9$\pm$0.7  &1117& 56    & 100 \\
    Kepler-138 c & 1.97$\pm$1.5 & 6.3$\pm$4.9  & 4.68$\pm$4.17 & 402 & 26    & 44 \\
    Kepler-105 c & 4.6$\pm$0.9 & 11.2$\pm$2.2 & 3.17$\pm$0.6   & 997 & 45    & 100 \\
    Kepler-345 c & 2.2$\pm$0.9 & 7.0$\pm$2.9   & 2.75$\pm1.7$  & 575&  42     & 83 \\
    Kepler-197 c & 5.3$\pm$3.1 &15.6$\pm$9.2  & 5.37$\pm$3.1   & 930 & 2     & 63 \\
    Trappist-1b & 1.374$\pm$0.07 &5.4$\pm$0.3& 7.6$\pm$2.2  & 401 & 1     & 5 \\
    Trappist-1g & 1.321$\pm$0.038  &5.0$\pm$0.1 & 7.6$\pm$2.2    & 199 & 1     & 5 \\
    Trappist-1c & 1.308$\pm$0.056 &5.4$\pm$0.1 & 7.6$\pm$2.2    & 341 &1     & 5 \\
    Trappist-1f & 1.039$\pm$0.031 & 5.0$\pm$0.1 & 7.6$\pm$2.2     & 219 &0     & 4 \\
    Kepler-36 b & 3.83$\pm$0.1  & 6.3$\pm$0.2 & 4.79$\pm$0.65   &  978 &0     & 85 \\
    HD 219134 b & 4.74$\pm$0.19 &6.3$\pm$0.03 & 11$\pm$2.2  &1015& 0     & 1 \\
    HD 219134 c & 4.36$\pm$0.22 &6.9$\pm$0.4  & 11$\pm$2.2     & 782&0     & 1 \\ 
    Kepler 93 b & 4.54$\pm$0.85 &6.5$\pm$1.2 & 6.6$\pm$0.9 &1037&0&17\\
    Kepler-68 c & 2.04$\pm$1.75 &14.4$\pm$12.3   & 6.31$\pm$0.82  &941& 0     & 6 \\
    HD 136352 b & 4.62$\pm$0.45 &7.2$\pm$0.7   & 8.2$\pm$3.2  &911& 5     & 22 \\
    \end{tabular} }
  \label{tab:planets}
\end{table}

\begin{table}[htbp]

\begin{threeparttable}
\label{tab:vals}
\centering
\caption{Model inputs for HPEs}

\begin{tabular}{c|c|c|c|c}
     &Half-life & Current Earth: $t$ = 12.5 Gyr & Earth: $t$ = 8 Gyr & Solar$^{\ddagger}$ \\
     Element & (Gyr) & abundance (ppb)$^{\dagger}$ & abundance (ppb) & $X$/Mg (by mol) \\
     \hline
     $^{40}$K & 1.25 & 30.5 & 382 & 3.8$\times10^{-3}$$^{*}$ \\
     $^{238}$U & 4.47 & 22 & 44 & 9.6$\times10^{-8}$$^{**}$ \\
     $^{235}$U & 0.704 & 0.16 & 14 & 9.6$\times10^{-8}$$^{**}$ \\
     $^{232}$Th & 14 & 83 & 104 & 3.5$\times10^{-8}$ \\
     \hline
\end{tabular}
\begin{tablenotes}
\small
\item $^{\dagger}$ ref. \citet{palme2003}
\item $^{\ddagger}$ ref. \citet{Lodd09}
\item $^{*}$ Bulk K 
\item $^{**}$ Bulk Eu
\end{tablenotes}

\end{threeparttable}

\end{table}

\begin{figure}[h!]
    \centering
    \begin{tabular}{c}
             \includegraphics[width=\linewidth]{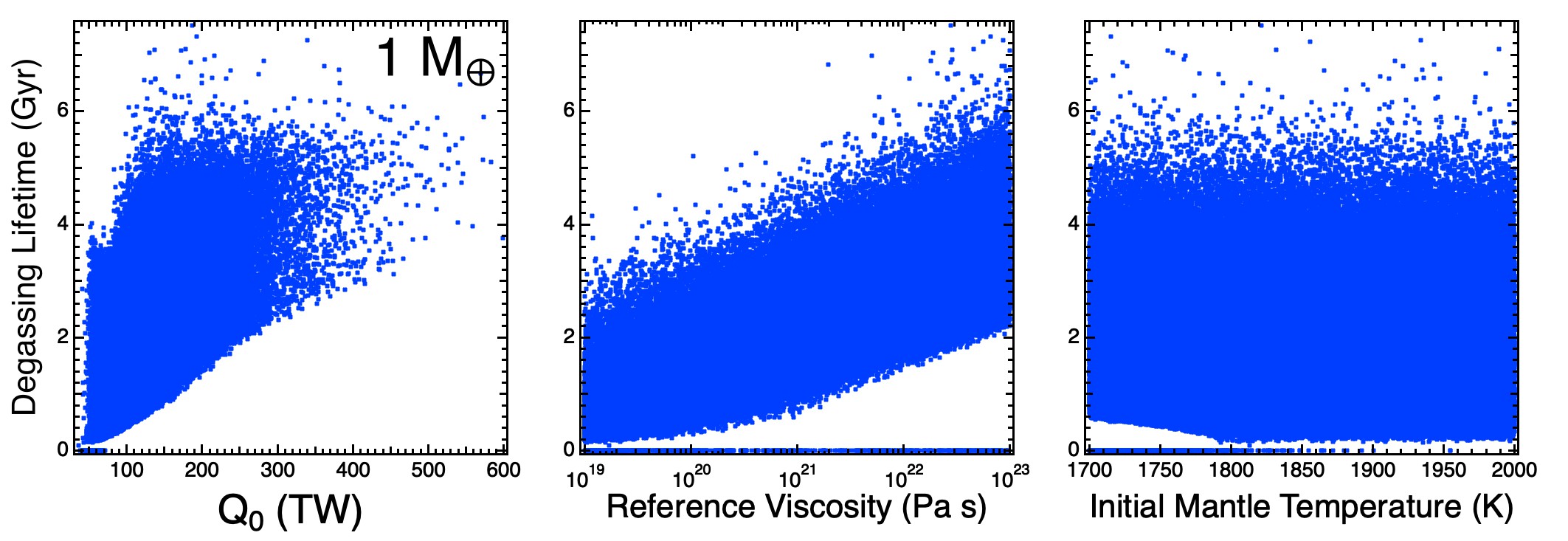}  \\
             \includegraphics[width=\linewidth]{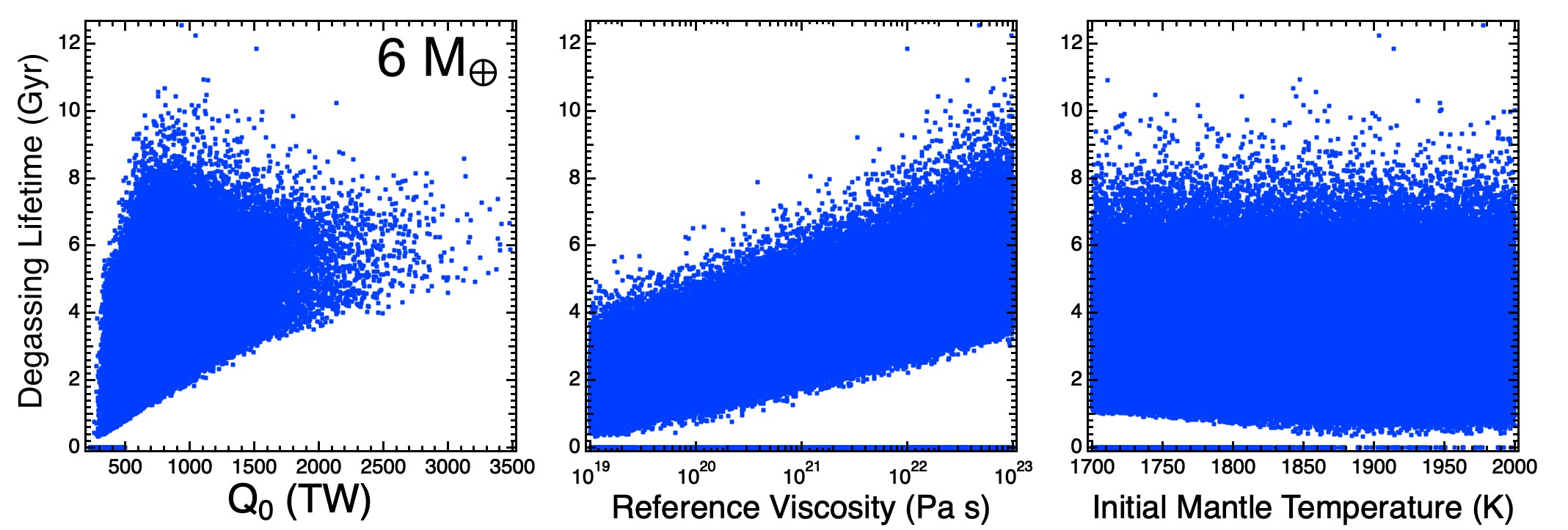}
    \end{tabular}

    \caption{Degassing lifetime as a function of initial radiogenic heat budget (left), reference viscosity (center), and initial mantle temperature (right) for 1 M$_\oplus$ (top) and 6 M$_\oplus$ planets with Earth-like core mass fraction (0.33) that formed at the same time as the Earth ($t$ = 8 Gyr).}
    \label{fig:all_params}
\end{figure}

\begin{figure}[h!]
    \centering
    \includegraphics[width=0.9\linewidth]{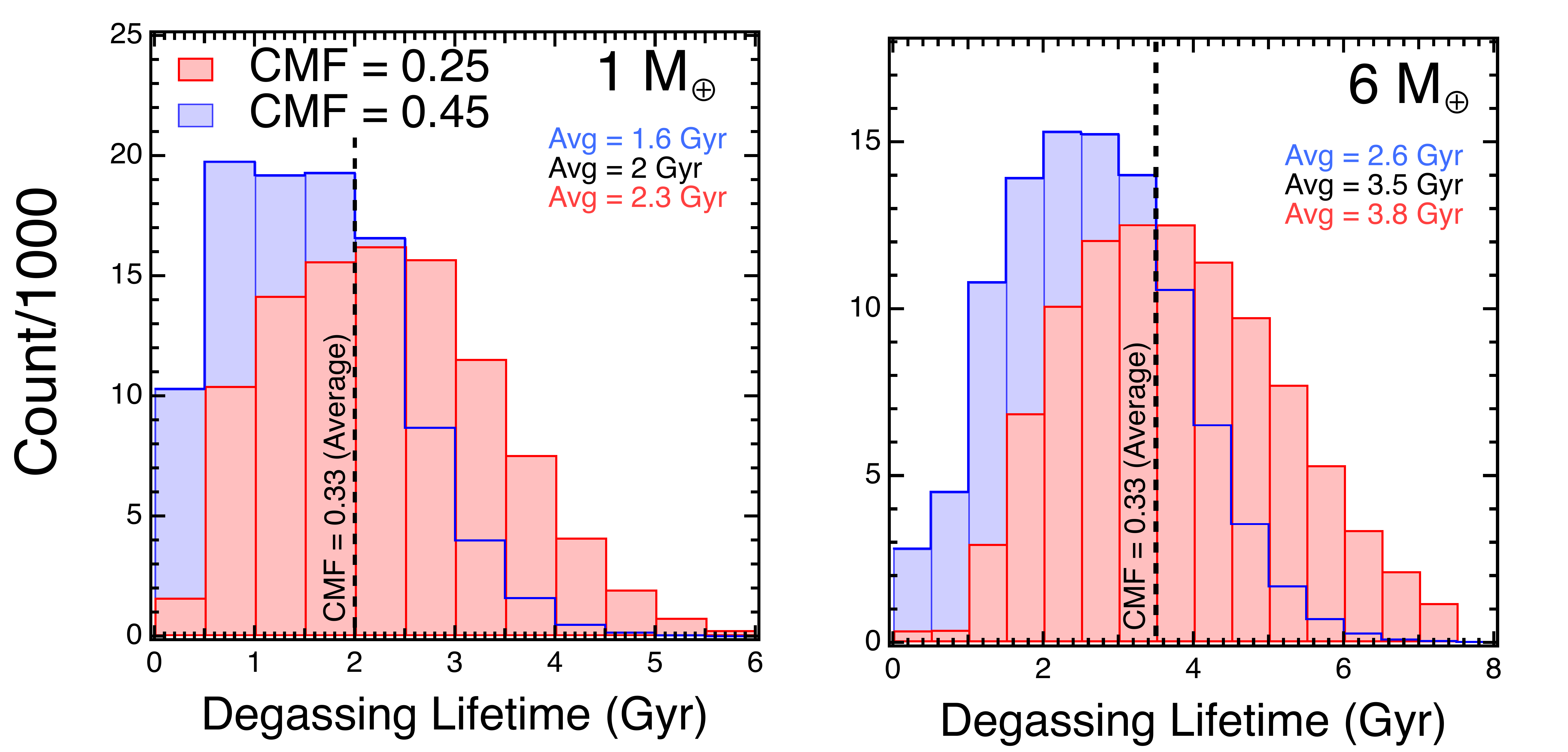}
    \caption{Histograms of the degassing lifetime for 1 (top) and 6 (bottom) M$_{\oplus}$ planets with core mass fractions of 0.25 (red) and 0.45 (blue) for a planet forming at the same time as the Earth ($t$ = 8 Gyr). The average of our CMF = 0.33 models (Figure 2 of main text) for each mass is shown as black dashed line. These histograms were derived from 200,000 random samplings of our parameters as described above.}
    \label{fig:core}
\end{figure}

\begin{figure}[t!]
    \centering
    \includegraphics[width=\linewidth]{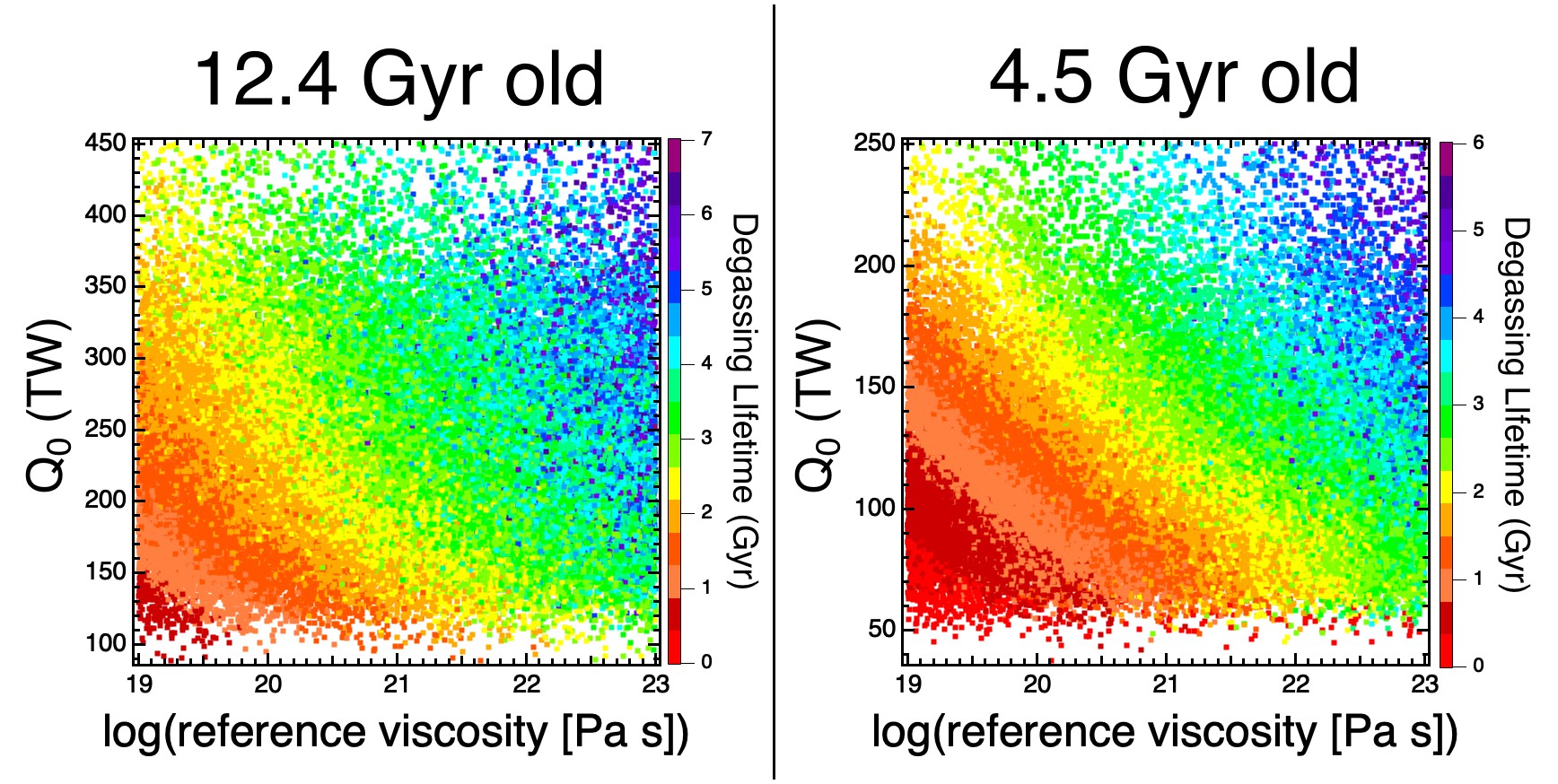}
    \caption{Age when degassing ends for 1 M$_\oplus$ (colors) as a function of reference viscosity and the initial radiogenic heat budget for formation times 12.4 (left) and 8 (right) Gyr into the past. Figure created using 50,000 random samples as described above. For comparison the Earth's reference viscosity is $\sim10^{21}$ Pa s.}
    \label{fig:h0visc}
\end{figure}

\begin{figure}[t!]
    \centering
    \includegraphics[width=\linewidth]{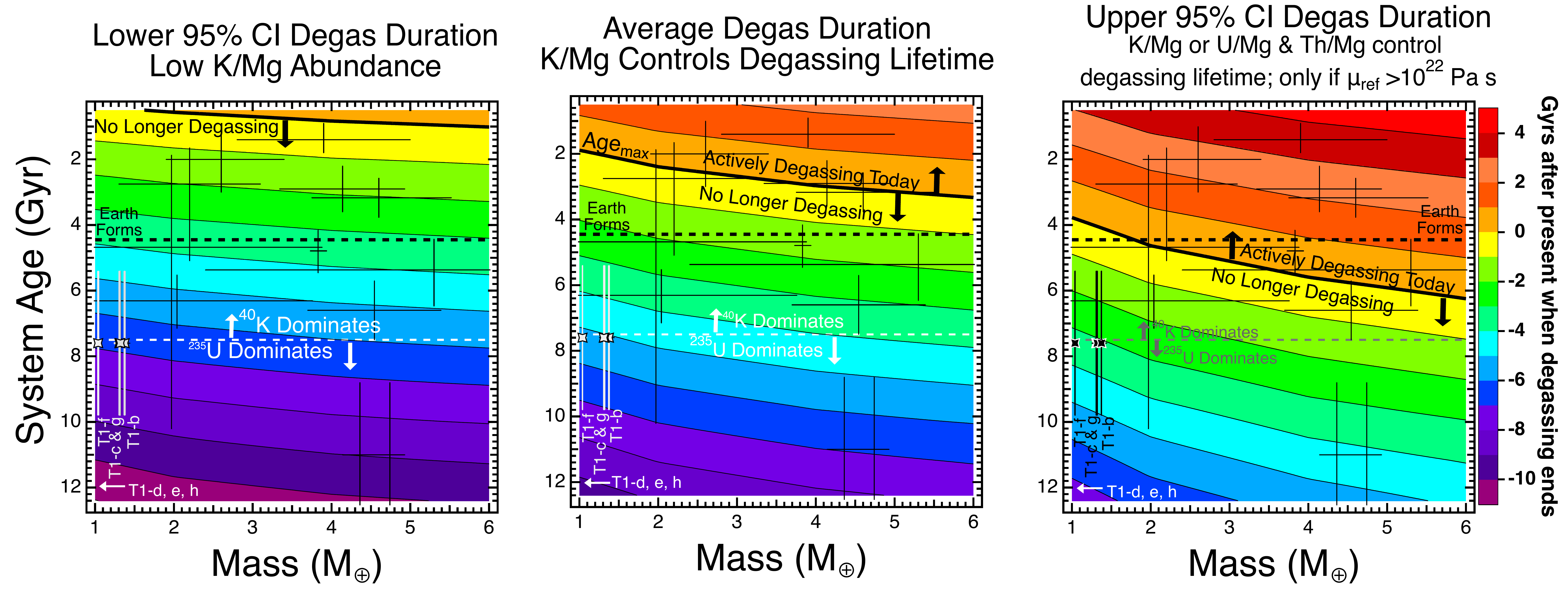}
    \caption{Planet age when degassing ceases (present = 12.5 Gyr) as a function of planetary mass and current system age adopting the average (center) and lower (left) and upper bounds of our 95\% confidence intervals of degassing lifetime for planets of a given mass and time of formation (Figure 2 of main text). Those 17 planets with measured mass, radius and system age and densities consistent with being rocky are included for reference (Table \ref{tab:planets}). Uncertainties represent the reported 1$\sigma$ error bars in both mass and age. The TRAPPIST-1 planets are highlighted adopting system ages of \citet{Burg17} and planet masses of \citet{Agol20}.}
    \label{fig:TimeMass}
\end{figure}

\begin{figure}[h!]
    \centering
    \includegraphics[width=\linewidth]{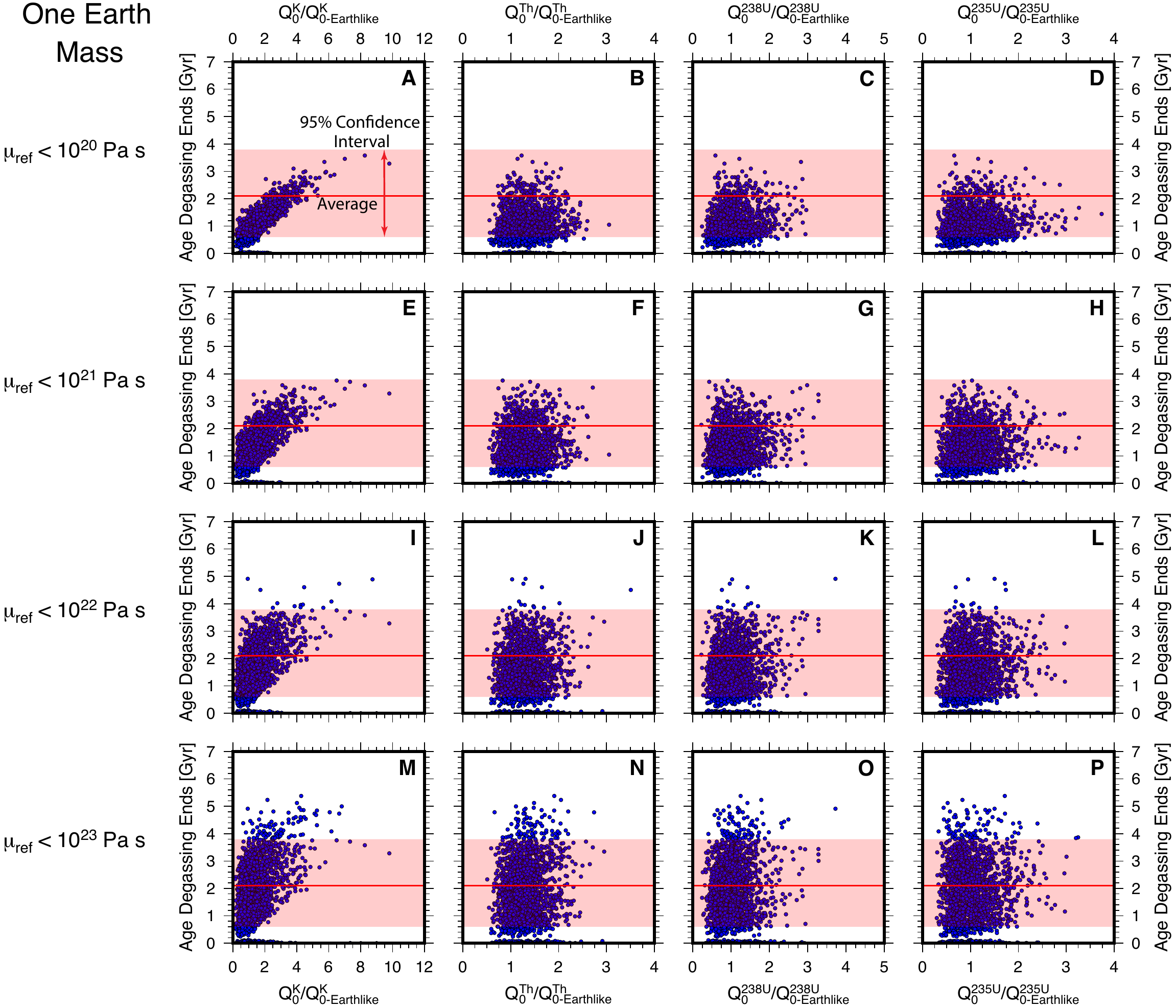}
    \caption{Age when degassing ends for a 1 M$_{\oplus}$ planet as a function of initial heat budget for an HPE $X$ ($Q^X_0$) relative to the initial heat budget for the same element for the Earth ($Q^X_{\rm{0-Earth}}$; see Table \ref{tab:vals}), for restricted ranges of mantle reference viscosity: $\mu_{\mathrm{ref}}<10^{20}$ Pa s (top row), $\mu_{\mathrm{ref}}<10^{21}$ Pa s (second from top row), $\mu_{\mathrm{ref}}<10^{22}$ Pa s (second from bottom row), and $\mu_{\mathrm{ref}}<10^{23}$ Pa s (bottom row). Models were run assuming planets formed at the same time as the Earth ($t$ = 8 Gyr). The average (red line) and 95\% confidence intervals in the degassing cessation age (red band) are included for reference. In each plot only 2,000 models are shown to reduce the density of data points. }
    \label{fig:hpe_plot2}
\end{figure}

\begin{figure}[h!]
    \centering
    \includegraphics[width=\linewidth]{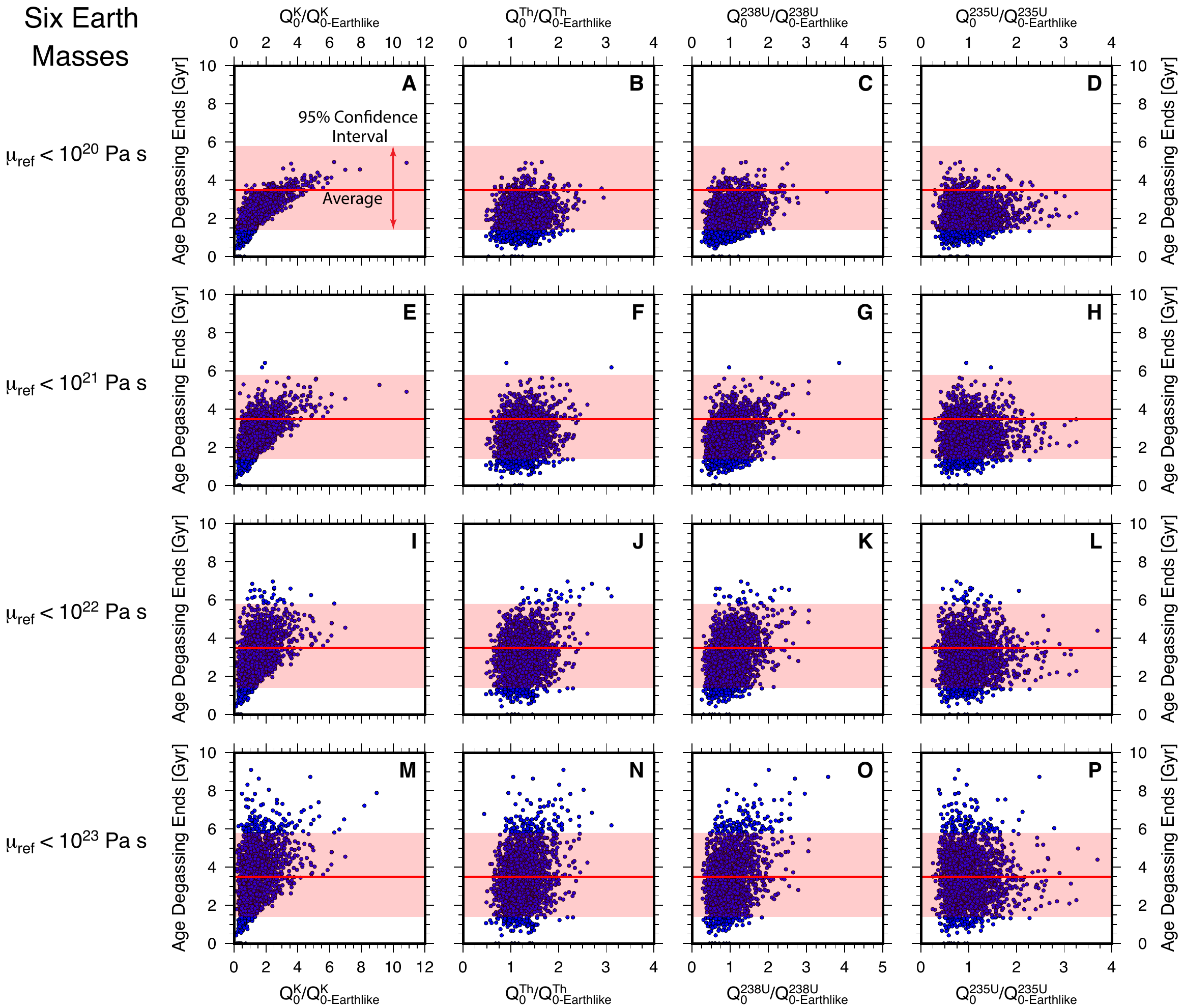}
    \caption{Same as Figure \ref{fig:hpe_plot2}, but for a 6 M$_{\oplus}$ planet.}
    \label{fig:sixME_hpe_plot2}
\end{figure}
\clearpage
\restartappendixnumbering
\section{Compositional changes to Age$_{\rm{max}}$}
\subsection{Effects of Galactic Chemical Evolution on Degassing Lifetime}
\label{sec:GCE}
As the Galaxy aged HPEs were produced via supernovae processes while those already existing decayed. This is all while the non-radioactive rocky-planet-building elements were continuously being produced, thus diluting the concentration of the HPEs as the Galaxy aged (Figure \ref{fig:rat_frank}). This, in turn, causes the average $Q_0$, and thus degassing lifetimes, to also decrease as the Galaxy aged (Figure \ref{fig:rat_frank}, center). The exact makeup of a planet's HPE budget too affected degassing lifetimes. Those planets older than $\sim$8 Gyr had initial heat budgets composed primarily of $^{235}$U. For planets younger than 8 Gyr, $^{40}$K instead became the dominant component of its initial HPE budget. The shorter half-life of $^{235}$U (704 Myr) compared to $^{40}$K (1.25 Gyr), causes older planets to release their internal heat more rapidly. As a result, the oldest, $^{235}$U-dominant planets degas for only $\sim$1 Gyr longer than $^{40}$K-dominant planets formed at the same time as the Earth, despite having nearly double the $Q_0$ and 5 times as much $^{235}$U.

These results suggest that as the Galaxy evolves, newer planets may form with a lower HPE budget than older ones (Figure \ref{fig:rat_frank}). While we do not extrapolate our results into the future, this potential decrease in HPE abundance would cause degassing lifetimes for stagnant-lid exoplanets to also decrease. In order to to counteract the decay and subsequent dilution of radionuclides as the Galaxy evolves, the likelihood of a rocky exoplanet supporting a temperate climate may be more reliant on the local emplacement of HPE-rich material from supernovae sources to boost its radiogenic heat budget above the average from Galactic chemical evolution. Exploring these effects are beyond the scope of this paper, but our results point to the need for more work exploring the spatial and temporal distribution of the long-lived HPEs due to Galactic chemical evolution. 

\subsection{Planetary K Concentration}
\label{sec:planet_K}
We first consider possible sources of error or bias in our estimated probability distributions for the mantle abundances of the four major HPEs in rocky exoplanets as derived from stellar chemistry observations. Of the four HPEs we considered, a rocky exoplanet's mantle concentration of $^{40}$K is dependent on both the disk's bulk abundance of $^{40}$K and depletion of K relative to the refractory rock-forming elements during planetary formation. Unlike Th and U, K is moderately volatile, meaning there could be significant differences between measured stellar abundances and resulting planetary abundances. In our Solar System, accretion of chondrites themselves cannot explain the depletion of K by the factor of 5 seen in Earth relative to the Sun. Moderately volatile elements within chondrules (igneous melt spherules), however, appear to be  depleted with respect to the matrix grains by factors comparable to the requisite factor of 5 depletion of K relative to Mg between Earth and Sun. Perhaps the preferential accretion of chondrules (rather than chondrites) by Earth may explain its depletion \cite{Desch20}, however, no first-principle models exist that quantitatively predict the degree of depletion of the volatile elements in planetary materials (e.g., see review by \cite{Alex05}). Whether preferential melting, vaporization, or differentiation play a role in the degree of K volatilization during planet formation, is an open, but critical, question in understanding whether a rocky exoplanet is likely to be temperate today. Until such time as a theory is developed to quantify the range of possible frationation, all that can be said is that unlike Th/Mg and U/Mg ratios, the K/Mg ratio of a planet cannot be expected to match its host-star and a volatility/fractionation factor must be assumed for exoplanetary systems. If planets were composed of purely chondritic material, stellar abundances matching planetary abundances might be the case, but based on our solar system, it is more likely that the planet has a lower K/Mg than its host-star. This degree of volatilization then remains unknown. Given its effect on the degassing lifetime of rocky exoplanets, work by both the meteoritics and planetary formation communities is critically needed to quantify this parameter in order to better constrain the evolution of rocky exoplanets.

Additionally, the $^{40}$K/K ratio for other stars, and hence planetary systems, is unconstrained. In our previous models, we simply assumed the same depletion factor between the Earth and Sun for all planets, and hence the same volatility, and the same $^{40}$K/K ratio as the Earth. However, system-to-system variation in these properties would lead to an even wider distribution of $^{40}$K compared to those in Figure 1 of the main text. The probability of a planet acquiring a larger $^{40}$K concentration could therefore be higher than we previously estimated.  

\begin{figure}
    \centering
    \includegraphics[width=0.6\linewidth]{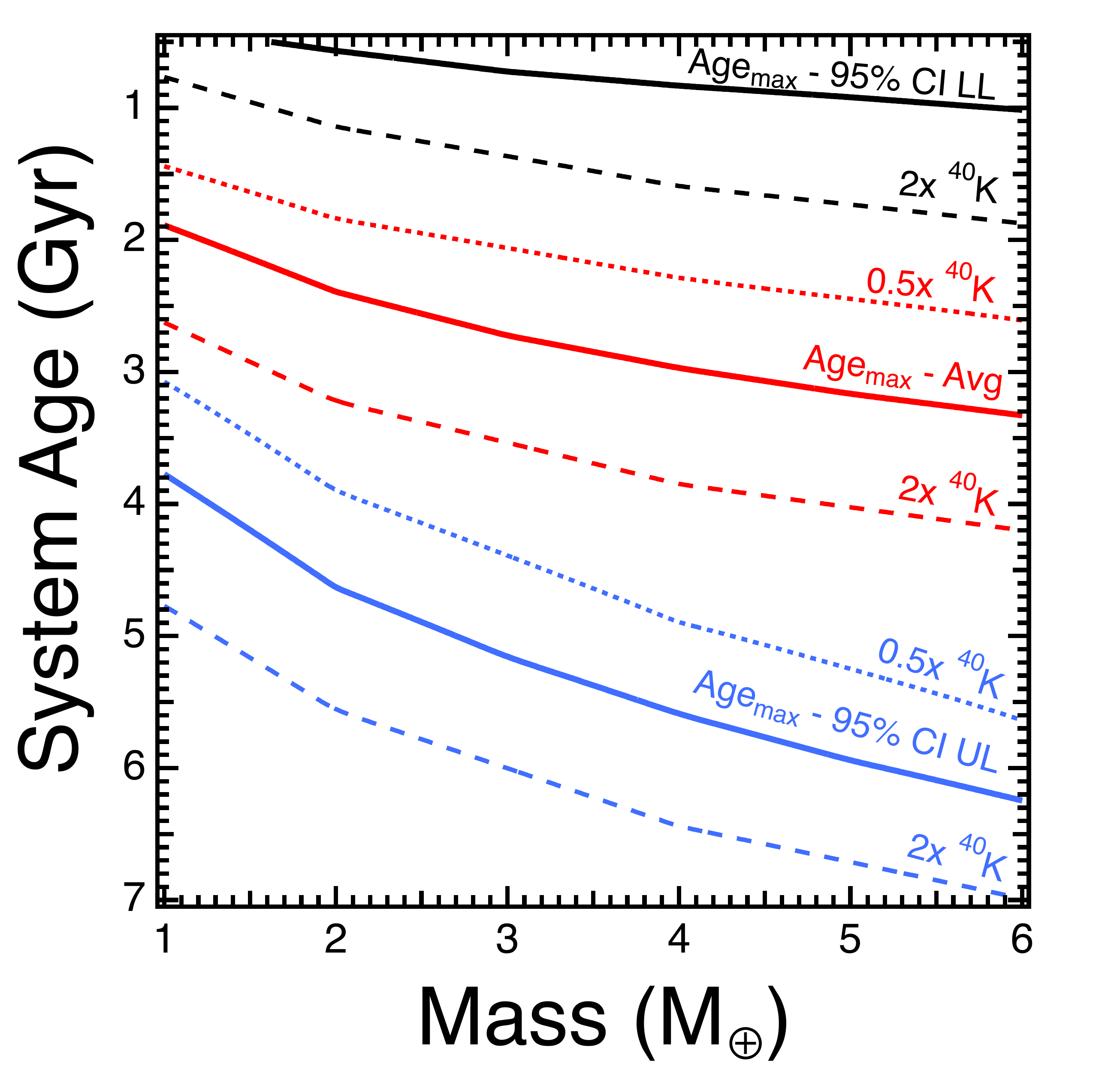}
    \caption{Best fit average (red) and upper- (blue) and lower (black) limit of the 95\% confidence interval of Age$_{\rm{max}}$ assuming that $^{40}$K is arbitrarily halved (dotted), left Earth-like (solid) and doubled (dashed) in equation \ref{eq:vol} for 25,000 model iterations for each curve. }
    \label{fig:Kchange}
\end{figure}

To assess how changes in volatility and $^{40}$K/K affect Age$_{\rm{max}}$ (Figure \ref{fig:TimeMass}), we reran our models for determining the degassing lifetime as a function of planet formation time and mass (Figure 2 of main text), and arbitrarily changed the abundance of $^{40}$K ($f_{x}^{K}$, equation \ref{eq:vol}) by a factor of 0.5 and 2. Halving a planet's $^{40}$K abundance decreases Age$_{\rm{max}}^{\rm{Avg}}$ by $\sim500-700$ Myr and Age$_{\rm{max}}^{\rm{UL95\%CI}}$ by $\sim600-750$ Myr, with this difference increasing as mass increases (Figure \ref{fig:Kchange}). Doubling the K abundance yields increases in Age$_{\rm{max}}^{\rm{Avg}}$ by $\sim700-850$ Myr and Age$_{\rm{max}}^{\rm{UL95\%CI}}$ by $0.7-1$ Gyr. While observations of a host-star's K/Mg abundance can allow us to estimate the degassing lifetime of a rocky exoplanet, differences in the volatilization and/or the system's $^{40}$K/K abundances compared to Earth will have a direct effect on the location of Age$_{\rm{max}}$. Next, we will explore the degree of potential K volatility and the likelihood of systems having non-Earth $^{40}$K/K ratios, for comparison to these estimates of $^{40}$K-dependent Age$_{\rm{max}}$.

\subsection{Enrichment of $^{40}$K via supernovae injections and background concentration}
\label{sec:K_super}
An efficient method of $^{40}$K enrichment might be supernova injection of material into the protoplanetary disk. Supernova remnants frequently exhibit large-scale bipolar or unipolar morphologies \citep[e.g., ][]{Leonard00, Ouellette07, Larsson13, Reilly17, MoranchelBasurto17}, as well as filaments and clumpy ``bullets'' of ejected material on smaller scales \citep[e.g., ][]{Willingale02, Fesen16, Zhou16, Garcia17}. These sorts of small-scale structures are seen in hydrodynamic simulations \citep[e.g., ][]{Hammer10, Ellinger12, Wongwathanarat15}. As a result, star and planet forming structures near the supernova can be enriched by bullets of material that represent the abundances in a specific part of the explosion rather than the average yields. This material can be injected as gas into a cloud core or as dust into a protoplanetary disk \citep{Young09, Ellinger12, Pan}. It is possible to estimate the maximum enhancement that such injection can yield by comparing the amount of $^{40}$K in an ejecta clump to the total inventory in the protoplanetary material. 

\begin{figure}[h!]
    \centering
    \includegraphics[width=0.8\linewidth]{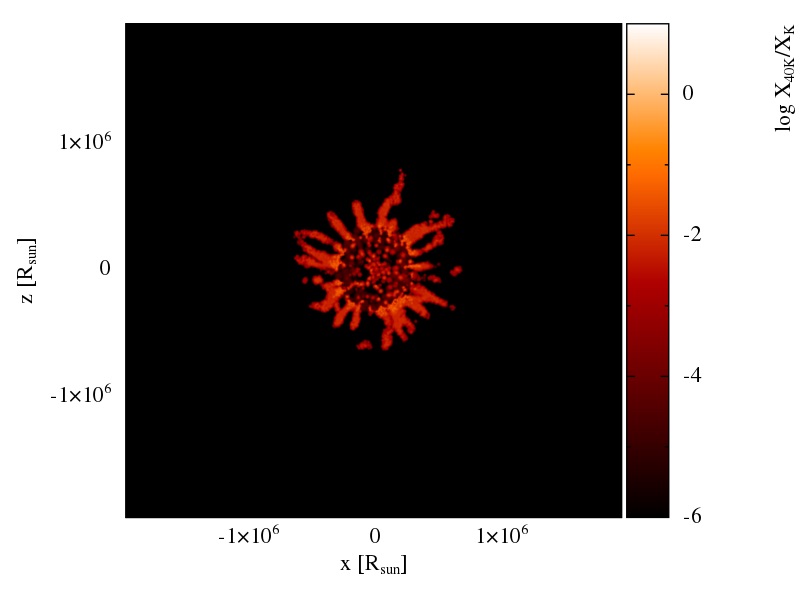}
    \caption{Cross section of ejecta in a mildly bipolar explosion of a 15  M$_{\odot}$ star. Axes are in units of solar radii. The color bar represents the log of the mass ratio of $^{40}$K to total K. The peak $^{40}$K/K$_{\rm{tot}}$ occurs in areas of large overdensity near the base of the Rayleigh-Taylor fingers. These clumps persist out to large radii during the remnant phase of the supernova.}
    \label{fig:warmjets}
\end{figure}

\begin{figure}[h!]  
    \centering
    \includegraphics[width=0.9\linewidth]{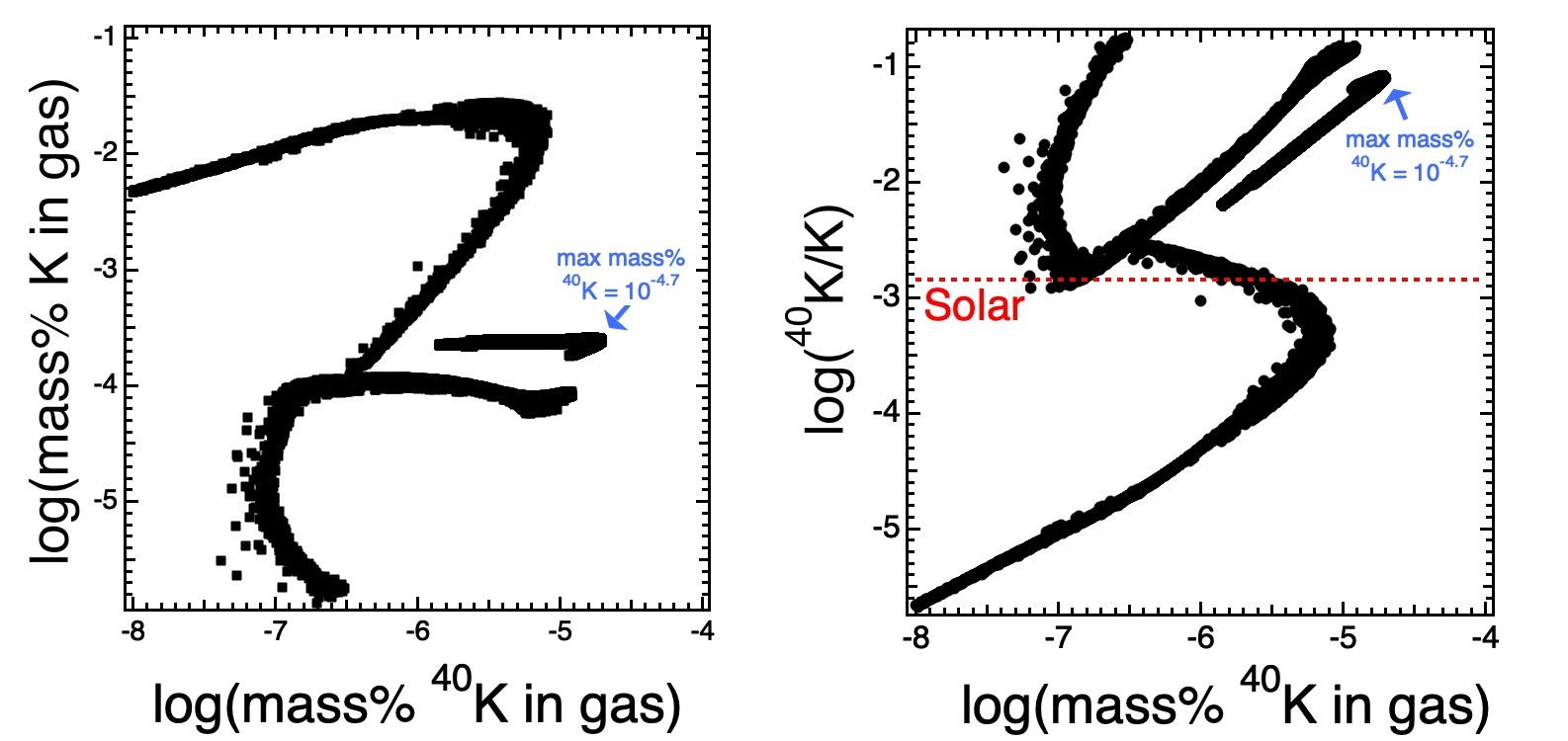}
    \caption{\textbf{Left:} Mass percentage of bulk K and $^{40}$K in the gas for a 15 M$_\odot$ progenitor. \textbf{Right:} $^{40}$K/K and versus mass fraction of $^{40}$K in the gas for the same progenitor \citep{Ellinger12}. The Sun's primordial $^{40}$K/K is shown for reference \citep{Lodd03}.} 
    \label{fig:40K}
\end{figure}

Based on ALMA observations, the dust mass in M dwarf disks is of order $10^{-5}$ M$_{\odot}$, and total disk mass for a dust-to-gas ratio of 1:100 is $10^{-3}$ M$_{\odot}$ \citep{wardduong18}. If a $10^{-3}$ M$_{\odot}$ disk initially contains a Solar K concentration of 3.7 ppm (3.7 $\mu$g K per g of disk) and the primordial Solar $^{40}$K/K (0.14\%) of \citet{Lodd03}, the mass of bulk K and $^{40}$K in the disk are $3.7\times10^{-9}$ and $5.17\times 10^{-12}$ M$_{\odot}$, respectively. For an individual 15 M$_{\odot}$ supernovae progenitor (Figure \ref{fig:warmjets}; ref. \citet{Ellinger12}), we calculate a peak mass fraction of bulk K and $^{40}$K potentially injected into the disk of $\sim 10^{-3.6}$ and $\sim 10^{-4.7}$, respectively (Figure \ref{fig:40K}). This corresponds to $^{40}$K/K$\sim10^{-1.1}$ (8\%). The average mass of a bullet in our model is $10^{-5}$ M$_{\odot}$. For a $10^{-5}$ M$_{\odot}$ bullet mass, then, we estimate $\sim$2.5$\times 10^{-11}$ M$_{\odot}$ of K and $2\times 10^{-12}$ M$_{\odot}$ of $^{40}$K are delivered to the disk via the injection of supernovae material. Injecting a single $10^{-5}$ M$_{\odot}$ bullet into our disk raises the total mass of K in the disk by $<$1\% and $^{40}$K by $\sim$39\%. This would raise the disk's primordial $^{40}$K/K to $\sim$0.19\%. The injection of this super-enriched supernovae-derived $^{40}$K material will only account for an increase in the protoplanetary disk's $^{40}$K/K by only $\sim$36\%. 

As disk mass increases above 10$^{-3}$ M$_{\odot}$, the relative mass contribution of high $^{40}$K/K material from these injections to the protoplanetary disk begins to diminish. For the minimum-mass Solar nebula of 0.013 M$_{\oplus}$ \citep{Hayashi81,Kuch04}, the contribution from a single injection increases the disk's $^{40}$K/K by $<1$\%. Thus, while $^{40}$K/K can increase somewhat due to the direct injection of supernova material, this mechanism is likely to only affect radiogenic heat budgets of planets forming in low-mass, M-dwarf disks, and only marginally, leading only to a $<$500 Myr increase in Age$_{\rm{max}}$ (Figure \ref{fig:40K}). Larger bullet masses on the order of 10$^{-4}$ M$_{\odot}$, however, are possible \citep{Young09, Ellinger12}, which would allow for a significant enrichment for higher mass disks. The relative occurrence rates of these high-mass bullets are unconstrained and beyond the scope of this paper. Thus, we offer only a conservative estimate of $^{40}$K enrichment via supernova injection. We argue then that supernovae events then are not likely to increase a planet's radiogenic heat budget and degassing lifetime enough to drastically change Age$_{\rm{max}}$ for all but those planets orbiting low-mass stars. Instead, a system's bulk K abundance and the degree of K volatilization are the most likely first-order sources for increasing a rocky exoplanet's radiogenic heat budget and degassing lifetime. In this sense then, observations of the host-star's K/Mg coupled with modeling of the range of volatility are the most likely avenues for quantifying whether a planet contains sufficient K to be actively degassing today. 

\subsection{Mantle Water Content}
\label{sec:water}
We estimate how increasing mantle water content would influence Age$_{\rm{Max}}$, taking into account the competing effects of water content on mantle viscosity and solidus. Viscosity has been found to scale as $(1/\chi_m)^r$, where $\chi_m$ is the mantle water concentration, expressed here as water mass fraction, and $r$ is an experimentally determined exponent: $r=0.7-1.2$ according to the compilation in \citet{hirth2003}. Drier mantles would thus have higher viscosities, up to a point (there is a threshold water content where the material is effectively dry, and further decreases in water content will not affect viscosity) and wetter mantles would have lower viscosities. At the same time, water content also affects the mantle solidus, with melting temperatures decreasing as mantle water content increases \citep[e.g., ][]{kushiro1968}. These two effects counteract each other: a drier mantle has a higher viscosity, acting to extend degassing lifetime and Age$_{\mathrm{max}}$, but also a higher solidus temperature, which inhibits melting and acts to lower degassing lifetime and Age$_{\mathrm{max}}$. 

\begin{figure}[h!]
    \centering
    \includegraphics[width=0.9\linewidth]{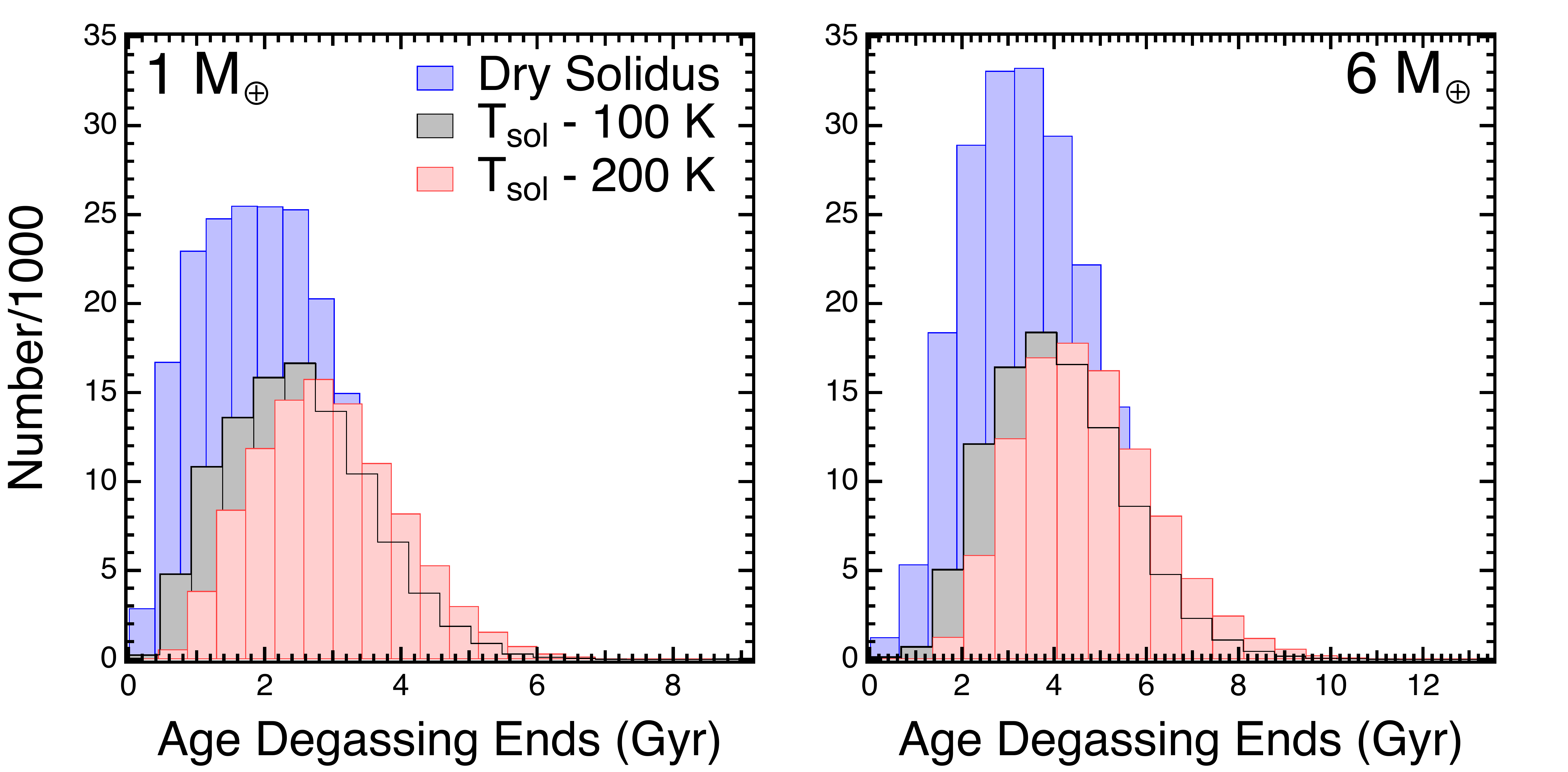}
    \caption{Histograms of the age when planetary degassing ends for 1 (left) and 6 (right) M$_{\oplus}$ planets with our default solidus (blue) and the solidus decreased arbitrarily by 100 (gray) and 200 (red) Kelvin. These histograms represent 200,000 random samplings of our parameters as described in Appendix \ref{sec:final_structure}.}
    \label{fig:solidus}
\end{figure}

\begin{figure}[h!]
    \centering
    \includegraphics[width=\linewidth]{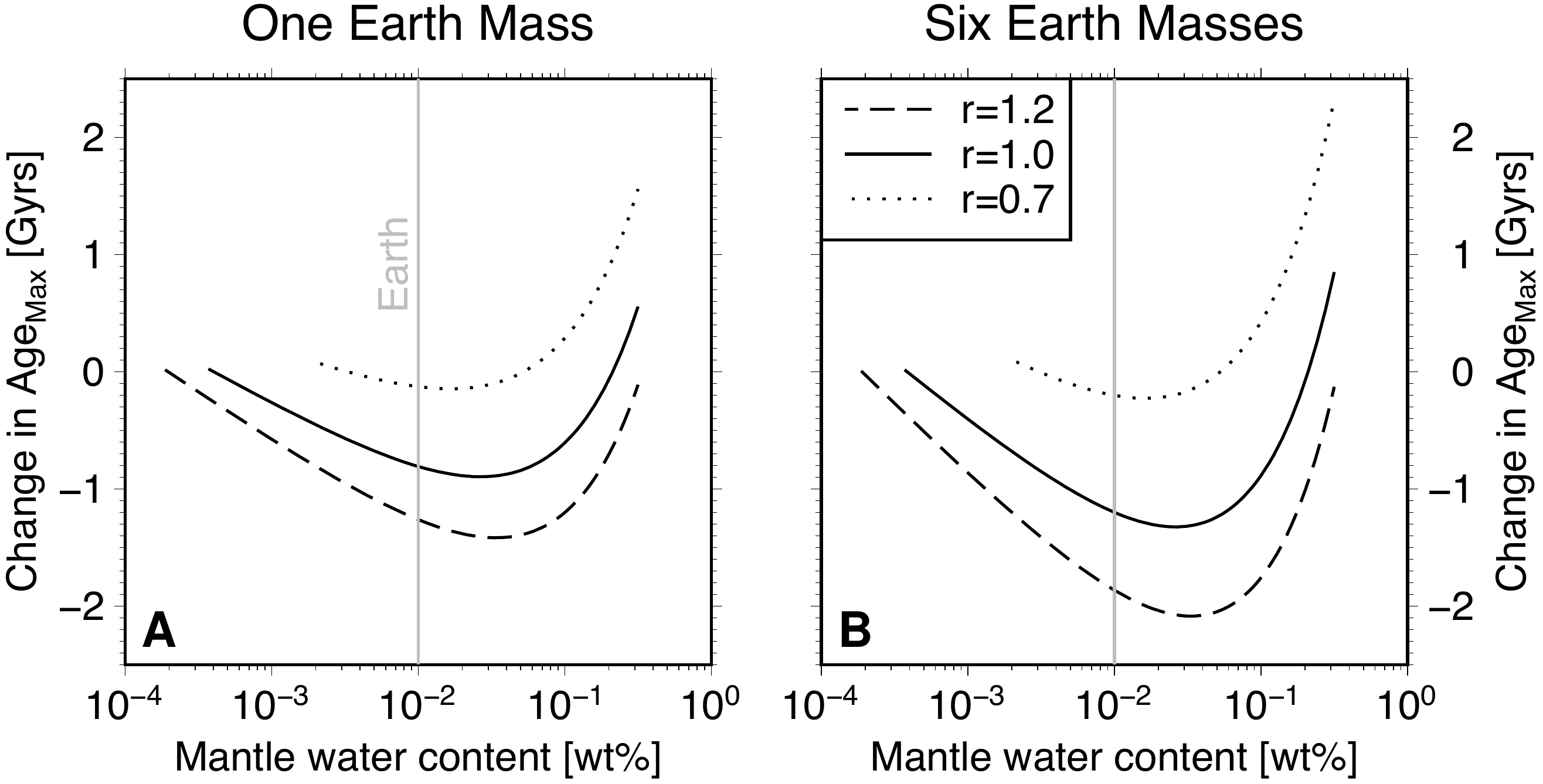}
    \caption{Change in  Age$^{\rm{Avg}}_{\rm{max}}$ in comparison to a dry planet baseline as a function of mantle water content. Three different values of the viscosity water dependence exponent, $r$, are used: $r=0.7$ (dotted line), $r=1$ (solid line), and $r=1.2$ (dashed line). Results for a 1 M$_{\oplus}$ (A) and a 6 M$_{\oplus}$ (B) are presented. The lower bound on mantle water concentration for each curve is determined by the mantle water concentration where the dry olivine and wet olivine flow laws are equal. This cross-over point depends on $r$, which is why curves with different $r$ do not start at the same point, in terms of mantle water concentration. Vertical gray line denotes Earth's mantle's estimated water content. } 
    \label{fig:water_effect}
\end{figure}

Ultimately, to fully capture these competing effects, new models incorporating the combined effects of water on viscosity and solidus, and how mantle water content evolves over time due to in- and outgassing from the mantle \citep[e.g., ][]{mcgovern1989,Crow11,Spaargaren2020}, would be needed. However, we can provide a first-order estimate using two additional model suites where solidus temperature is lowered first by 100 K everywhere, then by 200 K, and mantle reference viscosity is systematically varied as before. From the results of these model suites, we find that, on average, degassing lifetime increases by $\approx 0.43$ Gyr per 100 K decrease in solidus temperature for a 1 M$_{\oplus}$ planet, and by $\approx 0.60$ Gyr for 6 M$_{\oplus}$ (Figure \ref{fig:solidus}). We then use the diffusion creep viscosity flow laws from \citet{hirth2003} and the solidus parameterization from \citet{Katz2003} to estimate how Age$_{\rm{max}}$ would change with increasing water content. We start from the water content where the wet viscosity flow law is equal to the dry viscosity flow law, and increase water content from this point up to $\approx 0.3$ wt\% (Figure \ref{fig:water_effect}). For comparison, the Earth's mantle water content is estimated to be $10^{-2}$ wt\% \citep{Ohtani2019}. We do this for a range of water viscosity dependency exponents, $r$. 

For both 1 M$_{\oplus}$ and 6 M$_{\oplus}$ planets, increasing water content first decreases Age$^{\rm{Avg}}_{\rm{Max}}$ in comparison to a dry-planet baseline. Here, the effect of decreasing mantle reference viscosity due to a higher water content dominates, because solidus temperature does not decrease significantly from that of dry peridotite until water content approaches $\approx0.05-0.1$ wt\%. When water content increases beyond this point ($\chi_{m}\gtrsim0.05-0.1$ wt\%), the decrease in solidus temperature begins to dominate over the decrease in viscosity, and the change in Age$_{\mathrm{max}}$ compared to a dry planet baseline begins to increase. The higher $r$, the more the viscosity effect dominates, as mantle viscosity is more sensitive to water content in this case. For a 1 M$_{\oplus}$ planet, Age$^{\rm{Avg}}_{\rm{Max}}$ could decrease by up to $\approx 1.4$ Gyr when $\chi_m = 0.05$ wt\% and $r=1.2$. Meanwhile, Age$^{\rm{Avg}}_{\rm{Max}}$ could increase by up to $\approx 1.6$ Gyr for $\chi_m = 0.3$ wt\% and $r=0.7$. The effects are larger for a 6 M$_{\oplus}$ planet, with the biggest decrease in Age$^{\rm{Avg}}_{\rm{Max}}$ being $\approx 2.1$ Gyr and the biggest increase being $\approx 2.3$ Gyr. Note again that these estimates assume fixed mantle water concentrations, when in reality the water concentration would evolve over time as a result of in- and outgassing. On a stagnant-lid planet ingassing will be limited \citep[e.g., ][]{Spaargaren2020}, so mantle water content is likely to decrease over time; this means the changes to Age$_{\rm{max}}$ presented here are likely upper-limits. 

Mantle water content, then, is an important parameter that can have a large effect on $Age_{\rm{max}}$. Should a planet's mantle form with an Earth-like concentration ($10^{-2}$ wt\%), Age$^{\rm{Avg}}_{\rm{Max}}$ is roughly the same as, or lower than, the dry mantle model (Figure \ref{fig:TimeMass}). It is only those planets with mantles enriched in water relative to the Earth by a factor of 10-100 where Age$^{\rm{Avg}}_{\rm{Max}}$ increases, according to our simple estimates. Planets forming with this much water is a possibility \citep[e.g., ][]{raymond2004, Unter18}. However, despite a lower mantle solidus promoting melting on such water-rich planets, there are other factors that could hamper their habitability, and the possibility of detecting biosignatures on such planets, should they be inhabited. Exposed land may be essential for life on Earth, as weathering of subaerial land supplies critical nutrients to the oceans \citep[e.g.,  ][]{Maruyama2013,Dohm2015}, and some leading theories for the origin of life rely on wet-dry cycles, and therefore require exposed land \citep[e.g., ][]{Bada2018}. For an Earth-mass planet, complete flooding of the surface is expected to occur when the mass of surface water is roughly that of 8--40 oceans. This equates to $\sim0.2$\% of the mass of the planet, with this mass-fraction threshold decreasing with increasing planet mass due to higher gravity limiting surface topography \citep{Kite09, Cowan2014}. 

Furthermore, as surface water content increases, the pressure at the water-rock interface increases, forcing melting to occur at higher temperatures, limiting mantle degassing \citep{Kite09, Cowan2014}. This pressure effect would then counteract the effect of mantle water content on the solidus temperature, and therefore potentially limit the increase in Age$_{\rm{max}}$ at higher mantle water concentrations than in Figure \ref{fig:water_effect} (where ocean bottom pressure is assumed to be negligible for mantle melting). In fact, for a very thick ocean layer, pressures may be high enough to shut down degassing entirely \citep{Kite09, Kriss21}, by preventing melting or forcing melting to occur at such high pressures that the melt would no longer be buoyant. Moreover, even if life were to develop on an ocean-covered planet with mantle degassing, detectability of life on these planets is more difficult \citep{Glaser20}. This is because for planets with enough surface water to submerge continental crust, weathering rates of the biocritical element phosphorus is significantly reduced. With less bioavailable P, the biogenic flux of O$_2$ is reduced, becoming comparable to the abiotic O$_2$ flux \citep{Glaser20}. Even for those water-worlds where the pressure at the water-mantle boundary allows for mantle degassing, as these planets age degassing will wane. In this case, the source of reductants from the mantle will also decrease, allowing abiotic O$_2$ to build up in the atmosphere, potentially masking any biotic source of O$_2$ \citep{Kriss21}. Each of these factors will limit our ability to distinguish between biotic or abiotic O$_2$ through atmospheric observations, even though the planet may be habitable and hosting life.   

Therefore, it is unclear whether high water concentrations ($\gtrsim0.2$ wt\%) would actually promote long-lived mantle degassing, temperate climates and habitability. Instead, for water contents $\lesssim0.2$ wt\% where continents may still be exposed on small exoplanets, increasing water content generally leads to lower degassing lifetimes and a lower Age$_{\rm{max}}$. While more work is needed to quantify these differences as a function of planet mass, our results shown in Figure \ref{fig:TimeMass} may represent the most optimistic case for Age$_{\rm{max}}$ with regards to mantle water content.

\begin{thebibliography}{}
\expandafter\ifx\csname natexlab\endcsname\relax\def\natexlab#1{#1}\fi
\providecommand{\url}[1]{\href{#1}{#1}}
\providecommand{\dodoi}[1]{doi:~\href{http://doi.org/#1}{\nolinkurl{#1}}}
\providecommand{\doeprint}[1]{\href{http://ascl.net/#1}{\nolinkurl{http://ascl.net/#1}}}
\providecommand{\doarXiv}[1]{\href{https://arxiv.org/abs/#1}{\nolinkurl{https://arxiv.org/abs/#1}}}

\bibitem[{{Abbot} {et~al.}(2012){Abbot}, {Cowan}, \& {Ciesla}}]{Abbot2012}
{Abbot}, D.~S., {Cowan}, N.~B., \& {Ciesla}, F.~J. 2012, Astrophys. J., 756,
  178, \dodoi{10.1088/0004-637X/756/2/178}

\bibitem[{{Akeson} {et~al.}(2013){Akeson}, {Chen}, {Ciardi}, {Crane}, {Good},
  {Harbut}, {Jackson}, {Kane}, {Laity}, {Leifer}, {Lynn}, {McElroy}, {Papin},
  {Plavchan}, {Ram{\'\i}rez}, {Rey}, {von Braun}, {Wittman}, {Abajian}, {Ali},
  {Beichman}, {Beekley}, {Berriman}, {Berukoff}, {Bryden}, {Chan}, {Groom},
  {Lau}, {Payne}, {Regelson}, {Saucedo}, {Schmitz}, {Stauffer}, {Wyatt}, \&
  {Zhang}}]{Akeson13}
{Akeson}, R.~L., {Chen}, X., {Ciardi}, D., {et~al.} 2013, \pasp, 125, 989,
  \dodoi{10.1086/672273}

\bibitem[{{Ballmer} {et~al.}(2017){Ballmer}, {Houser}, {Hernlund},
  {Wentzcovitch}, \& {Hirose}}]{Ballmer2017}
{Ballmer}, M.~D., {Houser}, C., {Hernlund}, J.~W., {Wentzcovitch}, R.~M., \&
  {Hirose}, K. 2017, Nature Geoscience, 10, 236, \dodoi{10.1038/ngeo2898}

\bibitem[{Barnes {et~al.}(2009)Barnes, Jackson, Raymond, West, \&
  Greenberg}]{Barnes2009}
Barnes, R., Jackson, B., Raymond, S., West, A., \& Greenberg, R. 2009, The
  Astrophysical Journal, 695, 1006

\bibitem[{Bercovici {et~al.}(2015)Bercovici, Tackley, \&
  Ricard}]{Berco2015_treatise}
Bercovici, D., Tackley, P., \& Ricard, Y. 2015, in Treatise on Geophysics, ed.
  D.~Bercovici \& {G. Schubert (chief editor)}, Vol. 7, Mantle Dynamics (New
  York: Elsevier), 271--318

\bibitem[{{Borrelli} {et~al.}(2021){Borrelli}, {O'Rourke}, {Smrekar}, \&
  {Ostberg}}]{Borrelli2021}
{Borrelli}, M.~E., {O'Rourke}, J.~G., {Smrekar}, S.~E., \& {Ostberg}, C.~M.
  2021, Journal of Geophysical Research (Planets), 126, e06756,
  \dodoi{10.1029/2020JE006756}

\bibitem[{{Botelho} {et~al.}(2019){Botelho}, {Milone}, {Mel{\'e}ndez},
  {Bedell}, {Spina}, {Asplund}, {dos Santos}, {Bean}, {Ram{\'{\i}}rez}, {Yong},
  {Dreizler}, {Alves-Brito}, \& {Yana Galarza}}]{Bote18}
{Botelho}, R.~B., {Milone}, A.~d.~C., {Mel{\'e}ndez}, J., {et~al.} 2019,
  \mnras, 482, 1690, \dodoi{10.1093/mnras/sty2791}

\bibitem[{Breuer \& Moore(2015)}]{Breuer2015}
Breuer, D., \& Moore, W. 2015, in Treatise on Geophysics (Second Edition),
  second edition edn., ed. G.~Schubert (Oxford: Elsevier), 255 -- 305,
  \dodoi{https://doi.org/10.1016/B978-0-444-53802-4.00173-1}

\bibitem[{{Byrne} {et~al.}(2021){Byrne}, {Ghail}, {Cel{\^a}l {\c{S}}eng{\"o}r},
  {James}, {Klimczak}, \& {Solomon}}]{Byrne2021}
{Byrne}, P.~K., {Ghail}, R.~C., {Cel{\^a}l {\c{S}}eng{\"o}r}, A.~M., {et~al.}
  2021, Proceedings of the National Academy of Science, 118, 2025919118,
  \dodoi{10.1073/pnas.2025919118}

\bibitem[{Byrne \& Krishnamoorthy(2022)}]{Byrne2022}
Byrne, P.~K., \& Krishnamoorthy, S. 2022, Journal of Geophysical Research:
  Planets, 127, e2021JE007040, \dodoi{10.1029/2021JE007040}

\bibitem[{{Cowan} \& {Abbot}(2014)}]{Cowan2014}
{Cowan}, N.~B., \& {Abbot}, D.~S. 2014, Astrophys. J., 781, 27,
  \dodoi{10.1088/0004-637X/781/1/27}

\bibitem[{{Crisp}(1984)}]{Crisp1984}
{Crisp}, J.~A. 1984, J. Volcanol. Geotherm. Res., 20, 177,
  \dodoi{10.1016/0377-0273(84)90039-8}

\bibitem[{Crowley {et~al.}(2011)Crowley, G\'{e}rault, \& O'Connell}]{Crow11}
Crowley, J.~W., G\'{e}rault, M., \& O'Connell, R.~J. 2011, EPSL, 310, 380,
  \dodoi{10.1016/j.epsl.2011.08.035}

\bibitem[{{Davaille} {et~al.}(2017){Davaille}, {Smrekar}, \&
  {Tomlinson}}]{Davaille2017}
{Davaille}, A., {Smrekar}, S.~E., \& {Tomlinson}, S. 2017, Nature Geoscience,
  10, 349, \dodoi{10.1038/ngeo2928}

\bibitem[{{Dorn} {et~al.}(2018){Dorn}, {Noack}, \& {Rozel}}]{Dorn2018}
{Dorn}, C., {Noack}, L., \& {Rozel}, A.~B. 2018, Astron. Astrophys., 614, A18,
  \dodoi{10.1051/0004-6361/201731513}

\bibitem[{Driscoll \& Barnes(2015)}]{driscoll2015b}
Driscoll, P.~E., \& Barnes, R. 2015, Astrobiology, 15, 739

\bibitem[{{Fegley} \& {Prinn}(1989)}]{Fegl89}
{Fegley}, B., \& {Prinn}, R.~G. 1989, \nat, 337, 55, \dodoi{10.1038/337055a0}

\bibitem[{{Filiberto} {et~al.}(2020){Filiberto}, {Trang}, {Treiman}, \&
  {Gilmore}}]{Filiberto2020}
{Filiberto}, J., {Trang}, D., {Treiman}, A.~H., \& {Gilmore}, M.~S. 2020,
  Science Advances, 6, eaax7445, \dodoi{10.1126/sciadv.aax7445}

\bibitem[{{Foley}(2019)}]{Foley2019_stag}
{Foley}, B.~J. 2019, Astrophys. J., 875, 72, \dodoi{10.3847/1538-4357/ab0f31}

\bibitem[{Foley {et~al.}(2012)Foley, Bercovici, \& Landuyt}]{Foley2012}
Foley, B.~J., Bercovici, D., \& Landuyt, W. 2012, Earth Planet. Sci. Lett.,
  331–332, 281 , \dodoi{10.1016/j.epsl.2012.03.028}

\bibitem[{Foley \& Driscoll(2016)}]{Foley2016_review}
Foley, B.~J., \& Driscoll, P.~E. 2016, Geochem., Geophys., Geosyst., 17,
  \dodoi{10.1002/2015GC006210}

\bibitem[{Foley \& Smye(2018)}]{Foley2018_stag}
Foley, B.~J., \& Smye, A.~J. 2018, Astrobiology, 18, 873,
  \dodoi{10.1089/ast.2017.1695}

\bibitem[{{Frank} {et~al.}(2014){Frank}, {Meyer}, \& {Mojzsis}}]{Frank14}
{Frank}, E.~A., {Meyer}, B.~S., \& {Mojzsis}, S.~J. 2014, \icarus, 243, 274,
  \dodoi{10.1016/j.icarus.2014.08.031}

\bibitem[{{G{\"u}lcher} {et~al.}(2020){G{\"u}lcher}, {Gerya}, {Mont{\'e}si}, \&
  {Munch}}]{Gulcher2020}
{G{\"u}lcher}, A. J.~P., {Gerya}, T.~V., {Mont{\'e}si}, L. G.~J., \& {Munch},
  J. 2020, Nature Geoscience, 13, 547, \dodoi{10.1038/s41561-020-0606-1}

\bibitem[{{Haqq-Misra} {et~al.}(2016){Haqq-Misra}, {Kopparapu}, {Batalha},
  {Harman}, \& {Kasting}}]{Haqq2016}
{Haqq-Misra}, J., {Kopparapu}, R.~K., {Batalha}, N.~E., {Harman}, C.~E., \&
  {Kasting}, J.~F. 2016, Astrophys. J., 827, 120,
  \dodoi{10.3847/0004-637X/827/2/120}

\bibitem[{Hinkel {et~al.}(2014)Hinkel, Timmes, Young, Pagano, \&
  Turnbull}]{Hink14}
Hinkel, N.~R., Timmes, F., Young, P.~A., Pagano, M.~D., \& Turnbull, M.~C.
  2014, \aj, 148, 54.
\newblock \url{http://stacks.iop.org/1538-3881/148/i=3/a=54}

\bibitem[{Hirth \& Kohlstedt(1996)}]{hirth1996}
Hirth, G., \& Kohlstedt, D. 1996, Earth Planet. Sci. Lett., 144, 93

\bibitem[{Hirth \& Kohlstedt(2003)}]{hirth2003}
---. 2003, in Subduction Factory Mongraph, ed. J.~Eiler, Vol. 138 (Washington,
  DC: Am. Geophys. Union), 83--105

\bibitem[{{Kadoya} \& {Tajika}(2014)}]{Kadoya2014}
{Kadoya}, S., \& {Tajika}, E. 2014, Astrophys. J., 790, 107,
  \dodoi{10.1088/0004-637X/790/2/107}

\bibitem[{Karato(2011)}]{Karato2010}
Karato, S. 2011, Icarus, 212, 14 , \dodoi{10.1016/j.icarus.2010.12.005}

\bibitem[{Karato \& Wu(1993)}]{karato1993}
Karato, S., \& Wu, P. 1993, Science, 260, 771,
  \dodoi{10.1126/science.260.5109.771}

\bibitem[{{Kasting} \& {Catling}(2003)}]{Kasting2003}
{Kasting}, J.~F., \& {Catling}, D. 2003, Annu. Rev. Astron. Astrophys., 41,
  429, \dodoi{10.1146/annurev.astro.41.071601.170049}

\bibitem[{{Kasting} {et~al.}(1993){Kasting}, {Whitmire}, \&
  {Reynolds}}]{Kasting1993}
{Kasting}, J.~F., {Whitmire}, D.~P., \& {Reynolds}, R.~T. 1993, Icarus, 101,
  108, \dodoi{10.1006/icar.1993.1010}

\bibitem[{{Katz} {et~al.}(2003){Katz}, {Spiegelman}, \& {Langmuir}}]{Katz2003}
{Katz}, R.~F., {Spiegelman}, M., \& {Langmuir}, C.~H. 2003, Geochem., Geophys.,
  Geosyst., 4, 1073, \dodoi{10.1029/2002GC000433}

\bibitem[{{Kislyakova} {et~al.}(2017){Kislyakova}, {Noack}, {Johnstone},
  {Zaitsev}, {Fossati}, {Lammer}, {Khodachenko}, {Odert}, \&
  {G{\"u}del}}]{Kislyakova2017}
{Kislyakova}, K.~G., {Noack}, L., {Johnstone}, C.~P., {et~al.} 2017, Nature
  Astronomy, 1, 878, \dodoi{10.1038/s41550-017-0284-0}

\bibitem[{{Kite} \& {Ford}(2018)}]{Kite18}
{Kite}, E.~S., \& {Ford}, E.~B. 2018, \apj, 864, 75,
  \dodoi{10.3847/1538-4357/aad6e0}

\bibitem[{{Kite} {et~al.}(2009){Kite}, {Manga}, \& {Gaidos}}]{Kite09}
{Kite}, E.~S., {Manga}, M., \& {Gaidos}, E. 2009, \apj, 700, 1732,
  \dodoi{10.1088/0004-637X/700/2/1732}

\bibitem[{Kopparapu {et~al.}(2013)Kopparapu, Ramirez, Kasting, Eymet, Robinson,
  Mahadevan, Terrien, Domagal-Goldman, Meadows, \& Deshpande}]{kopparapu2013}
Kopparapu, R.~K., Ramirez, R., Kasting, J.~F., {et~al.} 2013, The Astrophysical
  Journal, 765, 131

\bibitem[{{Krissansen-Totton} {et~al.}(2021){Krissansen-Totton}, {Galloway},
  {Wogan}, {Dhaliwal}, \& {Fortney}}]{Kriss21}
{Krissansen-Totton}, J., {Galloway}, M.~L., {Wogan}, N., {Dhaliwal}, J.~K., \&
  {Fortney}, J.~J. 2021, \apj, 913, 107, \dodoi{10.3847/1538-4357/abf560}

\bibitem[{Kushiro {et~al.}(1968)Kushiro, Syono, \& Akimoto}]{kushiro1968}
Kushiro, I., Syono, Y., \& Akimoto, S. 1968, J. Geophys. Res., 73, 6023

\bibitem[{Marty \& Tolstikhin(1998)}]{Marty1998}
Marty, B., \& Tolstikhin, I.~N. 1998, Chem. Geol., 145, 233

\bibitem[{McGovern \& Schubert(1989)}]{mcgovern1989}
McGovern, P., \& Schubert, G. 1989, Earth and planetary science letters, 96, 27

\bibitem[{Menou(2015)}]{Menou2015}
Menou, K. 2015, Earth Planet. Sci. Lett., 429, 20,
  \dodoi{10.1016/j.epsl.2015.07.046}

\bibitem[{{Nimmo} {et~al.}(2020){Nimmo}, {Primack}, {Faber}, {Ramirez-Ruiz}, \&
  {Safarzadeh}}]{Nimmo2020}
{Nimmo}, F., {Primack}, J., {Faber}, S.~M., {Ramirez-Ruiz}, E., \&
  {Safarzadeh}, M. 2020, \apjl, 903, L37, \dodoi{10.3847/2041-8213/abc251}

\bibitem[{Noack \& Breuer(2014)}]{noack2014}
Noack, L., \& Breuer, D. 2014, Planetary and Space Science, 98, 41

\bibitem[{{O'Neill} \& {Lenardic}(2007)}]{ONeill2007}
{O'Neill}, C., \& {Lenardic}, A. 2007, Geophys. Res. Lett., 34, 19204,
  \dodoi{10.1029/2007GL030598}

\bibitem[{{Oosterloo} {et~al.}(2021){Oosterloo}, {H{\"o}ning}, {Kamp}, \& {van
  der Tak}}]{Ooster2021}
{Oosterloo}, M., {H{\"o}ning}, D., {Kamp}, I.~E.~E., \& {van der Tak}, F.~F.~S.
  2021, arXiv e-prints, arXiv:2103.09505.
\newblock \doarXiv{2103.09505}

\bibitem[{{O'Rourke} \& {Korenaga}(2015)}]{ORourke2015}
{O'Rourke}, J.~G., \& {Korenaga}, J. 2015, Icarus, 260, 128,
  \dodoi{10.1016/j.icarus.2015.07.009}

\bibitem[{{Sandwell} \& {Schubert}(1992)}]{Sandwell1992}
{Sandwell}, D.~T., \& {Schubert}, G. 1992, Science, 257, 766,
  \dodoi{10.1126/science.257.5071.766}

\bibitem[{{Schaefer} \& {Sasselov}(2015)}]{Schaefer2015}
{Schaefer}, L., \& {Sasselov}, D. 2015, Astrophys. J., 801, 40,
  \dodoi{10.1088/0004-637X/801/1/40}

\bibitem[{{Schulze} {et~al.}(2021){Schulze}, {Wang}, {Johnson}, {Gaudi},
  {Unterborn}, \& {Panero}}]{Schulze21}
{Schulze}, J.~G., {Wang}, J., {Johnson}, J.~A., {et~al.} 2021, \psj, 2, 113,
  \dodoi{10.3847/PSJ/abcaa8}

\bibitem[{{Shen} {et~al.}(1998){Shen}, {Solomon}, {Bjarnason}, \&
  {Wolfe}}]{Yang1998}
{Shen}, Y., {Solomon}, S.~C., {Bjarnason}, I.~T., \& {Wolfe}, C.~J. 1998,
  Nature, 395, 62, \dodoi{10.1038/25714}

\bibitem[{{Sleep} \& {Zahnle}(2001)}]{Sleep2001b}
{Sleep}, N.~H., \& {Zahnle}, K. 2001, J. Geophys. Res., 106, 1373,
  \dodoi{10.1029/2000JE001247}

\bibitem[{{Smrekar} {et~al.}(2010){Smrekar}, {Stofan}, {Mueller}, {Treiman},
  {Elkins-Tanton}, {Helbert}, {Piccioni}, \& {Drossart}}]{Smrekar2010}
{Smrekar}, S.~E., {Stofan}, E.~R., {Mueller}, N., {et~al.} 2010, Science, 328,
  605, \dodoi{10.1126/science.1186785}

\bibitem[{{Spaargaren} {et~al.}(2020){Spaargaren}, {Ballmer}, {Bower}, {Dorn},
  \& {Tackley}}]{Spaargaren2020}
{Spaargaren}, R.~J., {Ballmer}, M.~D., {Bower}, D.~J., {Dorn}, C., \&
  {Tackley}, P.~J. 2020, \aap, 643, A44, \dodoi{10.1051/0004-6361/202037632}

\bibitem[{Stamenkovic {et~al.}(2011)Stamenkovic, Breuer, \&
  Spohn}]{Stamenkovic2011}
Stamenkovic, V., Breuer, D., \& Spohn, T. 2011, Icarus, 216, 572 ,
  \dodoi{10.1016/j.icarus.2011.09.030}

\bibitem[{{Tackley} {et~al.}(2013){Tackley}, {Ammann}, {Brodholt}, {Dobson}, \&
  {Valencia}}]{Tackley2013}
{Tackley}, P.~J., {Ammann}, M., {Brodholt}, J.~P., {Dobson}, D.~P., \&
  {Valencia}, D. 2013, Icarus, 225, 50, \dodoi{10.1016/j.icarus.2013.03.013}

\bibitem[{{Thompson} \& {Gibson}(2000)}]{Thompson2000}
{Thompson}, R.~N., \& {Gibson}, S.~A. 2000, Nature, 407, 502,
  \dodoi{10.1038/35035058}

\bibitem[{{Unterborn} {et~al.}(2015){Unterborn}, {Johnson}, \&
  {Panero}}]{Unte15}
{Unterborn}, C.~T., {Johnson}, J.~A., \& {Panero}, W.~R. 2015, \apj, 806, 139,
  \dodoi{10.1088/0004-637X/806/1/139}

\bibitem[{{Unterborn} \& {Panero}(2019)}]{Unter19}
{Unterborn}, C.~T., \& {Panero}, W.~R. 2019, Journal of Geophysical Research
  (Planets), 124, 1704, \dodoi{10.1029/2018JE005844}

\bibitem[{{Valencia} {et~al.}(2007){Valencia}, {O'Connell}, \&
  {Sasselov}}]{Valencia2007b}
{Valencia}, D., {O'Connell}, R.~J., \& {Sasselov}, D.~D. 2007, Astrophys. J.,
  670, L45, \dodoi{10.1086/524012}

\bibitem[{{van Heck} \& {Tackley}(2011)}]{vanheck2011}
{van Heck}, H.~J., \& {Tackley}, P.~J. 2011, Earth Planet. Sci. Lett., 310,
  252, \dodoi{10.1016/j.epsl.2011.07.029}

\bibitem[{Walker {et~al.}(1981)Walker, Hayes, \& Kasting}]{walker1981}
Walker, J., Hayes, P., \& Kasting, J. 1981, J. Geophys. Res., 86, 9776,
  \dodoi{10.1029/JC086iC10p09776}

\bibitem[{{White} \& {McKenzie}(1995)}]{White1995}
{White}, R.~S., \& {McKenzie}, D. 1995, J. Geophys. Res., 100, 17,543,
  \dodoi{10.1029/95JB01585}

\bibitem[{{Zhao} {et~al.}(2009){Zhao}, {Zimmerman}, \& {Kohlstedt}}]{Zhao2009}
{Zhao}, Y.-H., {Zimmerman}, M.~E., \& {Kohlstedt}, D.~L. 2009, Earth Planet.
  Sci. Lett., 287, 229, \dodoi{10.1016/j.epsl.2009.08.006}

\end{thebibliography}

\clearpage

\begin{thebibliography}{}
\expandafter\ifx\csname natexlab\endcsname\relax\def\natexlab#1{#1}\fi
\providecommand{\url}[1]{\href{#1}{#1}}
\providecommand{\dodoi}[1]{doi:~\href{http://doi.org/#1}{\nolinkurl{#1}}}
\providecommand{\doeprint}[1]{\href{http://ascl.net/#1}{\nolinkurl{http://ascl.net/#1}}}
\providecommand{\doarXiv}[1]{\href{https://arxiv.org/abs/#1}{\nolinkurl{https://arxiv.org/abs/#1}}}

\bibitem[{{Agol} {et~al.}(2021){Agol}, {Dorn}, {Grimm}, {Turbet}, {Ducrot},
  {Delrez}, {Gillon}, {Demory}, {Burdanov}, {Barkaoui}, {Benkhaldoun},
  {Bolmont}, {Burgasser}, {Carey}, {de Wit}, {Fabrycky}, {Foreman-Mackey},
  {Haldemann}, {Hernandez}, {Ingalls}, {Jehin}, {Langford}, {Leconte},
  {Lederer}, {Luger}, {Malhotra}, {Meadows}, {Morris}, {Pozuelos}, {Queloz},
  {Raymond}, {Selsis}, {Sestovic}, {Triaud}, \& {Van Grootel}}]{Agol20}
{Agol}, E., {Dorn}, C., {Grimm}, S.~L., {et~al.} 2021, The Planetary Science
  Journal, 2, 1, \dodoi{10.3847/PSJ/abd022}

\bibitem[{Alexander(2005)}]{Alex05}
Alexander, C. M.~O. 2005, Meteoritics \& Planetary Science, 40, 943,
  \dodoi{10.1111/j.1945-5100.2005.tb00166.x}

\bibitem[{Bada \& Korenaga(2018)}]{Bada2018}
Bada, J.~L., \& Korenaga, J. 2018, Life, 8, \dodoi{10.3390/life8040055}

\bibitem[{{Barbuy} {et~al.}(2011){Barbuy}, {Spite}, {Hill}, {Primas}, {Plez},
  {Cayrel}, {Spite}, {Wanajo}, {Siqueira Mello}, {Andersen}, {Nordstr{\"o}m},
  {Beers}, {Bonifacio}, {Fran{\c c}ois}, \& {Molaro}}]{Barb11}
{Barbuy}, B., {Spite}, M., {Hill}, V., {et~al.} 2011, \aap, 534, A60,
  \dodoi{10.1051/0004-6361/201117450}

\bibitem[{{Beattie}(1993)}]{Beattie1993}
{Beattie}, P. 1993, Nature, 363, 63, \dodoi{10.1038/363063a0}

\bibitem[{{Beers} \& {Christlieb}(2005)}]{Beer05}
{Beers}, T.~C., \& {Christlieb}, N. 2005, \araa, 43, 531,
  \dodoi{10.1146/annurev.astro.42.053102.134057}

\bibitem[{{Botelho} {et~al.}(2019){Botelho}, {Milone}, {Mel{\'e}ndez},
  {Bedell}, {Spina}, {Asplund}, {dos Santos}, {Bean}, {Ram{\'{\i}}rez}, {Yong},
  {Dreizler}, {Alves-Brito}, \& {Yana Galarza}}]{Bote18}
{Botelho}, R.~B., {Milone}, A.~d.~C., {Mel{\'e}ndez}, J., {et~al.} 2019,
  \mnras, 482, 1690, \dodoi{10.1093/mnras/sty2791}

\bibitem[{Breuer \& Moore(2015)}]{Breuer2015}
Breuer, D., \& Moore, W. 2015, in Treatise on Geophysics (Second Edition),
  second edition edn., ed. G.~Schubert (Oxford: Elsevier), 255 -- 305,
  \dodoi{https://doi.org/10.1016/B978-0-444-53802-4.00173-1}

\bibitem[{{Burgasser} \& {Mamajek}(2017)}]{Burg17}
{Burgasser}, A.~J., \& {Mamajek}, E.~E. 2017, \apj, 845, 110,
  \dodoi{10.3847/1538-4357/aa7fea}

\bibitem[{{Cowan} \& {Abbot}(2014)}]{Cowan2014}
{Cowan}, N.~B., \& {Abbot}, D.~S. 2014, Astrophys. J., 781, 27,
  \dodoi{10.1088/0004-637X/781/1/27}

\bibitem[{Crowley {et~al.}(2011)Crowley, G\'{e}rault, \& O'Connell}]{Crow11}
Crowley, J.~W., G\'{e}rault, M., \& O'Connell, R.~J. 2011, EPSL, 310, 380,
  \dodoi{10.1016/j.epsl.2011.08.035}

\bibitem[{Desch {et~al.}(2020)Desch, Abbot, Krijt, Unterborn, Morard, \&
  Hartnett}]{Desch20}
Desch, S.~J., Abbot, D., Krijt, S., {et~al.} 2020, in Planetary Diversity,
  2514-3433 (IOP Publishing), 6--1 to 6--40,
  \dodoi{10.1088/2514-3433/abb4d9ch6}

\bibitem[{Desch {et~al.}(2018)Desch, Kalyaan, \& Alexander}]{Desc18}
Desch, S.~J., Kalyaan, A., \& Alexander, C. M.~O. 2018, The Astrophysical
  Journal Supplement Series, 238, 11, \dodoi{10.3847/1538-4365/aad95f}

\bibitem[{Dohm \& Maruyama(2015)}]{Dohm2015}
Dohm, J.~M., \& Maruyama, S. 2015, Geoscience Frontiers, 6, 95,
  \dodoi{https://doi.org/10.1016/j.gsf.2014.01.005}

\bibitem[{{Dorn} {et~al.}(2018){Dorn}, {Noack}, \& {Rozel}}]{Dorn2018}
{Dorn}, C., {Noack}, L., \& {Rozel}, A.~B. 2018, Astron. Astrophys., 614, A18,
  \dodoi{10.1051/0004-6361/201731513}

\bibitem[{{Driscoll} \& {Bercovici}(2014)}]{Driscoll2014}
{Driscoll}, P., \& {Bercovici}, D. 2014, Phys. Earth Planet. Inter., 236, 36,
  \dodoi{10.1016/j.pepi.2014.08.004}

\bibitem[{Duffy {et~al.}(2015)Duffy, Madhusudhan, \& Lee}]{Duffy2015_treatise}
Duffy, T., Madhusudhan, N., \& Lee, K. 2015, in Treatise on Geophysics (Second
  Edition), second edition edn., ed. G.~Schubert (Oxford: Elsevier), 149--178,
  \dodoi{https://doi.org/10.1016/B978-0-444-53802-4.00053-1}

\bibitem[{{Ellinger} {et~al.}(2012){Ellinger}, {Young}, {Fryer}, \&
  {Rockefeller}}]{Ellinger12}
{Ellinger}, C.~I., {Young}, P.~A., {Fryer}, C.~L., \& {Rockefeller}, G. 2012,
  \apj, 755, \dodoi{10.1088/0004-637X/755/2/160}

\bibitem[{{Fesen} \& {Milisavljevic}(2016)}]{Fesen16}
{Fesen}, R.~A., \& {Milisavljevic}, D. 2016, \apj, 818,
  \dodoi{10.3847/0004-637X/818/1/17}

\bibitem[{{Foley}(2019)}]{Foley2019_stag}
{Foley}, B.~J. 2019, Astrophys. J., 875, 72, \dodoi{10.3847/1538-4357/ab0f31}

\bibitem[{Foley {et~al.}(2020)Foley, Houser, Noack, \& Tosi}]{Foley2020_div}
Foley, B.~J., Houser, C., Noack, L., \& Tosi, N. 2020, in Planetary Diversity:
  Rocky Planet Processes and their Observational Signatures, ed. E.~J. Tasker,
  C.~Unterborn, M.~Laneuville, Y.~Fuji, S.~J. Desch, \& H.~E. Hartnett
  (Bristol, UK: IOP Publishing), 4--1 -- 4--60,
  \dodoi{doi:10.1088/2514-3433/abb4d9ch4}

\bibitem[{Foley \& Smye(2018{\natexlab{a}})}]{Foley2018_stag}
Foley, B.~J., \& Smye, A.~J. 2018{\natexlab{a}}, Astrobiology, 18, 873,
  \dodoi{10.1089/ast.2017.1695}

\bibitem[{Foley \& Smye(2018{\natexlab{b}})}]{Foley17}
---. 2018{\natexlab{b}}, Astrobiology, 18, 873, \dodoi{10.1089/ast.2017.1695}

\bibitem[{{Fraeman} \& {Korenaga}(2010)}]{Fraeman2010}
{Fraeman}, A.~A., \& {Korenaga}, J. 2010, Icarus, 210, 43,
  \dodoi{10.1016/j.icarus.2010.06.030}

\bibitem[{{Frank} {et~al.}(2014){Frank}, {Meyer}, \& {Mojzsis}}]{Frank14}
{Frank}, E.~A., {Meyer}, B.~S., \& {Mojzsis}, S.~J. 2014, \icarus, 243, 274,
  \dodoi{10.1016/j.icarus.2014.08.031}

\bibitem[{{Frebel} {et~al.}(2007){Frebel}, {Christlieb}, {Norris}, {Thom},
  {Beers}, \& {Rhee}}]{Freb07}
{Frebel}, A., {Christlieb}, N., {Norris}, J.~E., {et~al.} 2007, Astrophys. J.
  Lett., 660, L117, \dodoi{10.1086/518122}

\bibitem[{{Garc{\'\i}a} {et~al.}(2017){Garc{\'\i}a}, {Su{\'a}rez}, {Miceli},
  {Bocchino}, {Combi}, {Orlando}, \& {Sasaki}}]{Garcia17}
{Garc{\'\i}a}, F., {Su{\'a}rez}, A.~E., {Miceli}, M., {et~al.} 2017, \aap, 604,
  \dodoi{10.1051/0004-6361/201731418}

\bibitem[{{Glaser} {et~al.}(2020){Glaser}, {Hartnett}, {Desch}, {Unterborn},
  {Anbar}, {Buessecker}, {Fisher}, {Glaser}, {Kane}, {Lisse}, {Millsaps},
  {Neuer}, {O'Rourke}, {Santos}, {Walker}, \& {Zolotov}}]{Glaser20}
{Glaser}, D.~M., {Hartnett}, H.~E., {Desch}, S.~J., {et~al.} 2020, \apj, 893,
  163, \dodoi{10.3847/1538-4357/ab822d}

\bibitem[{Goriely \& Arnould(2001)}]{Gori01}
Goriely, S., \& Arnould, M. 2001, A\&A, 379, 1113,
  \dodoi{10.1051/0004-6361:20011368}

\bibitem[{{Grewal} {et~al.}(2020){Grewal}, {Dasgupta}, \& {Farnell}}]{Grew20}
{Grewal}, D.~S., {Dasgupta}, R., \& {Farnell}, A. 2020, \gca, 280, 281,
  \dodoi{10.1016/j.gca.2020.04.023}

\bibitem[{{Hammer} {et~al.}(2010){Hammer}, {Janka}, \& {M{\"u}ller}}]{Hammer10}
{Hammer}, N.~J., {Janka}, H.~T., \& {M{\"u}ller}, E. 2010, \apj, 714, 1371,
  \dodoi{10.1088/0004-637X/714/2/1371}

\bibitem[{{Hansen} {et~al.}(2017){Hansen}, {Simon}, {Marshall}, {Li},
  {Carollo}, {DePoy}, {Nagasawa}, {Bernstein}, {Drlica-Wagner}, {Abdalla},
  {Allam}, {Annis}, {Bechtol}, {Benoit-L{\'e}vy}, {Brooks}, {Buckley-Geer},
  {Carnero Rosell}, {Carrasco Kind}, {Carretero}, {Cunha}, {da Costa}, {Desai},
  {Eifler}, {Fausti Neto}, {Flaugher}, {Frieman}, {Garc{\'{\i}}a-Bellido},
  {Gaztanaga}, {Gerdes}, {Gruen}, {Gruendl}, {Gschwend}, {Gutierrez}, {James},
  {Krause}, {Kuehn}, {Kuropatkin}, {Lahav}, {Miquel}, {Plazas}, {Romer},
  {Sanchez}, {Santiago}, {Scarpine}, {Smith}, {Soares-Santos}, {Sobreira},
  {Suchyta}, {Swanson}, {Tarle}, {Walker}, \& {DES Collaboration}}]{hansen17}
{Hansen}, T.~T., {Simon}, J.~D., {Marshall}, J.~L., {et~al.} 2017, \apj, 838,
  44, \dodoi{10.3847/1538-4357/aa634a}

\bibitem[{{Hart} \& {Brooks}(1974)}]{Hart1974}
{Hart}, S.~E., \& {Brooks}, C. 1974, Geochim. Cosmchim. Acta, 38, 1799,
  \dodoi{10.1016/0016-7037(74)90163-X}

\bibitem[{{Hauck} \& {Phillips}(2002)}]{Hauck2002}
{Hauck}, S.~A., \& {Phillips}, R.~J. 2002, J. Geophys. Res. Planets, 107, 6,
  \dodoi{10.1029/2001JE001801}

\bibitem[{{Hauri} {et~al.}(2006){Hauri}, {Gaetani}, \& {Green}}]{Hauri2006}
{Hauri}, E.~H., {Gaetani}, G.~A., \& {Green}, T.~H. 2006, Earth Planet. Sci.
  Lett., 248, 715, \dodoi{10.1016/j.epsl.2006.06.014}

\bibitem[{{Hayashi}(1981)}]{Hayashi81}
{Hayashi}, C. 1981, Progress of Theoretical Physics Supplement, 70, 35,
  \dodoi{10.1143/PTPS.70.35}

\bibitem[{{Hayworth} \& {Foley}(2020)}]{Hayworth2020}
{Hayworth}, B. P.~C., \& {Foley}, B.~J. 2020, Astrophys. J. Lett., 902, L10,
  \dodoi{10.3847/2041-8213/abb882}

\bibitem[{Hinkel {et~al.}(2014)Hinkel, Timmes, Young, Pagano, \&
  Turnbull}]{Hink14}
Hinkel, N.~R., Timmes, F., Young, P.~A., Pagano, M.~D., \& Turnbull, M.~C.
  2014, \aj, 148, 54.
\newblock \url{http://stacks.iop.org/1538-3881/148/i=3/a=54}

\bibitem[{{Hinkel} \& {Unterborn}(2018)}]{Hink18}
{Hinkel}, N.~R., \& {Unterborn}, C.~T. 2018, \apj, 853, 83,
  \dodoi{10.3847/1538-4357/aaa5b4}

\bibitem[{Hirose {et~al.}(2013)Hirose, Labrosse, \& Hernlund}]{Hiro13}
Hirose, K., Labrosse, S., \& Hernlund, J. 2013, Annual Review of Earth and
  Planetary Sciences, 41, 657, \dodoi{10.1146/annurev-earth-050212-124007}

\bibitem[{Hirth \& Kohlstedt(2003)}]{hirth2003}
Hirth, G., \& Kohlstedt, D. 2003, in Subduction Factory Mongraph, ed. J.~Eiler,
  Vol. 138 (Washington, DC: Am. Geophys. Union), 83--105

\bibitem[{Karato \& Wu(1993)}]{karato1993}
Karato, S., \& Wu, P. 1993, Science, 260, 771,
  \dodoi{10.1126/science.260.5109.771}

\bibitem[{{Katz} {et~al.}(2003){Katz}, {Spiegelman}, \& {Langmuir}}]{Katz2003}
{Katz}, R.~F., {Spiegelman}, M., \& {Langmuir}, C.~H. 2003, Geochem., Geophys.,
  Geosyst., 4, 1073, \dodoi{10.1029/2002GC000433}

\bibitem[{{Kite} {et~al.}(2009){Kite}, {Manga}, \& {Gaidos}}]{Kite09}
{Kite}, E.~S., {Manga}, M., \& {Gaidos}, E. 2009, \apj, 700, 1732,
  \dodoi{10.1088/0004-637X/700/2/1732}

\bibitem[{{Korenaga}(2009)}]{korenaga2009}
{Korenaga}, J. 2009, Geophys. J. Int., 179, 154,
  \dodoi{10.1111/j.1365-246X.2009.04272.x}

\bibitem[{{Krissansen-Totton} \& {Catling}(2017)}]{Krissansen-Totton2017}
{Krissansen-Totton}, J., \& {Catling}, D.~C. 2017, Nature Communications, 8,
  15423, \dodoi{10.1038/ncomms15423}

\bibitem[{{Krissansen-Totton} {et~al.}(2021){Krissansen-Totton}, {Galloway},
  {Wogan}, {Dhaliwal}, \& {Fortney}}]{Kriss21}
{Krissansen-Totton}, J., {Galloway}, M.~L., {Wogan}, N., {Dhaliwal}, J.~K., \&
  {Fortney}, J.~J. 2021, \apj, 913, 107, \dodoi{10.3847/1538-4357/abf560}

\bibitem[{{Kuchner}(2004)}]{Kuch04}
{Kuchner}, M.~J. 2004, \apj, 612, 1147, \dodoi{10.1086/422577}

\bibitem[{Kushiro {et~al.}(1968)Kushiro, Syono, \& Akimoto}]{kushiro1968}
Kushiro, I., Syono, Y., \& Akimoto, S. 1968, J. Geophys. Res., 73, 6023

\bibitem[{{Larsson} {et~al.}(2013){Larsson}, {Fransson}, {Kjaer}, {Jerkstrand},
  {Kirshner}, {Leibundgut}, {Lundqvist}, {Mattila}, {McCray}, {Sollerman},
  {Spyromilio}, \& {Wheeler}}]{Larsson13}
{Larsson}, J., {Fransson}, C., {Kjaer}, K., {et~al.} 2013, \apj, 768,
  \dodoi{10.1088/0004-637X/768/1/89}

\bibitem[{{Leonard} {et~al.}(2000){Leonard}, {Filippenko}, {Barth}, \&
  {Matheson}}]{Leonard00}
{Leonard}, D.~C., {Filippenko}, A.~V., {Barth}, A.~J., \& {Matheson}, T. 2000,
  \apj, 536, 239, \dodoi{10.1086/308910}

\bibitem[{Lodders(2003)}]{Lodd03}
Lodders, K. 2003, \apj, 1220, \dodoi{10.1086/375492}

\bibitem[{Lodders {et~al.}(2009)Lodders, Palme, \& Gail}]{Lodd09}
Lodders, K., Palme, H., \& Gail, H.~P. 2009, {4.4 Abundances of the elements in
  the Solar System}, Vol.~4B (Berlin, Heidelberg: Springer Berlin Heidelberg),
  712--770, \dodoi{10.1007/978-3-540-88055-4_34}

\bibitem[{Marty \& Tolstikhin(1998)}]{Marty1998}
Marty, B., \& Tolstikhin, I.~N. 1998, Chem. Geol., 145, 233

\bibitem[{Maruyama {et~al.}(2013)Maruyama, Ikoma, Genda, Hirose, Yokoyama, \&
  Santosh}]{Maruyama2013}
Maruyama, S., Ikoma, M., Genda, H., {et~al.} 2013, Geoscience Frontiers, 4, 141

\bibitem[{McDonough(2003)}]{McD03}
McDonough, W.~F. 2003, {Compositional Model for the Earth's Core} (Elsevier),
  547--568, \dodoi{10.1016/b0-08-043751-6/02015-6}

\bibitem[{McGovern \& Schubert(1989)}]{mcgovern1989}
McGovern, P., \& Schubert, G. 1989, Earth and planetary science letters, 96, 27

\bibitem[{{Mitrovica} \& {Forte}(2004)}]{Mitrovica2004}
{Mitrovica}, J.~X., \& {Forte}, A.~M. 2004, Earth Planet. Sci. Lett., 225, 177,
  \dodoi{10.1016/j.epsl.2004.06.005}

\bibitem[{{Moranchel-Basurto} {et~al.}(2017){Moranchel-Basurto},
  {Vel{\'a}zquez}, {Giacani}, {Toledo-Roy}, {Schneiter}, {De Colle}, \&
  {Esquivel}}]{MoranchelBasurto17}
{Moranchel-Basurto}, A., {Vel{\'a}zquez}, P.~F., {Giacani}, E., {et~al.} 2017,
  \mnras, 472, 2117, \dodoi{10.1093/mnras/stx2086}

\bibitem[{{Morschhauser} {et~al.}(2011){Morschhauser}, {Grott}, \&
  {Breuer}}]{Morschhauser2011}
{Morschhauser}, A., {Grott}, M., \& {Breuer}, D. 2011, Icarus, 212, 541,
  \dodoi{10.1016/j.icarus.2010.12.028}

\bibitem[{{Noack} {et~al.}(2016){Noack}, {H{\"o}ning}, {Rivoldini},
  {Heistracher}, {Zimov}, {Journaux}, {Lammer}, {Van Hoolst}, \&
  {Bredeh{\"o}ft}}]{Noack2016}
{Noack}, L., {H{\"o}ning}, D., {Rivoldini}, A., {et~al.} 2016, Icarus, 277,
  215, \dodoi{10.1016/j.icarus.2016.05.009}

\bibitem[{Ohtani(2019)}]{Ohtani2019}
Ohtani, E. 2019, National Science Review, 7, 224, \dodoi{10.1093/nsr/nwz071}

\bibitem[{{Ouellette} {et~al.}(2007){Ouellette}, {Desch}, \&
  {Hester}}]{Ouellette07}
{Ouellette}, N., {Desch}, S.~J., \& {Hester}, J.~J. 2007, \apj, 662, 1268,
  \dodoi{10.1086/518102}

\bibitem[{Palme \& O'Neill(2003)}]{palme2003}
Palme, H., \& O'Neill, H. S.~C. 2003, Treatise on geochemistry, 2, 1

\bibitem[{{Pan} {et~al.}(2012){Pan}, {Desch}, {Scannapieco}, \& {Timmes}}]{Pan}
{Pan}, L., {Desch}, S.~J., {Scannapieco}, E., \& {Timmes}, F.~X. 2012, \apj,
  756, 102, \dodoi{10.1088/0004-637X/756/1/102}

\bibitem[{{Raymond} {et~al.}(2004){Raymond}, {Quinn}, \&
  {Lunine}}]{raymond2004}
{Raymond}, S.~N., {Quinn}, T., \& {Lunine}, J.~I. 2004, Icarus, 168, 1,
  \dodoi{10.1016/j.icarus.2003.11.019}

\bibitem[{{Reese} {et~al.}(1998){Reese}, {Solomatov}, \& {Moresi}}]{Reese1998}
{Reese}, C.~C., {Solomatov}, V.~S., \& {Moresi}, L.-N. 1998, J. Geophys. Res.,
  103, 13643, \dodoi{10.1029/98JE01047}

\bibitem[{{Reese} {et~al.}(1999){Reese}, {Solomatov}, \& {Moresi}}]{Reese1999}
---. 1999, Icarus, 139, 67, \dodoi{10.1006/icar.1999.6088}

\bibitem[{{Reese} {et~al.}(2007){Reese}, {Solomatov}, \& {Orth}}]{Reese2007}
{Reese}, C.~C., {Solomatov}, V.~S., \& {Orth}, C.~P. 2007, J. Geophys. Res.
  Planets, 112, E04S04, \dodoi{10.1029/2006JE002782}

\bibitem[{{Reilly} {et~al.}(2017){Reilly}, {Maund}, {Baade}, {Wheeler},
  {H{\"o}flich}, {Spyromilio}, {Patat}, \& {Wang}}]{Reilly17}
{Reilly}, E., {Maund}, J.~R., {Baade}, D., {et~al.} 2017, \mnras, 470, 1491,
  \dodoi{10.1093/mnras/stx1228}

\bibitem[{{Roederer} {et~al.}(2009){Roederer}, {Kratz}, {Frebel}, {Christlieb},
  {Pfeiffer}, {Cowan}, \& {Sneden}}]{Roed09}
{Roederer}, I.~U., {Kratz}, K.-L., {Frebel}, A., {et~al.} 2009, \apj, 698,
  1963, \dodoi{10.1088/0004-637X/698/2/1963}

\bibitem[{{Schubert} {et~al.}(1979){Schubert}, {Cassen}, \&
  {Young}}]{Schubert1979}
{Schubert}, G., {Cassen}, P., \& {Young}, R.~E. 1979, Icarus, 38, 192,
  \dodoi{10.1016/0019-1035(79)90178-7}

\bibitem[{{Shimansky} {et~al.}(2003){Shimansky}, {Bikmaev}, {Galeev},
  {Shimanskaya}, {Ivanova}, {Sakhibullin}, {Musaev}, \&
  {Galazutdinov}}]{Shimansky03}
{Shimansky}, V.~V., {Bikmaev}, I.~F., {Galeev}, A.~I., {et~al.} 2003, Astronomy
  Reports, 47, 750, \dodoi{10.1134/1.1611216}

\bibitem[{Sleep \& Zahnle(2001)}]{sleep2001}
Sleep, N., \& Zahnle, K. 2001, Journal of Geophysical Research, 106, 1

\bibitem[{{Solomatov} \& {Moresi}(2000)}]{Solomatov2000b}
{Solomatov}, V.~S., \& {Moresi}, L.-N. 2000, J. Geophys. Res., 105, 21795,
  \dodoi{10.1029/2000JB900197}

\bibitem[{{Spaargaren} {et~al.}(2020){Spaargaren}, {Ballmer}, {Bower}, {Dorn},
  \& {Tackley}}]{Spaargaren2020}
{Spaargaren}, R.~J., {Ballmer}, M.~D., {Bower}, D.~J., {Dorn}, C., \&
  {Tackley}, P.~J. 2020, \aap, 643, A44, \dodoi{10.1051/0004-6361/202037632}

\bibitem[{{Spohn}(1991)}]{Spohn1991}
{Spohn}, T. 1991, Icarus, 90, 222, \dodoi{10.1016/0019-1035(91)90103-Z}

\bibitem[{{Stevenson} {et~al.}(1983){Stevenson}, {Spohn}, \&
  {Schubert}}]{Stevenson1983}
{Stevenson}, D.~J., {Spohn}, T., \& {Schubert}, G. 1983, Icarus, 54, 466,
  \dodoi{10.1016/0019-1035(83)90241-5}

\bibitem[{Takahashi \& Kushiro(1983)}]{Takahashi1983}
Takahashi, E., \& Kushiro, I. 1983, American Mineralogist, 68, 859

\bibitem[{Taylor \& McLennan(1985)}]{taylor1985}
Taylor, S., \& McLennan, S. 1985, The Continental Crust: Its Composition and
  Evolution (Oxford,UK: Blackwell), 312 pp

\bibitem[{Tosi {et~al.}(2017)Tosi, Godolt, Stracke, Ruedas, Grenfell,
  H{\"o}ning, Nikolaou, Plesa, Breuer, \& Spohn}]{Tosi2017}
Tosi, N., Godolt, M., Stracke, B., {et~al.} 2017, Astron. Astrophys., 605, A71,
  \dodoi{10.1051/0004-6361/201730728}

\bibitem[{{Tosi} {et~al.}(2017){Tosi}, {Godolt}, {Stracke}, {Ruedas},
  {Grenfell}, {H{\"o}ning}, {Nikolaou}, {Plesa}, {Breuer}, \& {Spohn}}]{Tosi17}
{Tosi}, N., {Godolt}, M., {Stracke}, B., {et~al.} 2017, \aap, 605, A71,
  \dodoi{10.1051/0004-6361/201730728}

\bibitem[{Turcotte \& Schubert(2002)}]{turc1982}
Turcotte, D., \& Schubert, G. 2002, Geodynamics, 2nd edn. (New York: Cambridge
  Univ. Press)

\bibitem[{{Unterborn} {et~al.}(2018){Unterborn}, {Desch}, {Hinkel}, \&
  {Lorenzo}}]{Unter18}
{Unterborn}, C.~T., {Desch}, S.~J., {Hinkel}, N.~R., \& {Lorenzo}, A. 2018,
  Nature Astronomy, 2, 297, \dodoi{10.1038/s41550-018-0411-6}

\bibitem[{{Unterborn} {et~al.}(2015){Unterborn}, {Johnson}, \&
  {Panero}}]{Unterborn2015}
{Unterborn}, C.~T., {Johnson}, J.~A., \& {Panero}, W.~R. 2015, Astrophys. J.,
  806, 139, \dodoi{10.1088/0004-637X/806/1/139}

\bibitem[{Valencia {et~al.}(2006)Valencia, O'Connell, \& Sasselov}]{Vale06}
Valencia, D., O'Connell, R.~J., \& Sasselov, D. 2006, Icarus, 181, 545,
  \dodoi{10.1016/j.icarus.2005.11.021}

\bibitem[{{Valencia} {et~al.}(2007){Valencia}, {O'Connell}, \&
  {Sasselov}}]{Valencia2007b}
{Valencia}, D., {O'Connell}, R.~J., \& {Sasselov}, D.~D. 2007, Astrophys. J.,
  670, L45, \dodoi{10.1086/524012}

\bibitem[{{Ward-Duong} {et~al.}(2018){Ward-Duong}, {Patience}, {Bulger}, {van
  der Plas}, {M{\'e}nard}, {Pinte}, {Jackson}, {Bryden}, {Turner}, {Harvey},
  {Hales}, \& {De Rosa}}]{wardduong18}
{Ward-Duong}, K., {Patience}, J., {Bulger}, J., {et~al.} 2018, \aj, 155, 54,
  \dodoi{10.3847/1538-3881/aaa128}

\bibitem[{{Warren}(2016)}]{Warren2016}
{Warren}, J.~M. 2016, Lithos, 248, 193, \dodoi{10.1016/j.lithos.2015.12.023}

\bibitem[{Wasson {et~al.}(1988)Wasson, Kallemeyn, Runcorn, Turner, \&
  Woolfson}]{Wass88}
Wasson, J.~T., Kallemeyn, G.~W., Runcorn, S.~K., Turner, G., \& Woolfson, M.~M.
  1988, Philosophical Transactions of the Royal Society of London. Series A,
  Mathematical and Physical Sciences, 325, 535, \dodoi{10.1098/rsta.1988.0066}

\bibitem[{{Willingale} {et~al.}(2002){Willingale}, {Bleeker}, {van der Heyden},
  {Kaastra}, \& {Vink}}]{Willingale02}
{Willingale}, R., {Bleeker}, J.~A.~M., {van der Heyden}, K.~J., {Kaastra},
  J.~S., \& {Vink}, J. 2002, \aap, 381, 1039,
  \dodoi{10.1051/0004-6361:20011614}

\bibitem[{{Wongwathanarat} {et~al.}(2015){Wongwathanarat}, {M{\"u}ller}, \&
  {Janka}}]{Wongwathanarat15}
{Wongwathanarat}, A., {M{\"u}ller}, E., \& {Janka}, H.~T. 2015, \aap, 577,
  \dodoi{10.1051/0004-6361/201425025}

\bibitem[{{Young} {et~al.}(2009){Young}, {Ellinger}, {Arnett}, {Fryer}, \&
  {Rockefeller}}]{Young09}
{Young}, P.~A., {Ellinger}, C.~I., {Arnett}, D., {Fryer}, C.~L., \&
  {Rockefeller}, G. 2009, \apj, 699, 938, \dodoi{10.1088/0004-637X/699/2/938}

\bibitem[{{Zhang} {et~al.}(2006){Zhang}, {Gehren}, {Butler}, {Shi}, \&
  {Zhao}}]{Zhang06}
{Zhang}, H.~W., {Gehren}, T., {Butler}, K., {Shi}, J.~R., \& {Zhao}, G. 2006,
  \aap, 457, 645, \dodoi{10.1051/0004-6361:20064909}

\bibitem[{{Zhou} {et~al.}(2016){Zhou}, {Chen}, {Safi-Harb}, {Zhou}, {Sun},
  {Zhang}, \& {Zhang}}]{Zhou16}
{Zhou}, P., {Chen}, Y., {Safi-Harb}, S., {et~al.} 2016, \apj, 831,
  \dodoi{10.3847/0004-637X/831/2/192}

\end{thebibliography}
\end{document}